%% file: qsym2.tex
\pdfoutput=1
\def\notikzex{1}
\def\twocolumnmode{1}

\documentclass[%
  journal=jctcce,%
  manuscript=article,%
  layout=twocolumn%
]{achemso}

\input{preamble}
\usepackage[T1]{fontenc}

\DeclareMathOperator{\tr}{tr}

\DeclareMathOperator{\hash}{hash}
\DeclareMathOperator{\rank}{rank}
\newcommand*{\T}{\mathsf{T}}
\newcommand*{\D}{\mathrm{d}}

\newcommand*{\qsymsq}{\textsc{QSym\textsuperscript{2}}}

\author{Bang C. Huynh}
\email{bang.huynh@nottingham.ac.uk}
\affiliation{School of Chemistry, University of Nottingham, Nottingham NG7 2RD, United Kingdom}
\author{Meilani Wibowo-Teale}
\affiliation{School of Chemistry, University of Nottingham, Nottingham NG7 2RD, United Kingdom}
\author{Andrew M. Wibowo-Teale}
\affiliation{School of Chemistry, University of Nottingham, Nottingham NG7 2RD, United Kingdom}
\alsoaffiliation{Hylleraas Centre for Quantum Molecular Sciences, Department of Chemistry, University of Oslo, P.O. Box 1033 Blindern, N-0315 Oslo, Norway}

\title{\qsymsq{}: A Quantum Symbolic Symmetry Analysis Program for Electronic Structure}

\begin{document}

  \input{abstract/abstract}
  \input{introduction/introduction}
  \input{theory/theory}
  \input{illustrations/illustrations}
  \input{conclusion/conclusion}

  \section*{Acknowledgments}
    We acknowledge financial support from the European Research Council under the European Union’s H2020 research and innovation program / ERC Consolidator Grant topDFT [grant number 772259], and also financial support from the Norwegian Research Council through the CoE Hylleraas Center for Quantum Molecular Sciences [grant number 262695].

  \section*{Supporting Information}
    \paragraph{Source Code.}
      The main repository for \qsymsq{} can be accessed at \url{https://gitlab.com/bangconghuynh/qsym2} (accessed November 16, 2023).
      The documentations for \qsymsq{} can be found at \url{https://qsym2.dev/} (accessed November 16, 2023).
      The source code for \qsymsq{} \texttt{v0.7.0} released at the time of publication of this article was deposited permanently at \url{https://doi.org/10.5281/zenodo.8384779} (accessed November 16, 2023).

    \paragraph{Supporting Information.}
      This information is available free of charge via the Internet at \url{http://pubs.acs.org}.

  \bibliography{bib/qsym2}

  \clearpage

  \begin{figure*}
    \centering
    \includegraphics[width=.6\textwidth]{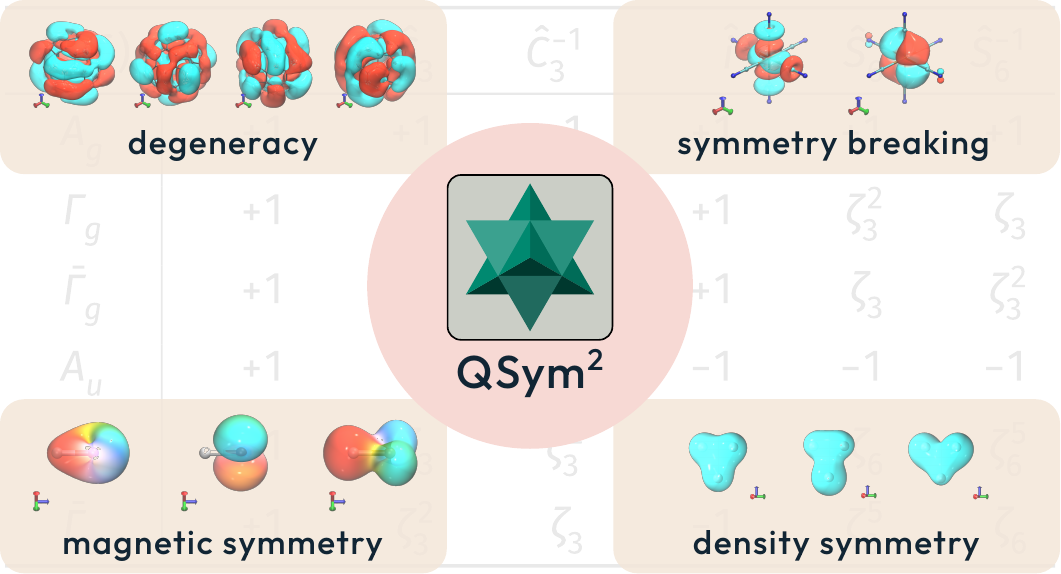}
    \caption*{Table of Contents Graphic.}
  \end{figure*}

\end{document}


\tableofcontents
  \clearpage
  \subimport{suppinfo/}{symopstruct/symopstruct.tex}
  \subimport{suppinfo/}{chartabgen/chartabgen.tex}

  \subimport{suppinfo/}{densym/densym.tex}

  \bibliography{suppinfo/bib/qsym2}

%% file: preamble.tex
\usepackage[version=4]{mhchem}

\usepackage{siunitx}
\DeclareSIUnit{\atomicunit}{a.u.}
\DeclareSIUnit\atomicmassunit{u}
\DeclareSIUnit\angstrom{\text {Å}}

\usepackage{%
  amsthm,%
  amssymb,%
  mathrsfs,%
  mathtools,%
  gensymb,%
  braket,%
  mleftright,%
  nicefrac,%
  interval%
}
\intervalconfig{%
  soft open fences%
}

\mleftright

\usepackage{bm}

\usepackage[usenames, svgnames, x11names, rgb]{xcolor}

\usepackage{graphicx}
\usepackage{grffile}

\usepackage{pgfplots}

\usetikzlibrary{external}
\newif\iftikzex
\ifdefined\notikzex
  \tikzexfalse
\else
  \tikzextrue
\fi

\iftikzex
  \pgfplotsset{compat=1.18}

  \usetikzlibrary{
    calc,
    luamath,
    positioning,
    decorations.pathreplacing,
    backgrounds,
    arrows.meta,
    pgfplots.groupplots,
    pgfplots.patchplots,
    3d,
  }

  \usepgfplotslibrary[colormaps]

  \pgfplotscreateplotcyclelist{coloronly}{
    {red},
    {blue},
    {black!60!green},
    {black!20!orange},
    {black!30!blue},
    {green!30!brown},
    {blue!40!red},
    {IndianRed},
    {black!40!yellow},
    {Magenta4},
    {DodgerBlue4},
    {Maroon4},
    {Firebrick4},
    {Cyan4},
  }
\fi
\tikzexfalse

\usepackage{xifthen}

\usepackage{caption, subcaption}

\usepackage{%
  booktabs,%
  array,%
  adjustbox,%
  makecell,%
  multirow,%
  bigdelim%
}
\newcolumntype{M}[1]{>{\centering\arraybackslash}m{#1}}
\newcolumntype{R}[1]{>{\raggedleft\arraybackslash}m{#1}}

\usepackage[inline]{enumitem}

\theoremstyle{plain}

\theoremstyle{remark}

\usepackage[%
  hyperfootnotes=false,%
  hypertexnames=false,%
  hidelinks,%
]{hyperref}
\hypersetup{%
  bookmarksnumbered=true,%
  colorlinks=true,%
  citecolor=teal,%
  linkcolor=Blue4,%
}

\usepackage{mciteplus}


%% file: abstract/abstract.tex
\begin{abstract}
  Symmetry provides a powerful machinery to classify, interpret, and understand quantum-mechanical theories and results.
  However, most contemporary quantum chemistry packages lack the ability to handle degeneracy and symmetry breaking effects, especially in non-Abelian groups, nor are they able to characterize symmetry in the presence of external magnetic or electric fields.
  In this article, a program written in Rust entitled \qsymsq{} that makes use of group and representation theories to provide symmetry analysis for a wide range of quantum-chemical calculations is introduced.
  With its ability to generate character tables symbolically on-the-fly, and by making use of a generic symmetry-orbit-based representation analysis method formulated in this work, \qsymsq{} is able to address all of these shortcomings.
  To illustrate these capabilities of \qsymsq{}, four sets of case studies are examined in detail in this article: (i) high-symmetry \ce{C84H64}, \ce{C60}, and \ce{B9-} to demonstrate the analysis of degenerate molecular orbitals (MOs); (ii) octahedral \ce{Fe(CN)6^{3-}} to demonstrate the analysis of symmetry-broken determinants and MOs; (iii) linear hydrogen fluoride in a magnetic field to demonstrate the analysis of magnetic symmetry; and (iv) equilateral \ce{H3+} to demonstrate the analysis of density symmetries.
\end{abstract}

%% file: introduction/introduction.tex

\tikzsetexternalprefix{./introduction/tikz/}

\section{Introduction}
\label{sec:intro}

  Symmetry provides a systematic framework to categorize and classify various mathematical quantities that are of interest to quantum chemists, such as electronic wavefunctions and densities, via the lenses of group and representation theories.
  The ability to examine these quantities based on their symmetry enhances one's arsenal of analysis tools that facilitate the assignment of such quantities calculated from approximate numerical methods to true eigenfunctions of the electronic Hamiltonian of the system.
  In such studies, having a robust method to unambiguously identify and label the symmetries of the quantities being investigated ensures that their properties can be correctly tracked and assigned to known or expected ground and excited electronic states of the system.
  This is especially true when the underlying equations that govern such quantities yield multiple solutions with differing degrees of physical relevance, thus making the task of understanding them much more challenging.
  The simplest and most familiar examples of such equations are the non-linear self-consistent-field (SCF) Hartree--Fock (HF) and Kohn--Sham (KS) density-functional theory (DFT) equations\cite{article:Mjolsness1968, article:Fukutome1971, article:Stanton1968, article:King1969, article:Redondo1989, article:Pulay1990}.

  In both SCF HF and KS theories, spin-orbitals are one-electron wavefunctions that form the cornerstones upon which relevant quantities of interest, \textit{e.g.}, single-determinantal wavefunctions in HF\cite{book:Szabo1996} and electron densities in KS\cite{book:Parr1989}, are constructed.
  The spin-orbitals themselves have long been deemed to be of great importance, for they provide chemists with a useful means to interpret the underlying multi-electron quantities which are often too complicated to examine directly.
  In fact, in HF theory, the spin-orbitals that result from the variational optimization of a single-determinantal \textit{ansatz} form the starting point for many families of post-HF correlated methods such as configuration interaction (CI)\cite{article:Nesbet1955}, coupled cluster (CC)\cite{book:Shavitt2009}, and complete active space (CAS)\cite{book:Helgaker2000,article:Olsen2011}.
  On the other hand, in KS theory, there have long been discussions that the KS spin-orbitals are just as useful as their HF counterparts in chemical theories based on molecular-orbital (MO) models [see Refs.~\citenum{article:Baerends1997, article:Stowasser1999} and also contributions (2.2.4)--(2.2.7) in Ref.~\citenum{article:Teale2022}].
  In either case, it is imperative that the shape and symmetry properties of spin-orbitals be identified so that they can be used effectively in the qualitative investigations of chemical phenomena\cite{article:Stowasser1999} and the quantitative calculations of physical properties such as correlation energies (via post-HF correlated treatments), ionization potentials\cite{article:Levy1984}, and vertical excitation energies\cite{article:Savin1998}.

  However, it is not only the symmetry of spin-orbitals that is important, since spin-orbitals are only one-electron functions and hence do not fully represent electronic states in any multi-electron system.
  In fact, in wavefunction theories, one often needs to obtain a good understanding of symmetry properties of multi-electron wavefunctions before one can confidently attribute them to actual electronic states of the system, especially when one is interested in more than just the ground state, such as in the computation of electronic spectra\cite{article:Rubio1999,article:Rubio2003}.
  A few studies in which symmetry is used to assist the interpretation of ground- and excited-state correlated wavefunctions can be found in Refs.~\citenum{article:Tran2019,article:Tran2020,article:Hanscam2022}.
  In addition, the multiple, generally non-orthogonal, SCF solutions that arise from the HF equations may interact with each other in a CI expansion, if their symmetries are compatible, to give improved multi-determinantal wavefunctions describing certain electronic states with definitive symmetries.
  Some examples of this include the examinations of low-lying HF solutions in the \ce{NO2} radical\cite{article:Jackels1976}, in various classes of hydrocarbons\cite{article:Sundstrom2014b}, in octahedral transition-metal complexes\cite{article:Huynh2020}, and in avoided crossings in LiF\cite{article:Mayhall2014}.
  Furthermore, a thorough insight into the symmetry properties of wavefunctions and densities proves necessary to ensure formal correctness in the fundamental development of DFT and the interpretation of DFT calculation results.
  This is particularly important in degenerate systems where care must be taken to handle any symmetry breaking in the densities correctly to avoid the well-known symmetry dilemma that often arises in the KS formalism where the KS effective potential has a different symmetry from that of the physical external potential, as discussed in great detail by many authors including G\"orling\cite{article:Gorling1993, article:Gorling2000}, Savin\cite{incollection:Savin1996}, and Chowdhury and Perdew\cite{article:Chowdhury2021}.

  In addition, since chemistry is hardly ever static, it is often of great interest to follow electronic states as the symmetry of the system is varied.
  Such a variation can be brought about by various factors such as distortions under vibronic coupling (\textit{i.e.}, Jahn--Teller distortions and related phenomena\cite{incollection:Bersuker2009}), mere applications of external magnetic or electric fields\cite{book:Ceulemans2013,article:Wibowo2021,article:Irons2022}, and structural distortions induced by external fields\cite{article:Bischoff2020, article:Irons2021, article:Wibowo2023}.
  As the symmetry of the system changes, degeneracies might be lifted and broken symmetry (\textit{i.e.}, when a function and its symmetry partners span multiple irreducible representations of the full symmetry group of the system) might be restored\cite{article:Huynh2020}.
  A knowledge of wavefunction and density symmetry allows one to correlate electronic states from low-symmetry configurations to high-symmetry configurations, thus gaining additional insight into their behaviors and properties.

  Unfortunately, to the best of our knowledge, despite the importance of symmetry in quantum-chemical theory and computation, there does not yet exist any implementation for a general analysis of symmetry properties of electronic wavefunctions, densities, and potentials.
  In fact, many existing general-purpose quantum chemistry packages such as \textsc{Q-Chem}\cite{article:Epifanovsky2021}, \textsc{Orca}\cite{article:Neese2022}, \textsc{PySCF}\cite{article:Sun2018}, \textsc{Dalton}\cite{article:Aidas2014}, \textsc{[Open]Molcas}\cite{article:Aquilante2020}, \textsc{Psi4}\cite{article:Smith2020}, \textsc{CFOUR}\cite{article:Matthews2020}, and \textsc{TURBOMOLE}\cite{article:Balasubramani2020} come with features to carry out symmetry analysis to some extent, but most (with \textsc{Q-Chem} and \textsc{TURBOMOLE} being exceptions) opt to work in $\mathcal{D}_{2h}$ or one of its subgroups, all of which are Abelian groups whose irreducible representations are real and one-dimensional, and hence are unable to take into account any spatial degeneracy in wavefunctions properly.
  Moreover, none of these packages is able to cope with symmetry breaking, nor are they programmed to examine symmetry properties of quantities other than wavefunctions, and as far as we are aware, no existing software provides options to analyze symmetry in the presence of external fields.

  In this article, a framework for general symmetry analysis is introduced.
  This framework is implemented in a Rust\cite{article:Matsakis2014, book:Klabnik2018} program named \qsymsq{}, which stands for \textbf{Q}uantum \textbf{Sym}bolic \textbf{Sym}metry, and which seeks to address some of the needs for symmetry in electronic-structure theory and computation that are currently not fulfilled by existing quantum chemistry packages.
  In particular, \qsymsq{} is designed to work with \emph{all} possible finite point groups, Abelian or not, for which necessary character tables are automatically and symbolically generated on-the-fly, so that degeneracies and symmetry breaking can be represented correctly.
  In addition, this framework is sufficiently general to be applicable to any linear-space quantities and not just wavefunctions or densities.
  Furthermore, \qsymsq{} is capable of performing symmetry analysis in the presence of external fields, particularly ensuring that complex irreducible representations, which occur frequently when a magnetic field is present, are handled explicitly.
  In addition, \qsymsq{} is able to provide transformation matrices that enable the generation of symmetry-equivalent partners of any linear-space quantities, as long as they can be expanded in terms of atomic-orbital (AO) basis functions or products thereof.
  All of this is possible thanks to one governing design principle that \qsymsq{} undertakes, which insists that all of its computational elements (\textit{e.g.}, symmetry operations and irreducible representation characters) are treated \emph{symbolically} as much as possible, so that defining properties of groups such as closure and existence of inverses are respected and utilized to guarantee accuracy and efficiency.

  The article is organized as follows.
  In Section~\ref{sec:theory}, the theoretical foundation for the symmetry analysis framework implemented in \qsymsq{} is laid out.
  In particular, the various aspects of group and representation theories involved in the determination of molecular symmetry groups, the management of symmetry operations, and the \textit{in situ} generation of character tables are explained.
  This is followed by the formulation of a general method for representation symmetry analysis applicable to any linear space.
  Then, Section~\ref{sec:illustrations} presents several case studies to illustrate the usefulness of symmetry analysis via \qsymsq{} in interpreting and understanding electronic-structure calculations.
  Finally, Section~\ref{sec:conclusion} concludes the article with a few remarks on the capabilities and limitations of the symmetry analysis framework implemented in \qsymsq{}, and also charts possible directions for \qsymsq{} to be extended in the future.

%% file: theory/theory.tex

\section{Theory}
\label{sec:theory}

  \subsection{Symmetry group determination}
  \label{sec:symgroupdet}

    \subsubsection{Unitary symmetry of the electronic Hamiltonian}
    \label{sec:unitarysym}

      For a molecular system with $N_{\mathrm{e}}$ electrons and $N_{\mathrm{n}}$ nuclei in a uniform \emph{external} electric field $\boldsymbol{\mathcal{E}}$ and magnetic field $\mathbf{B} = \boldsymbol{\nabla} \times \mathbf{A}(\mathbf{r})$, where $\mathbf{A}(\mathbf{r})$ denotes the magnetic vector potential, the electronic Hamiltonian is given by
      \begin{equation}
        \hat{\mathscr{H}} =
          \hat{\mathscr{H}}_0 + \hat{\mathscr{H}}_{\mathrm{elec}} + \hat{\mathscr{H}}_{\mathrm{mag}}.
        \label{eq:H}
      \end{equation}
      In atomic units, the first contribution has the form
      \begin{equation}
        \begin{gathered}
          \hat{\mathscr{H}}_0 =
            \sum_i^{N_{\mathrm{e}}}
              -\frac{1}{2}\nabla_i^2
            + \sum_i^{N_{\mathrm{e}}} \sum_{j>i}^{N_{\mathrm{e}}}
              \frac{1}{\lvert \mathbf{r}_i - \mathbf{r}_j \rvert}
            + v_{\mathrm{ext}},
        \end{gathered}
        \label{eq:H0}
      \end{equation}
      and is the zero-field Hamiltonian which has an explicit dependence on the multiplicative external potential $v_{\mathrm{ext}}$ whose form is governed by the geometric arrangement of the nuclei,
      \begin{equation}
        \begin{gathered}
          v_{\mathrm{ext}} = \sum_i^{N_{\mathrm{e}}} \sum_A^{N_{\mathrm{n}}}
            \frac{-Z_A}{\lvert \mathbf{r}_i - \mathbf{R}_A \rvert}.
        \end{gathered}
        \label{eq:vext}
      \end{equation}
      In Equation~\eqref{eq:vext}, $\mathbf{r}_i$ denotes the position vector of the $i$\textsuperscript{th} electron and $\mathbf{R}_A$ that of the $A$\textsuperscript{th} nucleus.
      The second contribution,
      \begin{equation}
        \hat{\mathscr{H}}_{\mathrm{elec}} =
          \sum_i^{N_{\mathrm{e}}}
          \boldsymbol{\mathcal{E}} \cdot \mathbf{r}_i,
      \end{equation}
      describes the interaction between the electrons and the external electric field\cite{article:Aschi2001}, and the third contribution,
      \ifdefined\twocolumnmode
        \begin{multline}
          \hat{\mathscr{H}}_{\mathrm{mag}} \\=
            \sum_i^{N_{\mathrm{e}}} \mathbf{A}(\mathbf{r}_i) \cdot \hat{\mathbf{p}}_i
            + \frac{g_s}{2} \sum_i^{N_{\mathrm{e}}} \mathbf{B} \cdot \hat{\mathbf{s}}_i
            + \frac{1}{2} \sum_i^{N_{\mathrm{e}}} A^2(\mathbf{r}_i),
          \label{eq:Hmag}
        \end{multline}
      \else
        \begin{equation}
          \hat{\mathscr{H}}_{\mathrm{mag}} =
            \sum_i^{N_{\mathrm{e}}} \mathbf{A}(\mathbf{r}_i) \cdot \hat{\mathbf{p}}_i
            + \frac{g_s}{2} \sum_i^{N_{\mathrm{e}}} \mathbf{B} \cdot \hat{\mathbf{s}}_i
            + \frac{1}{2} \sum_i^{N_{\mathrm{e}}} A^2(\mathbf{r}_i),
          \label{eq:Hmag}
        \end{equation}
      \fi
      where $\hat{\mathbf{p}}_i$ is the linear momentum operator for the $i$\textsuperscript{th} electron, $\hat{\mathbf{s}}_i$ the spin angular momentum operator for the $i$\textsuperscript{th} electron, and $g_s$ the electron spin $g$-factor, gives the interaction of the electrons with the external magnetic field\cite{book:Weil2007, article:Tellgren2018}.
      The \emph{unitary symmetry group} $\mathcal{G}$ of the system consists of all unitary transformations $\hat{u}$ that leave $\hat{\mathscr{H}}$ invariant:
      \begin{equation}
        \hat{u} \hat{\mathscr{H}} \hat{u}^{-1} = \hat{\mathscr{H}}.
      \end{equation}
      Clearly, $\mathcal{G}$ is the intersection of the unitary symmetry groups of $\hat{\mathscr{H}}_0$, $\hat{\mathscr{H}}_{\mathrm{elec}}$, and $\hat{\mathscr{H}}_{\mathrm{mag}}$, which we shall denote $\mathcal{G}_0$, $\mathcal{G}_{\mathrm{elec}}$, and $\mathcal{G}_{\mathrm{mag}}$, respectively.
      We further restrict the elements in these groups to be \textit{point transformations} acting on the \textit{configuration space} where physical systems such as atoms, molecules, and fields are described.\cite{book:Altmann1986}
      Then, $\mathcal{G}_0$ is also commonly known as the \emph{point group} of the molecular system.

      A robust algorithm to determine the name and elements of $\mathcal{G}_0$ for any molecular system has already been described by Beruski and Vidal\cite{article:Beruski2014}.
      As shown formally in Appendix A of Ref.~\citenum{article:Irons2022}, the group $\mathcal{G}_{\mathrm{mag}}$ consists of orthogonal transformations in three dimensions [\textit{i.e.}, elements of the group $\mathsf{O}(3)$] that would map the uniform magnetic field $\mathbf{B}$ onto itself and is commonly known as $\mathcal{C}_{\infty h}$\cite{book:Ceulemans2013, article:Pausch2021}, which is an infinite Abelian group with principal axis parallel to $\mathbf{B}$.
      A similar approach can be used to show formally that $\mathcal{G}_{\mathrm{elec}}$ consists of three-dimensional orthogonal transformations that would leave the uniform electric field $\boldsymbol{\mathcal{E}}$ unchanged and is commonly recognised as $\mathcal{C}_{\infty v}$\cite{book:Ceulemans2013}, which is an infinite, but not Abelian, group with principal axis parallel to $\boldsymbol{\mathcal{E}}$.
      Hence, a na\"{\i}ve procedure to locate all elements of $\mathcal{G}$ is to first identify all elements of $\mathcal{G}_0$, and then to filter out only those elements that would leave $\boldsymbol{\mathcal{E}}$ and/or $\mathbf{B}$ invariant.
      However, this procedure is unnecessarily wasteful as it requires additional efforts to be spent on finding a large number of elements of $\mathcal{G}_0$ that would eventually be discarded, since the presence of external fields almost always leads to a reduction of unitary symmetry.
      In fact, for highly symmetric molecular systems where $\mathcal{G}_0$ is large, these additional efforts can be non-trivial.

    \subsubsection{Including external fields: method of fictitious special atoms}

      It is desirable to make use of the algorithm by Beruski and Vidal\cite{article:Beruski2014} as much as possible to locate all elements of $\mathcal{G}$ directly without having to go through the intermediary of $\mathcal{G}_0$ in the presence of external fields.
      To this end, we propose that fictitious special atoms be introduced to represent the external fields such that the combination of the molecule and fictitious atoms has the same unitary symmetry group $\mathcal{G}$ as the combination of the molecule and the external fields.
      Each fictitious special atom is characterized by a pair of parameters $(t, \mathbf{R}_t)$, where $t$ encodes its type and $\mathbf{R}_t$ denotes its position.

\begin{figure*}
        \centering
        \includegraphics[width=\linewidth]{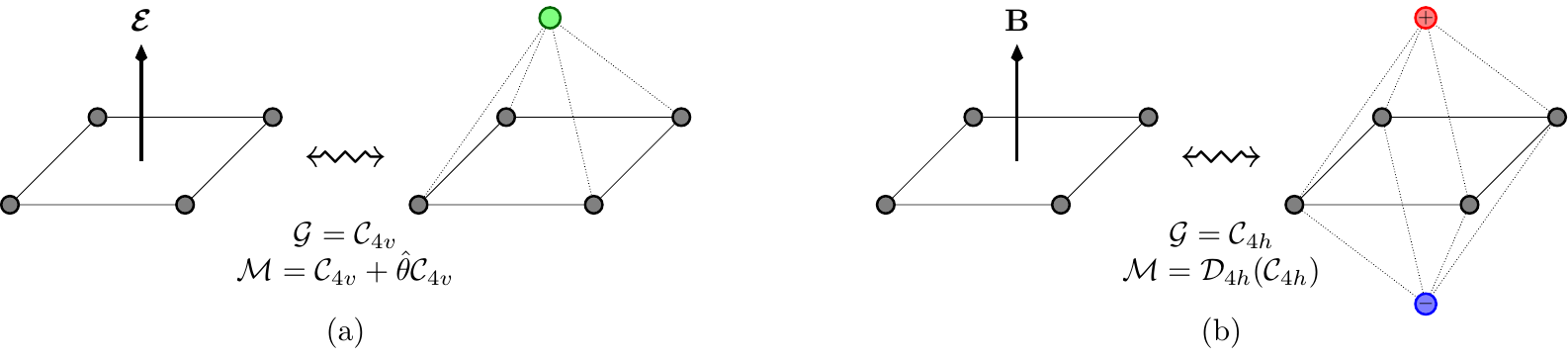}
        \caption{%
          Equivalence between systems in external fields and systems with fictitious special atoms.
          (a) A single fictitious special atom of type $e$ is placed at $\mathbf{R}_{\mathrm{com}} + k\boldsymbol{\mathcal{E}}$ to represent a uniform electric field.
          (b) Two fictitious special atoms, one of type $b+$ and the other of type $b-$, are placed at $\mathbf{R}_{\mathrm{com}} \pm k\mathbf{B}$ to represent a uniform magnetic field.
        }
        \label{fig:fictitiousatoms}
      \end{figure*}

      A uniform electric field $\boldsymbol{\mathcal{E}}$ is represented by one fictitious atom of type $t = e$ placed at $\mathbf{R}_e = \mathbf{R}_{\mathrm{com}} + k\boldsymbol{\mathcal{E}}$, where $\mathbf{R}_{\mathrm{com}}$ is the position vector of the center of mass of the molecule and $k$ a scalar factor chosen to ensure that this fictitious atom does not coincide with any actual atom in the molecule, and that the subsequent unitary symmetry group determination is numerically stable.
      The vector $\mathbf{R}_e - \mathbf{R}_{\mathrm{com}}$ is therefore parallel to $\boldsymbol{\mathcal{E}}$, and as $\boldsymbol{\mathcal{E}}$ is a polar vector\cite{book:Birss1966}, it is imposed that fictitious atoms of type $e$ transform under all operations in the group $\mathsf{O}(3)$ just as any ordinary atom does.
      It is easily seen that the combination of the molecule and the fictitious atom has the same unitary symmetry group as the molecule in the external $\boldsymbol{\mathcal{E}}$ field [Figure~\ref{fig:fictitiousatoms}(a)].

      On the other hand, a uniform magnetic field $\mathbf{B}$ is represented by two fictitious atoms, one of type $b+$ and the other of type $b-$, placed at $\mathbf{R}_{b\pm} = \mathbf{R}_{\mathrm{com}} \pm k\mathbf{B}$, where the $+/-$ signs in the type names signify the polarities of the fictitious atoms.
      The vector $\mathbf{R}_{b+} - \mathbf{R}_{b-}$ is parallel to $\mathbf{B}$, and these two fictitious atoms transform under all operations in the group $\mathsf{O}(3)$ almost like any ordinary atom, but since $\mathbf{B}$ is an axial vector\cite{book:Birss1966}, it is additionally required that the polarities of the fictitious atoms be reversed under improper transformations.
      This ensures that the combination of the molecule and the fictitious atoms has the same unitary symmetry group as the molecule in the external $\mathbf{B}$ field [Figure~\ref{fig:fictitiousatoms}(b)].

\begin{figure*}
        \centering
        \includegraphics[]{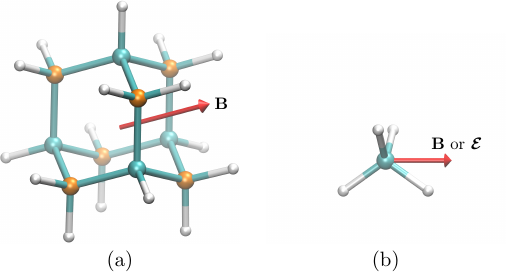}
        \caption{%
          Two special cases involving a uniform external field where the original Beruski--Vidal algorithm\cite{article:Beruski2014} needs to be modified.
          (a) A tetrahedral adamantane molecule placed in a uniform external magnetic field oriented along one of its $C_3$ axes.
          This illustrates a possible scenario in which a polyhedral SEA group (the six carbon atoms highlighted in orange) is arranged in a spherical top fashion (a regular octahedron).
          (b) A tetrahedral methane molecule placed in a uniform external magnetic or electric field oriented simultaneously parallel to one of the molecular mirror planes and perpendicular to another.
          This illustrates a possible scenario of the $\mathcal{C}_s$ unitary group arising from the symmetric top rotational symmetry.
        }
        \label{fig:unusualcases}
      \end{figure*}

      With the introduction of fictitious special atoms, external fields are no longer required to be treated separately in the unitary symmetry group determination.
      In fact, fictitious atoms can be incorporated directly into the Beruski--Vidal algorithm\cite{article:Beruski2014}, provided that the following modifications are taken into account:
      \begin{enumerate}[label=(\roman*)]
        \item Fictitious atoms must be included in the calculation of the principal moments of inertia of the system and the subsequent classification into four main rotational symmetry types: spherical top, symmetric top, asymmetric top, and linear.
        For this purpose, a mass of \SI{100.0}{\atomicmassunit} is chosen for the fictitious atoms: there is no physical significance to this value; it simply has been found to ensure numerical stability in all test cases.
        \item Fictitious atoms must be included in the determination of distance-based symmetrically-equivalent-atom (SEA) groups.
        This means that $b+$ and $b-$ can be in the same SEA group if they both have the same distance signature to all other atoms in the molecule, despite their different polarities.
        \item The possibility that polyhedral SEAs be arranged in a spherical top must also be taken into account.
        This additional possibility was not originally considered in Ref.~\citenum{article:Beruski2014} as it can only arise when a spherical top molecule is placed in an external magnetic field.
        Figure~\ref{fig:unusualcases}(a) shows an example where a magnetic field is applied along one of the $C_3$ axes of tetrahedral adamantane: this molecule-field combined system is now a symmetric top with the unique axis along the field direction, but the six carbon atoms highlighted in orange constitute a group of SEAs that are arranged in a regular octahedron.
        \item The symmetric top rotational symmetry may also result in the $\mathcal{C}_s$ group.
        This additional possibility was not considered in Ref.~\citenum{article:Beruski2014} either as it can only occur when an external field is applied to a spherical top in a manner that eliminates all symmetry elements of the system apart from a single mirror plane.
        Figure~\ref{fig:unusualcases}(b) illustrates an example of this when either a magnetic or an electric field is applied to tetrahedral \ce{CH4} such that it is simultaneously parallel to one of the molecular mirror planes and perpendicular to another.
      \end{enumerate}

    \subsubsection{Magnetic symmetry of the electronic Hamiltonian}
    \label{sec:magsym}

      When antiunitary operations are taken into account, the unitary symmetry group $\mathcal{G}$ might no longer be the largest symmetry group of the electronic Hamiltonian $\hat{\mathscr{H}}$.
      In fact, in many studies involving magnetic phenomena and magnetic materials\cite{article:Dimmock1962,article:Bradley1968,article:Lazzeretti1984,article:Keith1993,article:Pelloni2011}, it is necessary to consider a supergroup of $\mathcal{G}$ that also contains antiunitary symmetry operations that leave $\hat{\mathscr{H}}$ invariant.
      Such a group is called the \emph{magnetic symmetry group} $\mathcal{M}$ of the system, and it can easily be seen\cite{article:Bradley1968} that $\mathcal{M}$ must admit $\mathcal{G}$ as a normal subgroup of index $2$, so that we can write
      \begin{equation}
        \mathcal{M} = \mathcal{G} + \hat{a}_0\mathcal{G},
        \label{eq:maggroup}
      \end{equation}
      where $\hat{a}_0$ can be any of the antiunitary elements in $\mathcal{M}$ but must be fixed once chosen.
      The left coset $\hat{a}_0\mathcal{G}$ with respect to $\mathcal{G}$ contains all antiunitary elements of $\mathcal{M}$.

      Let us now consider the time-reversal operation $\hat{\theta}$, which is an archetype of antiunitary operations (see Chapter~26 of Ref.~\citenum{book:Wigner1959} for an in-depth discussion of the time-reversal operation in quantum mechanics).
      It turns out that, with respect to $\hat{\theta}$, magnetic symmetry groups can be classified into just two kinds\cite{article:Cracknell1965,article:Cracknell1966,article:Bradley1968}.
      The first kind are those that contain $\hat{\theta}$, in which case one can choose $\hat{a}_0 = \hat{\theta}$ so that
      \begin{equation}
        \mathcal{M} = \mathcal{G} + \hat{\theta}\mathcal{G}.
        \label{eq:greygroup}
      \end{equation}
      These are called \emph{magnetic grey groups}.
      The second kind are those that do not contain $\hat{\theta}$; however, one can always find a unitary operation $\hat{u}_0$ not in the group such that the product $\hat{\theta}\hat{u}_0$ is an antiunitary operation that occurs in the group.
      This then enables one to write
      \begin{equation}
        \mathcal{M} = \mathcal{G} + \hat{\theta}\hat{u}_0\mathcal{G},
        \label{eq:bwgroup}
      \end{equation}
      where $\hat{a}_0$ has been chosen to be $\hat{\theta}\hat{u}_0$.
      Such groups are called \emph{magnetic black-and-white groups}.
      It is then clear that, in the absence of an external magnetic field, $\hat{\theta}$ is a symmetry operation of the system.
      However, this ceases to be the case when an external magnetic field is applied: the magnetic field vector $\mathbf{B}$ is time-odd\cite{book:Birss1966,article:Bradley1968} and thus gives rise to terms in the electronic Hamiltonian [Equation~\eqref{eq:Hmag}] that do not commute with $\hat{\theta}$ (see Appendix A of Ref.~\citenum{article:Irons2022} for a detailed explanation).
      Therefore, the following general rules can be deduced:
      \begin{enumerate}[label=(\roman*)]
        \item in the absence of an external magnetic field, the system always has a magnetic symmetry group which must be one of the magnetic grey groups;
        \item in the presence of an external magnetic field, if the system exhibits any antiunitary symmetry, then it has a magnetic symmetry group that must be one of the magnetic black-and-white groups, but if the system exhibits no antiunitary symmetry, then it only has a unitary symmetry group.
      \end{enumerate}

      For both kinds of magnetic groups, it is often useful to consider a unitary group $\mathcal{M}'$ that is isomorphic to $\mathcal{M}$.
      In cases where $\mathcal{M}'$ is identifiable with a subgroup of the full rotation-inversion group in three dimensions $\mathsf{O}(3)$ and can thus be given a Sch\"onflies symbol, the magnetic group $\mathcal{M}$ can be written as $\mathcal{M}'(\mathcal{G})$ \cite{article:Cracknell1966,article:Pelloni2011}.
      When this is not possible, however, the antiunitary coset form with respect to the unitary symmetry group $\mathcal{G}$ and a representative antiunitary operation $\hat{a}_0$ [Equations~\eqref{eq:maggroup}--\eqref{eq:bwgroup}] can always be employed to uniquely denote $\mathcal{M}$ because it is always possible to assign a Sch\"onflies symbol to $\mathcal{G}$, which is guaranteed to be a subgroup of the molecular point group $\mathcal{G}_0$ (\textit{cf.} Section~\ref{sec:unitarysym}).
      Figure~\ref{fig:fictitiousatoms} depicts two examples of how $\mathcal{M}$ is typically denoted.

      To determine $\mathcal{M}$ and all of its elements given a molecular system in a uniform external field, the Beruski--Vidal algorithm\cite{article:Beruski2014} can once again be exploited with an additional modification that any unitary transformation considered in the algorithm can also be accompanied by the antiunitary action of time reversal.
      For all ordinary atoms and fictitious atoms of type $e$ representing an applied electric field, time reversal has no effects.
      However, for fictitious atoms of types $b+$ and $b-$ representing an applied magnetic field, their polarities must be reversed under time reversal due to the time-odd nature of the magnetic field vector $\mathbf{B}$\cite{book:Birss1966,article:Bradley1968}.

  \subsection{Abstract group construction}

    \subsubsection{Computational representation of symmetry operations}
    \label{sec:comprepofsymops}

      In \qsymsq{}, symmetry operations located using the method described in the previous Section are stored as instances of the \texttt{SymOp} structure.
      It shall henceforth be written ``$\texttt{SymOp}(\hat{g})$'' to denote an instance of the \texttt{SymOp} structure that represents the actual $\hat{g}$ symmetry operation computationally.
      In order for this representation to be efficient and to respect discrete-group-theoretic properties, most notably compositability and closure, of the underlying symmetry operations, it is imposed that the \texttt{SymOp} structure fulfill the following traits:
      \begin{enumerate}[label=(\roman*)]
        \item equality comparisons that are equivalence relations --- reflexivity, transitivity, and symmetry must be satisfied for the ``$=$'' relation between \texttt{SymOp} instances, which must take into account the $2\pi$-periodicity of spatial rotations;

        \item hashability --- each $\texttt{SymOp}(\hat{g})$ instance must be able to produce an integer hash value $\hash[\texttt{SymOp}(\hat{g})]$ that allows itself to be looked up from a hash table with an average constant time $O(1)$, and that must be compatible with equality comparisons:
        \ifdefined\twocolumnmode
          \begin{align*}
            \texttt{SymOp}(\hat{g}_1) &= \texttt{SymOp}(\hat{g}_2) \\ \implies \hash[\texttt{SymOp}(\hat{g}_1)] &= \hash[\texttt{SymOp}(\hat{g}_2)];
          \end{align*}
        \else
          \begin{equation*}
            \texttt{SymOp}(\hat{g}_1) = \texttt{SymOp}(\hat{g}_2) \implies \hash[\texttt{SymOp}(\hat{g}_1)] = \hash[\texttt{SymOp}(\hat{g}_2)];
          \end{equation*}
        \fi

        \item compositability --- $\texttt{SymOp}(\hat{g}_1) * \texttt{SymOp}(\hat{g}_2) = \texttt{SymOp}(\hat{g}_1\hat{g}_2)$ where ``$*$'' denotes the composition operation between two \texttt{SymOp} instances.
      \end{enumerate}
      The design of the \texttt{SymOp} structure in \qsymsq{} is detailed in Section~S1 of the Supporting Information to illustrate how the above traits are satisfied.

    \subsubsection{Unitary conjugacy class structure}

      Prior to generating the character table of the symmetry group, its conjugacy class structure must first be determined.
      The conjugacy class structure of a group in turn depends on how the conjugacy equivalence relation between group elements is defined.
      In this article, only the familiar \emph{unitary conjugacy equivalence relation} is considered:
      \ifdefined\twocolumnmode
        \begin{multline}
          \hat{g}_1, \hat{g}_2 \in \mathcal{G},\\
          \hat{g}_1 \sim \hat{g}_2 \ \ \Longleftrightarrow \ \
          \exists \hat{u} \in \mathcal{G} \ :\ \hat{g}_1 = \hat{u} \hat{g}_2 \hat{u}^{-1},
        \end{multline}
      \else
        \begin{equation}
          \hat{g}_1, \hat{g}_2 \in \mathcal{G}, \quad
          \hat{g}_1 \sim \hat{g}_2 \ \ \Longleftrightarrow \ \
          \exists \hat{u} \in \mathcal{G} \ :\ \hat{g}_1 = \hat{u} \hat{g}_2 \hat{u}^{-1},
        \end{equation}
      \fi
      which holds when all elements in the group are represented as \emph{mathematical} unitary operators on linear spaces, even if some of them are actually \emph{physical} antiunitary operators.
      A different conjugacy equivalence relation called \emph{magnetic conjugacy equivalence relation} holds if some of the elements in the group are represented on linear spaces as mathematical antiunitary operators\cite{article:Newmarch1981}, which leads to a different conjugacy class structure\cite{article:Newmarch1981,article:Newmarch1983}.
      Although magnetic conjugacy classes have also been implemented in \qsymsq{}, their uses in magnetic symmetry via corepresentation theory\cite{book:Wigner1959} will be examined in a future study.

      The classification of elements of finite molecular symmetry groups in \qsymsq{} is carried out via the \emph{Cayley table} $\mathbf{C}$ of the group:
      \begin{equation}
        C_{ij} = k \quad \textrm{where} \quad \hat{g}_i \hat{g}_j = \hat{g}_k,
        \label{eq:ctb}
      \end{equation}
      which encodes the group's multiplicative structure in a two-dimensional array of integers.
      The compositions $\hat{g}_i \hat{g}_j$ are effected computationally through the corresponding compositions $\texttt{SymOp}(\hat{g}_i) * \texttt{SymOp}(\hat{g}_j)$ of the \texttt{SymOp} structure.
      Once the Cayley table $\mathbf{C}$ has been computed and stored, any operations that call for the multiplicative structure of the group, such as the determination of the conjugacy class structure or the construction of the group's character table (Section~S2 of the Supporting Information), only need to make cheap queries to $\mathbf{C}$ without having to repeatedly recalculate group element compositions.

  \subsection{Generation of character tables of irreducible representations}

    Once an abstract group structure has been obtained for the underlying symmetry group, its character table then needs to be computed to allow for subsequent symmetry analysis.
    This can indeed be performed on-the-fly in \qsymsq{}.
    Algorithms for the automatic generation of symbolic character tables\cite{article:Dixon1967,article:Schneider1990,book:Grove1997} are well-known and have been implemented before, most notably in GAP\cite{misc:GAP2022}.
    However, no such implementations exist for molecular symmetry applications in quantum chemistry.
    These algorithms are thus re-implemented in \qsymsq{} with additional functionalities to ensure that the generated character tables respect conventions that are familiar to most chemists, \textit{e.g.}, the labeling of irreducible representations using Mulliken symbols\cite{article:Mulliken1955}.
    The details of these algorithms are recapitulated in Section~S2 of the Supporting Information.

    The ability to generate character tables automatically and symbolically enables \qsymsq{} to completely circumvent the need for hard-coded character tables which would limit the number and types of groups available for symmetry analysis.
    In particular, as demonstrated in Section~\ref{sec:degeneracyscf}, \qsymsq{} is able to handle degeneracy in non-Abelian symmetry groups which are encountered in highly symmetric molecular structures such as disk-like boron clusters ($\mathcal{D}_{nh}$)\cite{article:Fowler2007,article:Dordevic2022}, hydro-clusters of group-14 elements in quantum dots ($\mathcal{T}_d$)\cite{article:Karttunen2008,article:Foerster2022}, and buckminsterfullerenes ($\mathcal{I}_h$).
    \qsymsq{} is also capable of tackling complex-valued representations that frequently arise in the presence of an external magnetic field, \textit{e.g.}, eight out of the twelve one-dimensional irreducible representations in $\mathcal{C}_{6h}$, which is the unitary symmetry group of benzene in the presence of a uniform perpendicular magnetic field, are complex.

    In all cases, there is no requirement for the molecule and/or external fields to be in any predefined standard orientation for the character-table generation algorithm to work.
    In fact, as long as the symmetry operations of the system can be computationally represented and composited as described in Section~\ref{sec:comprepofsymops} and Section~S1 of the Supporting Information, the group structure can be abstracted away from these concrete representations of symmetry operations to allow for the character table to be computed entirely algebraically without recourse to any other knowledge exterior to the group structure.
    Only when \emph{labels} of computed irreducible representations are to be deduced is information about molecular structures and symmetry operation orientations required in order to satisfy Mulliken's conventions.
    This ensures that molecules and external fields can be placed in whatever orientation is the most sensible or convenient for chemical computation while still being able to benefit from the symmetry analysis offered by \qsymsq{}.

  \subsection{Representation analysis}
  \label{sec:repanalysis}

    An initial formulation of the method for representation analysis has been discussed by one of the authors in a previous article (Appendices B and C of Ref.~\citenum{article:Huynh2020}).
    However, this formulation only focuses on wavefunctions in Hilbert spaces and therefore leaves out other quantum-chemical quantities that are not wavefunctions but that still have symmetry properties.
    Examples of such quantities include electron densities, vibrational coordinates, and magnetically induced ring currents.
    In this Section, a more general formulation of this method will be presented in which \emph{all} linear-space quantities are covered.
    It will also be pointed out how the computational availability of group multiplicative structures via Cayley tables leads to a reduction in the representation analysis time complexity from $O(\lvert \mathcal{G} \rvert^3)$ to $O(\lvert \mathcal{G} \rvert)$ by taking advantage of group closure.

    \subsubsection{Formulation of linear-space representation analysis}
      \label{sec:linrepanalysis}

      \paragraph{Characters and representation matrices.}
        Let $V$ be a linear space and $\mathbf{w}$ an element in $V$ whose symmetry under a prevailing group $\mathcal{G}$ is to be computationally determined.
        To this end, the linear subspace $W \subseteq V$ that is spanned by the orbit
        \begin{equation}
          \mathcal{G} \cdot \mathbf{w} = \lbrace
            \hat{g}_i \mathbf{w}\ :\ g_i \in \mathcal{G}
          \rbrace,
          \label{eq:orbitGw}
        \end{equation}
        where $\hat{g}_i$ denotes the action of $g_i$ in $V$ is first determined.
        The symmetry of $\mathbf{w}$ in $\mathcal{G}$ is then given by the decomposition of $W$ into known irreducible representation spaces on $V$ of the group $\mathcal{G}$.

        Technically, $\hat{g}_i$ and $g_i$ are two very different quantities: the former is an operator acting on $V$ and thus a member of $\mathsf{GL}(V)$ (\textit{i.e.}, the group of all general linear operators acting on $V$), whereas the latter is a member of an abstract group $\mathcal{G}$.
        However, this distinction is unnecessarily pedantic for the purpose of this article and will therefore be ignored: we will use the hatted forms almost exclusively, refer to them as members of the group $\mathcal{G}$, and make no attempt to distinguish operators that represent actions of the same abstract element $g_i$ but on different linear spaces.

        To characterize $W$, we seek its character function $\chi^W$ whose value for each element $\hat{g}$ in the group is given by
        \begin{equation*}
          \chi^W(\hat{g}) = \tr \mathbf{D}^W(\hat{g}),
        \end{equation*}
        where $\mathbf{D}^W(\hat{g})$ is the representation matrix of $\hat{g}$ in some finite basis chosen for $W$.
        Let $\mathcal{B} = \left\{\mathbf{e}_m \ :\ 1 \le m \le \dim W \right\}$ be such a basis.
        The elements of $\mathbf{D}^W(\hat{g})$ satisfy the set of equations
        \begin{equation}
          \hat{g} \mathbf{e}_m =
            \sum_{n=1}^{\dim W} \mathbf{e}_n D^W_{nm}(\hat{g}),
          \label{eq:repmat}
        \end{equation}
        one for each element $\mathbf{e}_m$ in the basis.

      \paragraph{Representation matrix determination.}
        Equation~\eqref{eq:repmat} now needs to be solved in a suitably chosen basis to determine the diagonal $D^W_{nn}(\hat{g})$ elements so that the character value $\chi^W(\hat{g})$ can be computed.
        In principle, these equations can be viewed as a set of simultaneous equations that can be solved algebraically to give the required matrix elements.
        However, such an approach would be tedious and it is much more common for equations of this type to be solved using a projection operator, which requires the existence of an inner product.

        Thus far, no reference has been made to any inner products on $V$, because inner products are not required in the definition of representation symmetry.
        In fact, symmetry is a linear-space property rather than an inner-product-space property.
        This realization has an important implication: we are at liberty to \emph{define} any inner product that is the most convenient to compute for a given linear space $V$ in order to construct a projection operator solely for the purpose of inverting Equation~\eqref{eq:repmat}; the value of the character $\chi^W(\hat{g})$ must be independent of this choice of inner product, even if $V$ itself does not possess an intrinsic inner product.

        Let us now endow $V$ with an inner product $\braket{\cdot | \cdot}$ with which the overlap matrix $\mathbf{S}$ between elements in the orbit $\mathcal{G} \cdot \mathbf{w}$ is defined:
        \begin{equation}
          S_{ij} = \braket{\hat{g}_i \mathbf{w} |  \hat{g}_j \mathbf{w}}.
          \label{eq:smat}
        \end{equation}
        It would then be ideal to use the orbit $\mathcal{G} \cdot \mathbf{w}$ as a basis $\mathcal{B}$ for $W$ with which Equation~\eqref{eq:repmat} can be solved to give the character values.
        However, the elements in $\mathcal{G} \cdot \mathbf{w}$ are not necessarily linearly independent, and the matrix $\mathbf{S}$ is thus not necessarily of full rank.
        In this case, a tall rectangular matrix $\mathbf{X}$ can be constructed:
        \begin{equation}
          X_{im} = \frac{1}{\sqrt{\lambda_m}} U_{im},
          \quad 1 \le m \le \rank \mathbf{S} = \dim W,
          \label{eq:xmat}
        \end{equation}
        where $\lambda_m$ is a non-zero eigenvalue of $\mathbf{S}$ and $U_{im}$ the $i$\textsuperscript{th} component of the corresponding eigenvector.
        The matrix $\mathbf{X}$ allows a linearly independent basis for $W$ to be defined:
        \ifdefined\twocolumnmode
          \begin{multline}
            \mathcal{B} = \left\{
              \tilde{\mathbf{w}}_m = \sum_{i=1}^{\lvert \mathcal{G} \rvert}
              (\hat{g}_i \mathbf{w}) X_{im}
              \right.\\
              \left. \vphantom{\sum_{i=1}^{\lvert \mathcal{G} \rvert}} : \ 1 \le m \le \rank \mathbf{S} = \dim W
            \right\},
            \label{eq:wtransformed}
          \end{multline}
        \else
          \begin{equation}
            \mathcal{B} = \left\{
              \tilde{\mathbf{w}}_m = \sum_{i=1}^{\lvert \mathcal{G} \rvert}
              (\hat{g}_i \mathbf{w}) X_{im}
              \ : \ 1 \le m \le \rank \mathbf{S} = \dim W
            \right\},
            \label{eq:wtransformed}
          \end{equation}
        \fi
        such that the overlap matrix in this basis,
        \begin{equation*}
          \tilde{\mathbf{S}} = \mathbf{X}^{\dagger} \mathbf{S} \mathbf{X}
          \quad \textrm{where} \quad
          \tilde{S}_{mn} = \braket{\tilde{\mathbf{w}}_m | \tilde{\mathbf{w}}_n},
        \end{equation*}
        is of full rank and Equation~\eqref{eq:repmat} becomes
        \begin{equation}
          \hat{g} \tilde{\mathbf{w}}_{m'} =
            \sum_{n'=1}^{\rank \mathbf{S}} \tilde{\mathbf{w}}_{n'} D^W_{n'm'}(\hat{g}),
          \label{eq:xrepmat}
        \end{equation}
        where primed subscripts have been used for later convenience.
        If $\mathbf{S}$ is already of full rank, then we simply set $\mathcal{B} = \mathcal{G} \cdot \mathbf{w}$, and so $\tilde{\mathbf{S}} = \mathbf{S}$.
        In either case, the square matrix $\tilde{\mathbf{S}}$ is invertible, with which a non-orthogonal projection operator $\hat{P}_m$ can be constructed\cite{article:Soriano2014}:
        \begin{equation}
          \hat{P}_m = \sum_{n=1}^{\rank \mathbf{S}}
            \ket{\tilde{\mathbf{w}}_m} \tilde{S}^{-1}_{mn} \bra{\tilde{\mathbf{w}}_n},
          \label{eq:projdef}
        \end{equation}
        where $\tilde{S}^{-1}_{mn} = (\tilde{\mathbf{S}}^{-1})_{mn}$.
        This projection operator satisfies
        \begin{equation}
          \hat{P}_m \ket{\tilde{\mathbf{w}}_n} = \delta_{mn} \ket{\tilde{\mathbf{w}}_m}.
          \label{wq:projkrondelta}
        \end{equation}
        Applying $\hat{P}_m$ to both sides of Equation~\eqref{eq:xrepmat} and making use of Equation~\eqref{wq:projkrondelta} gives
        \begin{align*}
          \hat{P}_m \ket{\hat{g} \tilde{\mathbf{w}}_{m'}}
          &= \sum_{n'=1}^{\rank \mathbf{S}}
            \hat{P}_m\ket{\tilde{\mathbf{w}}_{n'}} D^W_{n'm'}(\hat{g})\\
          &= \sum_{n'=1}^{\rank \mathbf{S}}
            \ket{\tilde{\mathbf{w}}_{m}} \delta_{mn'} D^W_{n'm'}(\hat{g})\\
          &= \ket{\tilde{\mathbf{w}}_{m}} D^W_{mm'}(\hat{g}).
        \end{align*}
        Multiplying both sides by $\bra{\tilde{\mathbf{w}}_{m}}$ and using the definition of $\hat{P}_m$ in Equation~\eqref{eq:projdef} then yields
        \ifdefined\twocolumnmode
          \begin{multline*}
            \braket{\tilde{\mathbf{w}}_{m} | \tilde{\mathbf{w}}_{m}}
            \sum_{n=1}^{\rank \mathbf{S}}
              \tilde{S}^{-1}_{mn}
              \braket{\tilde{\mathbf{w}}_n | \hat{g} \tilde{\mathbf{w}}_{m'}}
            \\= \braket{\tilde{\mathbf{w}}_{m} | \tilde{\mathbf{w}}_{m}} D^W_{mm'}(\hat{g}),
          \end{multline*}
        \else
          \begin{equation*}
            \braket{\tilde{\mathbf{w}}_{m} | \tilde{\mathbf{w}}_{m}}
            \sum_{n=1}^{\rank \mathbf{S}}
              \tilde{S}^{-1}_{mn}
              \braket{\tilde{\mathbf{w}}_n | \hat{g} \tilde{\mathbf{w}}_{m'}}
            = \braket{\tilde{\mathbf{w}}_{m} | \tilde{\mathbf{w}}_{m}} D^W_{mm'}(\hat{g}),
          \end{equation*}
        \fi
        or equivalently, by canceling out the $\braket{\tilde{\mathbf{w}}_{m} | \tilde{\mathbf{w}}_{m}}$ term on both sides,
        \begin{equation*}
          \sum_{n=1}^{\rank \mathbf{S}}
            \tilde{S}^{-1}_{mn}
            \braket{\tilde{\mathbf{w}}_n | \hat{g} \tilde{\mathbf{w}}_{m'}}
          = D^W_{mm'}(\hat{g}).
        \end{equation*}
        By reintroducing the original terms in the orbit $\mathcal{G} \cdot \mathbf{w}$ using Equation~\eqref{eq:wtransformed}, we obtain
        \begin{equation*}
          D^W_{mm'}(\hat{g}) =
          \sum_{n=1}^{\rank \mathbf{S}}
          \sum_{i,j=1}^{\lvert \mathcal{G} \rvert}
            \tilde{S}^{-1}_{mn}
            X_{in}^{\lozenge}
            \braket{%
              \hat{g}_i\mathbf{w} |%
              \hat{g} |
              \hat{g}_j\mathbf{w}%
            }%
            X_{jm'},
        \end{equation*}
        where
        \begin{equation*}
          X_{in}^{\lozenge} = \begin{cases}
            X_{in} & \textrm{if $\braket{\cdot | \cdot}$ is bilinear},\\
            X^*_{in} & \textrm{if $\braket{\cdot | \cdot}$ is sesquilinear}.
          \end{cases}
        \end{equation*}
        The above result can be conveniently written in a matrix form:
        \begin{equation}
          \mathbf{D}^W(\hat{g}) =
            \tilde{\mathbf{S}}^{-1}
            \mathbf{X}^{\T\lozenge}
            \mathbf{T}(\hat{g})
            \mathbf{X},
          \label{eq:dmat}
        \end{equation}
        where
        \begin{equation}
          T_{ij}(\hat{g}) = \braket{%
            \hat{g}_i\mathbf{w} |%
            \hat{g} |
            \hat{g}_j\mathbf{w}%
          },
          \label{eq:tmat}
        \end{equation}
        which gives a closed-form expression for the representation matrix $\mathbf{D}^W(\hat{g})$ to be computed from elements in the orbit $\mathcal{G} \cdot \mathbf{w}$.

      \paragraph{Optimization by group closure.}
        It is clear from Equation~\eqref{eq:dmat} that the computation speed of $\mathbf{D}^W(\hat{g})$ is limited by the computation speed of the orbit overlap matrix $\mathbf{S}$ and the matrices $\mathbf{T}(\hat{g})$, for all $\hat{g} \in \mathcal{G}$.
        Na\"{\i}vely, explicit constructions of $\mathbf{S}$ based on Equation~\eqref{eq:smat} and of $\mathbf{T}(\hat{g})$ based on Equation~\eqref{eq:tmat} would incur time complexities of $O(\lvert \mathcal{G} \rvert^2)$ and $O(\lvert \mathcal{G} \rvert^3)$, respectively.
        However, closure of $\mathcal{G}$ under composition allows all matrix elements of $\mathbf{S}$ and $\mathbf{T}(\hat{g})$ to be identifiable with only $\lvert \mathcal{G} \rvert$ unique values:
        \ifdefined\twocolumnmode
          \begin{align}
            S_{ij}
            &= \braket{\hat{g}_i \mathbf{w} |  \hat{g}_j \mathbf{w}} \nonumber \\
            &= \begin{cases}
              \braket{\hat{g}_k \mathbf{w} |  \mathbf{w}} & \textrm{if $\hat{g}_j$ is unitary},\\
              \braket{\hat{g}_k \mathbf{w} |  \mathbf{w}}^* & \textrm{if $\hat{g}_j$ is antiunitary},
            \end{cases}
            \nonumber \\
            &(\hat{g}_k = \hat{g}_j^{-1} \hat{g}_i \in \mathcal{G})
            \label{eq:smatopt}
          \end{align}
        \else
          \begin{equation}
            S_{ij}
            = \braket{\hat{g}_i \mathbf{w} |  \hat{g}_j \mathbf{w}}
            = \begin{cases}
              \braket{\hat{g}_k \mathbf{w} |  \mathbf{w}} & \textrm{if $\hat{g}_j$ is unitary},\\
              \braket{\hat{g}_k \mathbf{w} |  \mathbf{w}}^* & \textrm{if $\hat{g}_j$ is antiunitary},
            \end{cases}
            \quad
            (\hat{g}_k = \hat{g}_j^{-1} \hat{g}_i \in \mathcal{G})
            \label{eq:smatopt}
          \end{equation}
        \fi
        and
        \ifdefined\twocolumnmode
          \begin{align}
            T_{ij}(\hat{g})
            &= \braket{\hat{g}_i \mathbf{w} | \hat{g} |  \hat{g}_j \mathbf{w}} \nonumber \\
            &= \begin{cases}
              \braket{\hat{g}_l \mathbf{w} |  \mathbf{w}} & \textrm{if $\hat{g}\hat{g}_j$ is unitary},\\
              \braket{\hat{g}_l \mathbf{w} |  \mathbf{w}}^* & \textrm{if $\hat{g}\hat{g}_j$ is antiunitary}.
            \end{cases}
            \nonumber \\
            &(\hat{g}_l = \hat{g}_j^{-1} \hat{g}^{-1} \hat{g}_i \in \mathcal{G})
            \label{eq:tmatopt}
          \end{align}
        \else
          \begin{equation}
            T_{ij}(\hat{g})
            = \braket{\hat{g}_i \mathbf{w} | \hat{g} |  \hat{g}_j \mathbf{w}}
            = \begin{cases}
              \braket{\hat{g}_l \mathbf{w} |  \mathbf{w}} & \textrm{if $\hat{g}\hat{g}_j$ is unitary},\\
              \braket{\hat{g}_l \mathbf{w} |  \mathbf{w}}^* & \textrm{if $\hat{g}\hat{g}_j$ is antiunitary}.
            \end{cases}
            \quad
            (\hat{g}_l = \hat{g}_j^{-1} \hat{g}^{-1} \hat{g}_i \in \mathcal{G})
            \label{eq:tmatopt}
          \end{equation}
        \fi
        In both cases, as long as the overlaps between the elements in the orbit $\mathcal{G} \cdot \mathbf{w}$ and the orbit origin $\mathbf{w}$ have been evaluated, which costs $O(\lvert\mathcal{G}\rvert)$ time, all matrix elements of $\mathbf{S}$ and $\mathbf{T}(\hat{g})$ can be deduced without any further involvement of the expensive overlap computation.
        However, this optimization is only possible if one can make the identifications $\hat{g}_k = \hat{g}_j^{-1} \hat{g}_i$ [Equation~\eqref{eq:smatopt}] and $\hat{g}_l = \hat{g}_j^{-1} \hat{g}^{-1} \hat{g}_i$ [Equation~\eqref{eq:tmatopt}] which require knowledge of the multiplicative structure of the group $\mathcal{G}$.
        The implementation of \texttt{SymOp} in \qsymsq{} that enables the construction of the Cayley table [Equation~\eqref{eq:ctb}] achieves exactly this and allows \qsymsq{} to perform extremely efficient representation analysis of linear-space quantities.

    \subsubsection{Examples of linear-space representation analysis}
    \label{sec:exampleslinearspacequantities}

      \paragraph{Single-determinantal wavefunctions and spin-orbitals.}
        Let us consider an $N_{\mathrm{e}}$-electron single-determinantal wavefunction:
        \begin{equation}
          \Psi^{\mathrm{det}}(\mathbf{x}_1, \ldots, \mathbf{x}_{N_{\mathrm{e}}}) =
            \sqrt{N_{\mathrm{e}}!} \hat{\mathscr{A}} \left[\prod_{i=1}^{N_{\mathrm{e}}} \chi_i(\mathbf{x}_i) \right].
          \label{eq:slaterdet}
        \end{equation}
        In the above expression,
        \begin{equation*}
          \hat{\mathscr{A}} = \frac{1}{N_{\mathrm{e}}!} \sum_{\hat{P} \in \mathsf{Sym}(N_{\mathrm{e}})} (-1)^{\pi(\hat{P})} \hat{P}
        \end{equation*}
        is the antisymmetrizer with $\hat{P}$ an element of $\mathsf{Sym}(N_{\mathrm{e}})$, the symmetric group of degree $N_{\mathrm{e}}$, and $\pi(\hat{P})$ the parity of $\hat{P}$.
        The antisymmetrizer acts on the electronic spin-spatial coordinates $\mathbf{x}_i$ in terms of which the spin-orbitals $\chi_i$ are written.
        In these cases, the linear space $V$ (\textit{cf.} paragraph ``Characters and representation matrices'' under Section~\ref{sec:linrepanalysis}) is chosen to be an $N_{\mathrm{e}}$-particle Hilbert space denoted $\mathcal{H}_{N_{\mathrm{e}}}$.
        It should be noted that the spin-orbitals $\chi_i$ are special cases of $\Psi^{\mathrm{det}}$ with $N_{\mathrm{e}} = 1$.
        Being a Hilbert space, $\mathcal{H}_{N_{\mathrm{e}}}$ comes equipped with the familiar inner product
        \ifdefined\twocolumnmode
          \begin{multline}
            \braket{\Psi_w^{\mathrm{det}} | \Psi_x^{\mathrm{det}}} =
            \int
              \Psi_w^{\mathrm{det}}(\mathbf{x}_1, \ldots, \mathbf{x}_{N_{\mathrm{e}}})^*\\ \Psi_x^{\mathrm{det}}(\mathbf{x}_1, \ldots, \mathbf{x}_{N_{\mathrm{e}}})
              \ \D\mathbf{x}_1\ldots\D\mathbf{x}_{N_{\mathrm{e}}}
            \label{eq:hilbertov}
          \end{multline}
        \else
          \begin{equation}
            \braket{\Psi_w^{\mathrm{det}} | \Psi_x^{\mathrm{det}}} =
            \int
              \Psi_w^{\mathrm{det}}(\mathbf{x}_1, \ldots, \mathbf{x}_{N_{\mathrm{e}}})^* \Psi_x^{\mathrm{det}}(\mathbf{x}_1, \ldots, \mathbf{x}_{N_{\mathrm{e}}})
              \ \D\mathbf{x}_1\ldots\D\mathbf{x}_{N_{\mathrm{e}}}
            \label{eq:hilbertov}
          \end{equation}
        \fi
        which can be leveraged to compute the orbit overlap matrix $\mathbf{S}$ [Equation~\eqref{eq:smat}] for $\Psi^{\mathrm{det}}$ under the action of the prevailing group $\mathcal{G}$.
        In particular, for single determinants, the inner product in Equation~\eqref{eq:hilbertov} has a particularly simple form:\cite{article:Plasser2016}
        \ifdefined\twocolumnmode
          \begin{multline}
            \braket{\Psi_w^{\mathrm{det}} | \Psi_x^{\mathrm{det}}} \\=
            \begin{vmatrix*}[c]
              \braket{\chi_{w,1} | \chi_{x,1}} & \cdots & \braket{\chi_{w,1} | \chi_{x,N_{\mathrm{e}}}} \\
              \vdots & \ddots & \vdots \\
              \braket{\chi_{w,N_{\mathrm{e}}} | \chi_{x,1}} & \cdots & \braket{\chi_{w,N_{\mathrm{e}}} | \chi_{x,N_{\mathrm{e}}}}
            \end{vmatrix*},
            \label{eq:detov}
          \end{multline}
        \else
          \begin{equation}
            \braket{\Psi_w^{\mathrm{det}} | \Psi_x^{\mathrm{det}}} =
            \begin{vmatrix*}[c]
              \braket{\chi_{w,1} | \chi_{x,1}} & \cdots & \braket{\chi_{w,1} | \chi_{x,N_{\mathrm{e}}}} \\
              \vdots & \ddots & \vdots \\
              \braket{\chi_{w,N_{\mathrm{e}}} | \chi_{x,1}} & \cdots & \braket{\chi_{w,N_{\mathrm{e}}} | \chi_{x,N_{\mathrm{e}}}}
            \end{vmatrix*},
            \label{eq:detov}
          \end{equation}
        \fi
        where $\chi_{w,i}$ denotes the $i$\textsuperscript{th} occupied spin-orbital of the $\Psi_w^{\mathrm{det}}$ determinant.
        Each spin-orbital can be expanded in terms of the AO basis functions according to
        \begin{equation*}
          \chi_{w,i}(\mathbf{x}) = \sum_{\mu} \varphi_{\mu}(\mathbf{x}) C_{\mu i}^{w},
        \end{equation*}
        where $\varphi_{\mu}(\mathbf{x})$ is an AO spin-spatial basis function and $\mu$ a composite spin-spatial index.
        The required spin-orbital overlaps can then be written as
        \begin{equation*}
          \braket{\chi_{w,i} | \chi_{x,j}} =
            \sum_{\mu\nu} \braket{\varphi_{\mu} | \varphi_{\nu}} C^x_{\nu j} (C^w_{\mu i})^*,
        \end{equation*}
        where the two-center overlap integrals
        \begin{equation}
          \braket{\varphi_{\mu} | \varphi_{\nu}} =
          \int
            \varphi_{\mu}^*(\mathbf{x}) \varphi_{\nu}(\mathbf{x})\ \D\mathbf{x}
          \label{eq:twoeov}
        \end{equation}
        can be easily obtained from many available integral packages for Gaussian AO basis functions (\textit{e.g.}, \textsc{Libint}\cite{software:libint2023} and \textsc{Libcint}\cite{article:Sun2015a}) or London AO basis functions (\textit{e.g.}, \textsc{QUEST}\cite{software:Quest2022}, \textsc{London}\cite{article:Tellgren2008}, \textsc{BAGEL}\cite{software:BAGEL,article:Shiozaki2018}, and \textsc{ChronusQ}\cite{article:Williams-Young2020}).
        \qsymsq{} also implements its own generic $n$-center overlap integral routine based on the recursive algorithm by Honda \textit{et al.}\cite{article:Honda1991} that is capable of handling both Gaussian and London AO basis functions.
          \bibnote{Strictly speaking, the integrals obtained from these packages only give the spatial part of the spin-spatial integral in Equation~\eqref{eq:twoeov}. The spin part is typically handled separately and implicitly, especially for the commonly used spin functions $\alpha$ and $\beta$ whose orthogonality is known.}

        The calculation of overlaps between single determinants [Equation~\eqref{eq:detov}] is available in \qsymsq{}, thus enabling the symmetry analysis of single-determinantal wavefunctions and spin-orbitals.
        Analogous overlap calculations for multi-determinantal wavefunctions are in principle possible, but currently not yet implemented in \qsymsq{}.

      \paragraph{Electron densities.}
        Let us consider next an $N_{\mathrm{e}}$-electron density for an $N_{\mathrm{e}}$-electron wavefunction $\Psi(\mathbf{x}_1, \ldots, \mathbf{x}_{N_{\mathrm{e}}})$:\cite{article:Lieb1983,book:Parr1989}
        \ifdefined\twocolumnmode
          \begin{align*}
            \rho(\mathbf{r})
            &= \sum_{i=1}^{N_{\mathrm{e}}}
              \braket{ \Psi | \delta(\mathbf{r} - \mathbf{r}_i) | \Psi } \\
              &= \begin{multlined}[t]
                N_{\mathrm{e}} \int
                \Psi(\mathbf{r}, s, \mathbf{x}_2, \ldots, \mathbf{x}_{N_{\mathrm{e}}})^*\\
                \Psi(\mathbf{r}, s, \mathbf{x}_2, \ldots, \mathbf{x}_{N_{\mathrm{e}}})
                \ \D s\ \D\mathbf{x}_2 \ldots \D\mathbf{x}_{N_{\mathrm{e}}},
              \end{multlined}
          \end{align*}
        \else
          \begin{align*}
            \rho(\mathbf{r})
            &= \sum_{i=1}^{N_{\mathrm{e}}}
              \braket{ \Psi | \delta(\mathbf{r} - \mathbf{r}_i) | \Psi } \\
            &= N_{\mathrm{e}} \int
              \Psi(\mathbf{r}, s, \mathbf{x}_2, \ldots, \mathbf{x}_{N_{\mathrm{e}}})^*
              \Psi(\mathbf{r}, s, \mathbf{x}_2, \ldots, \mathbf{x}_{N_{\mathrm{e}}})
              \ \D s\ \D\mathbf{x}_2 \ldots \D\mathbf{x}_{N_{\mathrm{e}}},
          \end{align*}
        \fi
        where the composite spin-spatial coordinate $\mathbf{x}_1$ has been relabeled and separated into a spin coordinate $s$ and a spatial coordinate $\mathbf{r}$ in the integrand.
        In an AO basis, $\rho(\mathbf{r})$ can be expanded as
        \begin{equation}
          \rho(\mathbf{r}) = \sum_{\gamma \delta}
            \phi_{\gamma}(\mathbf{r})
            \phi_{\delta}(\mathbf{r})
            P_{\delta\gamma},
          \label{eq:denmatform}
        \end{equation}
        where $\phi_{\gamma}(\mathbf{r})$ and $\phi_{\delta}(\mathbf{r})$ are spatial AO basis functions, $\gamma$ and $\delta$ spatial indices, and $P_{\delta\gamma}$ elements of the corresponding density matrix $\mathbf{P}$ in this basis.

        The containing linear space for $\rho(\mathbf{r})$ is well known to be the Banach space $\mathcal{X} = L^3(\mathbb{R}^3) \cap L^1(\mathbb{R}^3)$\cite{article:Lieb1983}.
        Being a Banach space, $\mathcal{X}$ does not have an intrinsic inner product, but, as explained in Section~\ref{sec:linrepanalysis}, it is possible to endow $\mathcal{X}$ with an inner product for the purpose of representation analysis.
        The simplest such inner product can be defined as follows:
        \begin{align}
          \braket{\cdot | \cdot}: \mathcal{X} \times \mathcal{X}
            &\to \mathbb{C} \nonumber \\
          (\rho_w, \rho_x) &\mapsto \braket{\rho_w | \rho_x} \equiv  \int \rho_w^*(\mathbf{r}) \rho_x(\mathbf{r})\ \D\mathbf{r}.
          \label{eq:denov}
        \end{align}
        It is straightforward to show that the above definition for $\braket{\cdot | \cdot}$ satisfies all required properties of an inner product:
        \begin{itemize}
          \item conjugate symmetry:
          \ifdefined\twocolumnmode
            \begin{align*}
              \braket{\rho_w | \rho_x}
              &= \int \rho_w^*(\mathbf{r}) \rho_x(\mathbf{r})\ \D\mathbf{r}\\
              &= \left[\int \rho_x^*(\mathbf{r}) \rho_w(\mathbf{r})\ \D\mathbf{r}\right]^*\\
              &= \braket{\rho_x | \rho_w}^*,
            \end{align*}
          \else
            \begin{equation*}
              \braket{\rho_w | \rho_x}
              = \int \rho_w^*(\mathbf{r}) \rho_x(\mathbf{r})\ \D\mathbf{r}
              = \left[\int \rho_x^*(\mathbf{r}) \rho_w(\mathbf{r})\ \D\mathbf{r}\right]^*
              = \braket{\rho_x | \rho_w}^*,
            \end{equation*}
          \fi
          \item linearity in the second argument:
          \ifdefined\twocolumnmode
            \begin{multline*}
              \braket{\rho_w | a \rho_x + b \rho_t}\\
              \begin{aligned}
                &= \int \rho_w^*(\mathbf{r}) \left[a \rho_x(\mathbf{r}) + b \rho_t(\mathbf{r})\right] \D\mathbf{r}\\
                &= a \int \rho_w^*(\mathbf{r}) \rho_x(\mathbf{r}) \D\mathbf{r} + b \int \rho_w^*(\mathbf{r}) \rho_t(\mathbf{r}) \D\mathbf{r}\\
                &= a\braket{\rho_w | \rho_x} + b\braket{\rho_w | \rho_t} \quad \textrm{for $a, b \in \mathbb{R}$},
              \end{aligned}
            \end{multline*}
          \else
            \begin{align*}
              \braket{\rho_w | a \rho_x + b \rho_t}
              &= \int \rho_w^*(\mathbf{r}) \left[a \rho_x(\mathbf{r}) + b \rho_t(\mathbf{r})\right] \D\mathbf{r}\\
              &= a \int \rho_w^*(\mathbf{r}) \rho_x(\mathbf{r}) \D\mathbf{r} + b \int \rho_w^*(\mathbf{r}) \rho_t(\mathbf{r}) \D\mathbf{r}\\
              &= a\braket{\rho_w | \rho_x} + b\braket{\rho_w | \rho_t} \quad \textrm{for $a, b \in \mathbb{R}$},
            \end{align*}
          \fi
          \item positive-definiteness:
          \ifdefined\twocolumnmode
            \begin{align*}
              \braket{\rho | \rho}
              &= \int \rho^*(\mathbf{r}) \rho(\mathbf{r}) \D\mathbf{r}\\
              &= \int \left\lvert\rho(\mathbf{r})\right\rvert^2 \D\mathbf{r}
                \ge 0 \ \textrm{since $\left\lvert\rho(\mathbf{r})\right\rvert^2 \ge 0$},
            \end{align*}
          \else
            \begin{equation*}
              \braket{\rho | \rho}
              = \int \rho^*(\mathbf{r}) \rho(\mathbf{r}) \D\mathbf{r}
              = \int \left\lvert\rho(\mathbf{r})\right\rvert^2 \D\mathbf{r}
              \ge 0 \quad \textrm{since $\left\lvert\rho(\mathbf{r})\right\rvert^2 \ge 0$},
            \end{equation*}
          \fi
          where $\braket{\rho | \rho} = 0$ if and only if $\rho(\mathbf{r}) = 0$ identically, otherwise there would exist regions in $\mathbb{R}^3$ where $\left\lvert\rho(\mathbf{r})\right\rvert^2 < 0$, which is not possible.
        \end{itemize}
        Using the basis-expanded form of the electron density in Equation~\eqref{eq:denmatform} in the inner product definition in Equation~\eqref{eq:denov} gives
        \begin{equation*}
          \braket{\rho_w | \rho_x} =
            \sum_{\gamma \gamma' \delta \delta'}
            \braket{\phi_{\gamma} \phi_{\delta} | \phi_{\gamma'} \phi_{\delta'}}
            P^x_{\delta'\gamma'} (P^w_{\delta\gamma})^*,
        \end{equation*}
        where
        \begin{equation*}
          \braket{\phi_{\gamma} \phi_{\delta} | \phi_{\gamma'} \phi_{\delta'}}
          = \int
            \phi_{\gamma}^*(\mathbf{r})  \phi_{\delta}^*(\mathbf{r})
            \phi_{\gamma'}(\mathbf{r}) \phi_{\delta'}(\mathbf{r})
            \ \D\mathbf{r}
        \end{equation*}
        are four-center overlap integrals computable using the generic $n$-center overlap integral routine as described earlier.

        The calculation of overlaps between electron densities [Equation~\eqref{eq:denov}] is available in \qsymsq{}, thus enabling the symmetry analysis of electron densities obtained from a wide range of electronic-structure methods from single- and multi-determinantal wavefunctions to DFT.

%% file: illustrations/illustrations.tex
\tikzsetexternalprefix{./illustrations/tikz/}

\section{Results and Discussion}
\label{sec:illustrations}

  In this Section, several case studies showcasing the capabilities and utility of \qsymsq{} are presented.
  Each case study is based around a distinct computational chemical problem whose results can be better understood by a detailed analysis of symmetry provided by \qsymsq{}.

  \subsection{Degeneracy in non-Abelian groups}
  \label{sec:degeneracyscf}

\begin{figure*}
        \centering
        \begin{subfigure}{.3\textwidth}
          \includegraphics[trim=0.0cm 0.2cm 0.0cm 0.2cm, clip, scale=3.0]{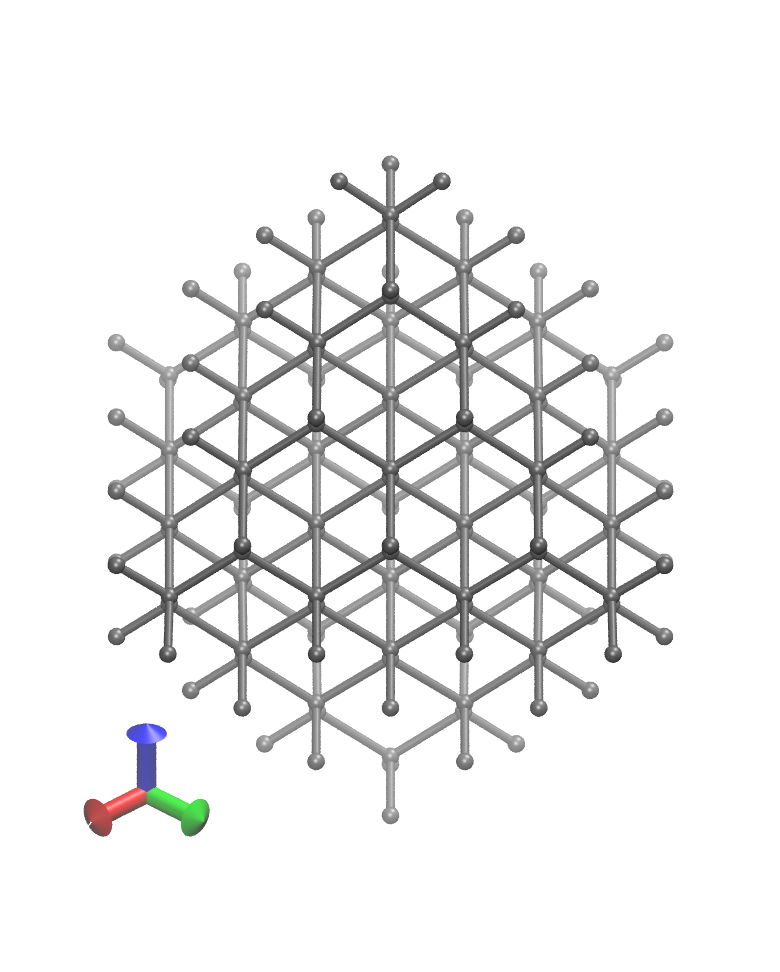}
          \caption{\ce{C84H64} ($\mathcal{T}_{d}$)}
          \label{fig:c84h64}
        \end{subfigure}
        \begin{subfigure}{.3\textwidth}
          \includegraphics[trim=0.0cm 0.2cm 0.0cm 0.2cm, clip, scale=3.0]{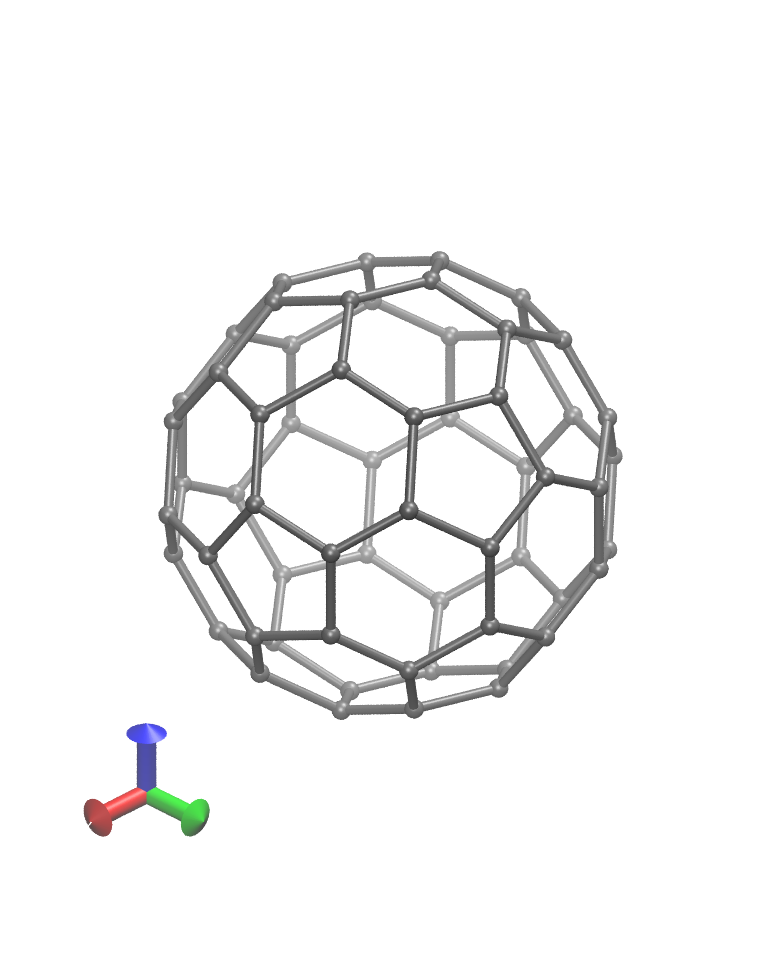}
          \caption{\ce{C60} ($\mathcal{I}_{h}$)}
          \label{fig:c60}
        \end{subfigure}
        \begin{subfigure}{.3\textwidth}
          \includegraphics[trim=0.0cm 0.30cm 0.0cm 0.35cm, clip, scale=3.0]{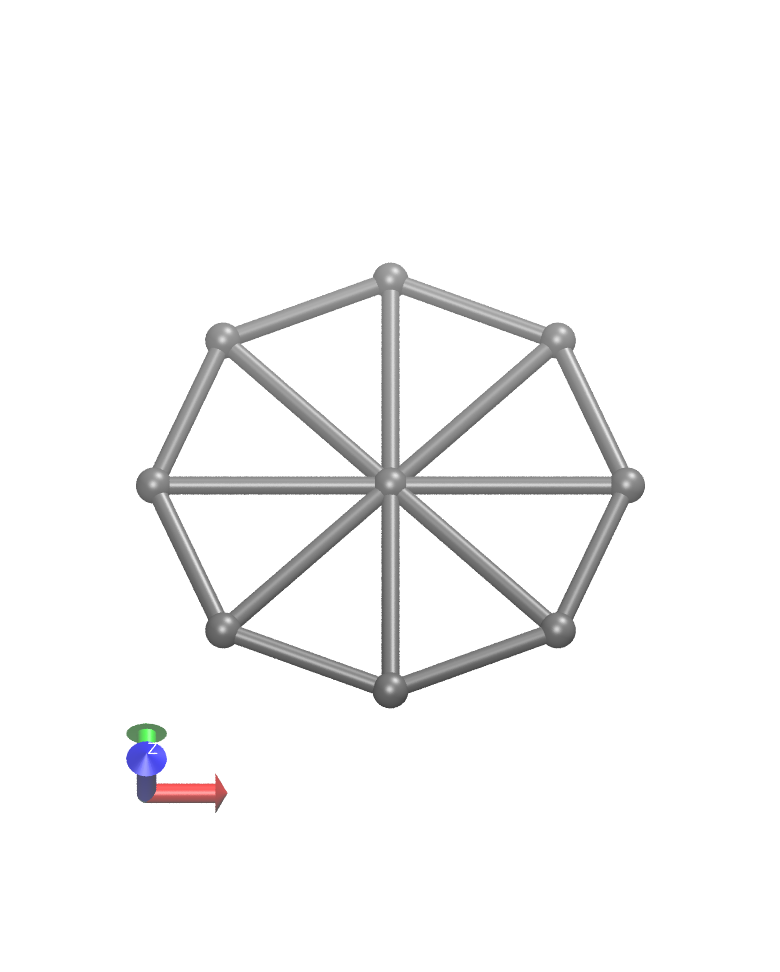}
          \caption{\ce{B9-} ($\mathcal{D}_{8h}$)}
          \label{fig:b9m}
        \end{subfigure}
        \caption{High-symmetry molecules whose ground-state KS MOs exhibit degeneracy.}
        \label{fig:degmols}
      \end{figure*}

    The first set of case studies consists of three molecules of various sizes and symmetries: tetrahedral \ce{C84H64} quantum dot (Figure~\ref{fig:c84h64}), icosahedral \ce{C60} (Figure~\ref{fig:c60}), and octagonal \ce{B9-} (Figure~\ref{fig:b9m}).
    The non-Abelian symmetry of these three molecules allows for degeneracy to occur in the SCF MOs that arise from either a HF or a KS-DFT description of the ground state of the system.
    By examining the symmetry of such degenerate MOs in the ground SCF solutions of these three molecules, we seek to demonstrate the capability of \qsymsq{} to determine degenerate irreducible representation labels accurately, irrespective of the size of the molecules or the complexity of the MOs.

    \subsubsection{Computational details}
      For each molecule, a ground-state KS-DFT calculation using an appropriate exchange-correlation functional and basis set was performed in \textsc{Q-Chem} 6.1.0.
      In all calculations, suitable symmetry thresholds were chosen to ensure that \textsc{Q-Chem} produced symmetry assignments for the computed MOs in the highest possible group.
      Afterwards, all geometry information, basis set information, and MO coefficients from these calculations were passed to \qsymsq{} where the unitary symmetry group $\mathcal{G}$ of the system was deduced, following which the representations of $\mathcal{G}$ spanned by the MOs were identified and analyzed according to the formulation given in Section~\ref{sec:repanalysis}.
      The MO symmetry assignments from \qsymsq{} were then compared with those from \textsc{Q-Chem}.

      \textsc{Q-Chem} was chosen as the benchmarking program for these case studies because of its ability to assign degenerate symmetry labels in certain non-Abelian groups.
      As stated in Section~\ref{sec:intro}, most other quantum-chemistry programs are only able to perform symmetry analysis in Abelian groups and are thus not suitable for this purpose.

    \subsubsection{Degenerate symmetry benchmarks}

      We begin with tetrahedral \ce{C84H64}, which proves to be a straightforward case.
      Using the geometry reported by Karttunen \textit{et al.}\cite{article:Karttunen2008}, a ground-state unrestricted CAM-B3LYP/6-31+G* calculation was performed and a set of KS MOs were obtained.
      Table~\ref{tab:c84h64mos} shows the symmetry assignments that have been produced by both \textsc{Q-Chem} and \qsymsq{} for the frontier MOs.
      These particular MOs have been chosen because they have been identified in Ref.~\citenum{article:Foerster2022} to be responsible for the most intense transition in the electronic absorption spectrum of this molecule, and are therefore the most interesting to examine from a symmetry perspective.
      For this system, \textsc{Q-Chem} is able to identify its symmetry group as $\mathcal{T}_d$, as is \qsymsq{}.
      Both programs are also able to agree on their symmetry assignments of the frontier MOs, be they degenerate or not, thus confirming that the representation analysis formulation implemented in \qsymsq{} (Section~\ref{sec:repanalysis}) is valid.

      \begin{table*}[t]
        \centering
        \caption{%
          Comparison of symmetry assignments of frontier canonical MOs in \ce{C84H64} calculated at the CAM-B3LYP/6-31+G* level of theory using the geometry reported by Karttunen \textit{et al.}\cite{article:Karttunen2008}
          For each MO $\chi(\mathbf{r})$, the isosurface is plotted at $\lvert \chi(\mathbf{r}) \rvert = 0.008$.
        }
        \label{tab:c84h64mos}
        \renewcommand*{\arraystretch}{0.5}
        \begin{subtable}[t]{.48\textwidth}
          \captionsetup{justification=centering}
          \centering
          \caption{Highest occupied MOs.}
          \begin{tabular}{%
            M{0.80cm} M{2.0cm} M{1.70cm} M{1.70cm}
          }
          \toprule
          MO
          & Isosurface
          & \textsc{Q-Chem} ($\mathcal{T}_{d}$)
          & \qsymsq{} ($\mathcal{T}_{d}$)\\
          \midrule
          $\chi^{\alpha}_{282}$
          & \includegraphics[trim=0.0cm 0.27cm 0.0cm 0.40cm, clip, scale=1.3]{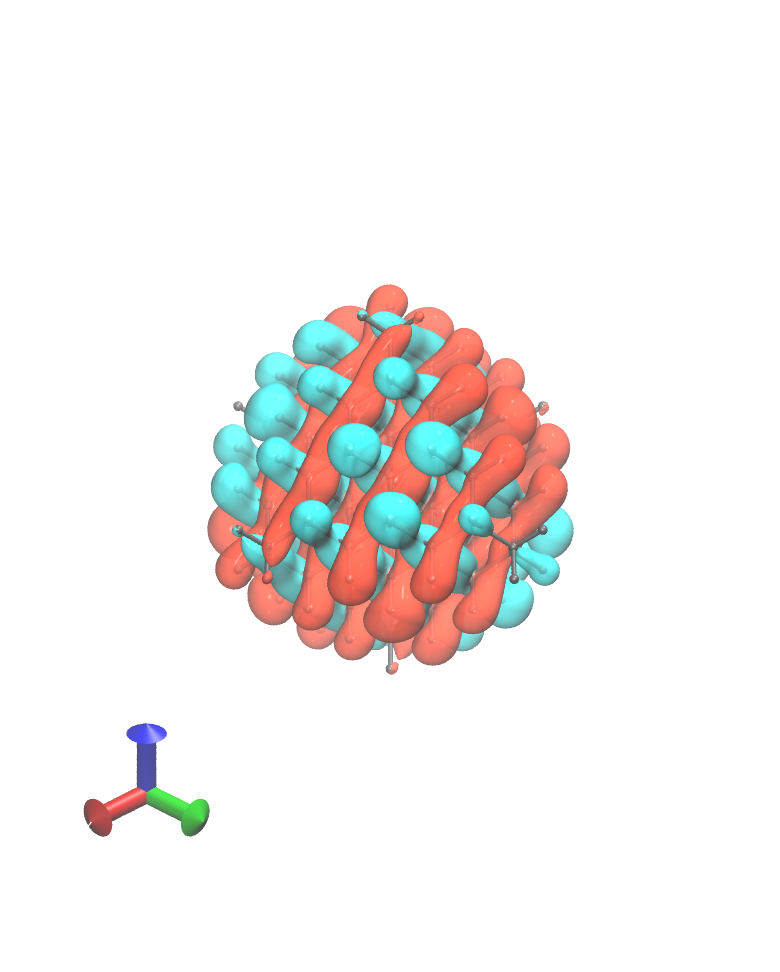}
          & $T_2$
          & $T_2$\\
          $\chi^{\alpha}_{283}$
          & \includegraphics[trim=0.0cm 0.27cm 0.0cm 0.40cm, clip, scale=1.3]{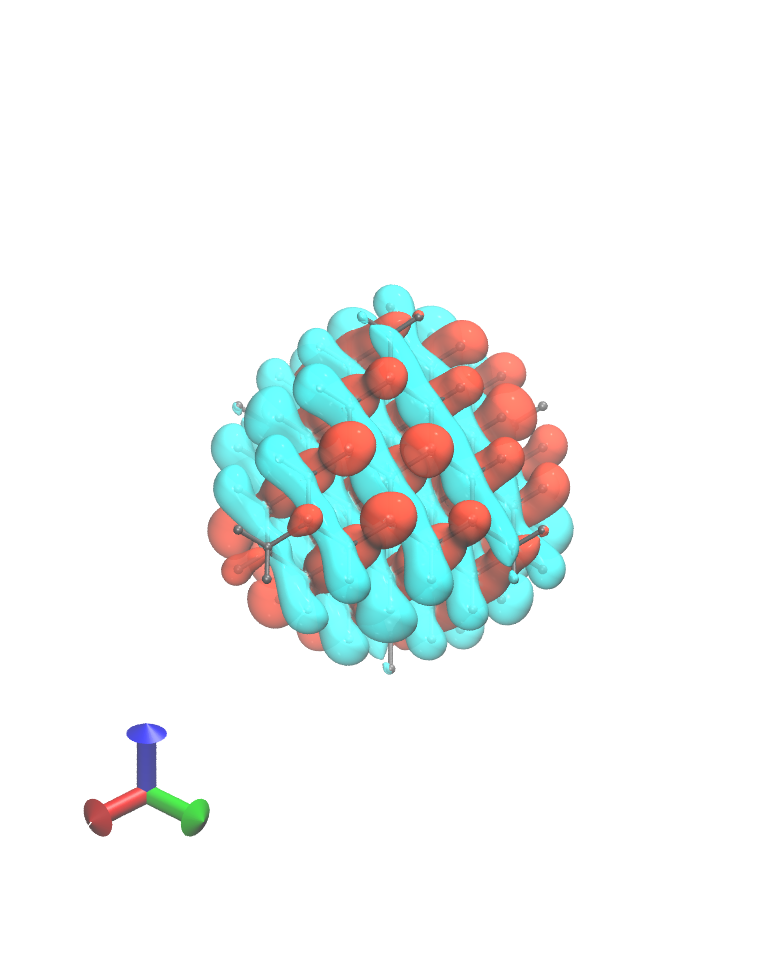}
          & $T_2$
          & $T_2$\\
          $\chi^{\alpha}_{284}$
          & \includegraphics[trim=0.0cm 0.27cm 0.0cm 0.40cm, clip, scale=1.3]{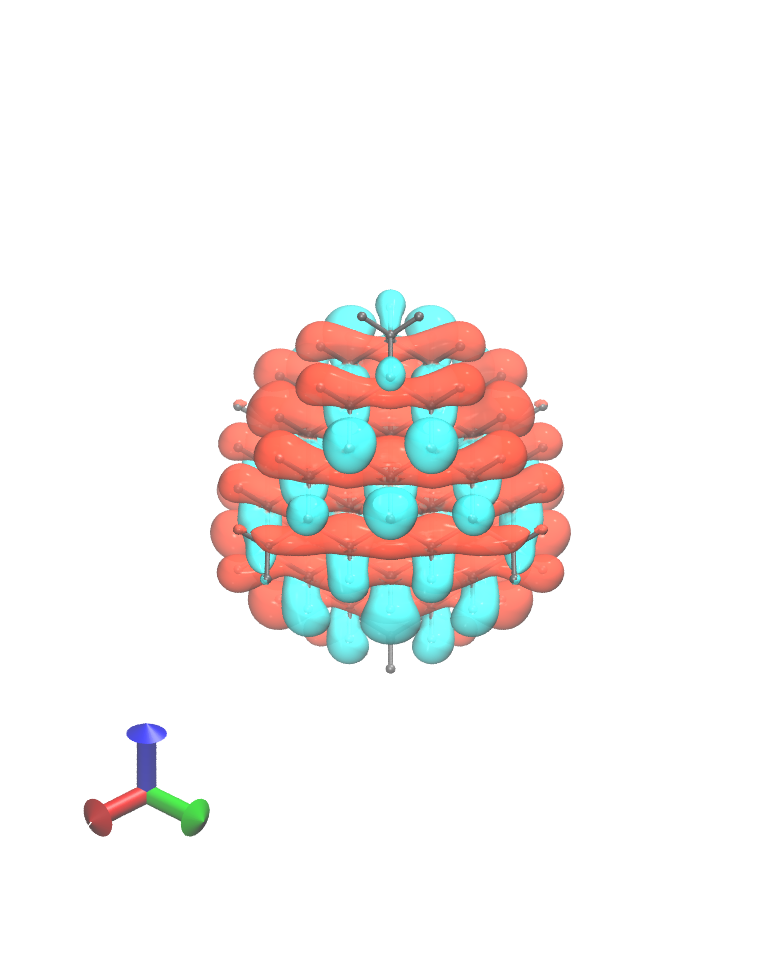}
          & $T_2$
          & $T_2$\\
          \bottomrule
          \end{tabular}
        \end{subtable}
        \hfill
        \begin{subtable}[t]{.48\textwidth}
          \captionsetup{justification=centering}
          \centering
          \caption{Lowest unoccupied MOs.}
          \begin{tabular}{%
            M{0.80cm} M{2.0cm} M{1.70cm} M{1.70cm}
          }
          \toprule
          MO
          & Isosurface
          & \textsc{Q-Chem} ($\mathcal{T}_{d}$)
          & \qsymsq{} ($\mathcal{T}_{d}$)\\
          \midrule
          $\chi^{\alpha}_{285}$
          & \includegraphics[trim=0.0cm 0.27cm 0.0cm 0.40cm, clip, scale=1.3]{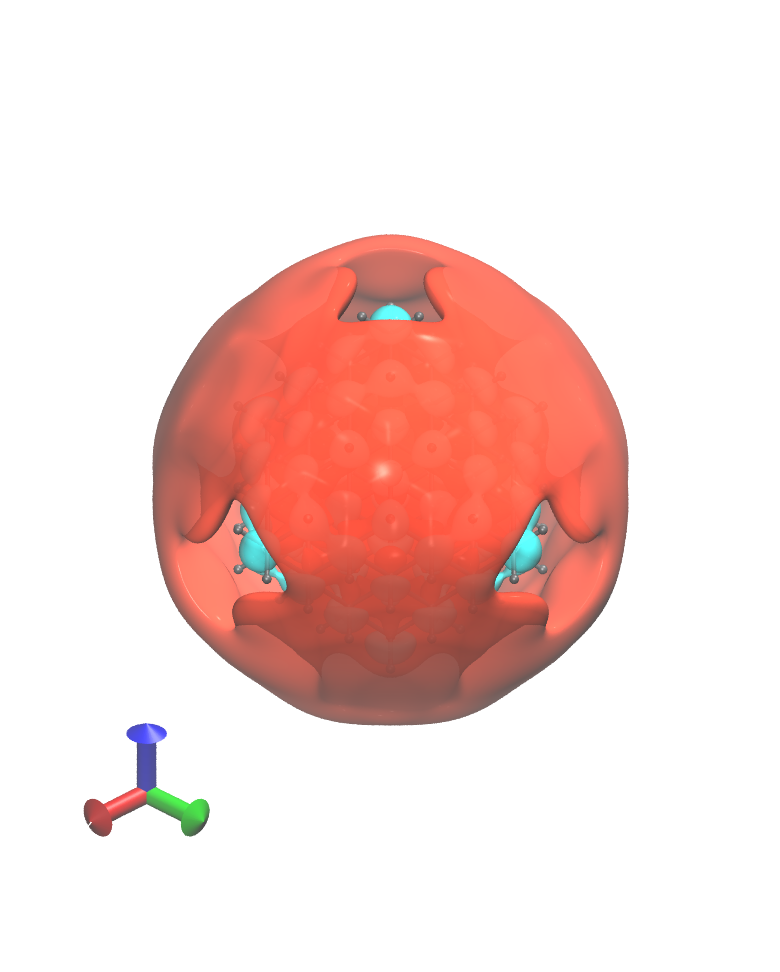}
          & $A$
          & $A$\\
          \midrule
          $\chi^{\alpha}_{286}$
          & \includegraphics[trim=0.0cm 0.27cm 0.0cm 0.40cm, clip, scale=1.3]{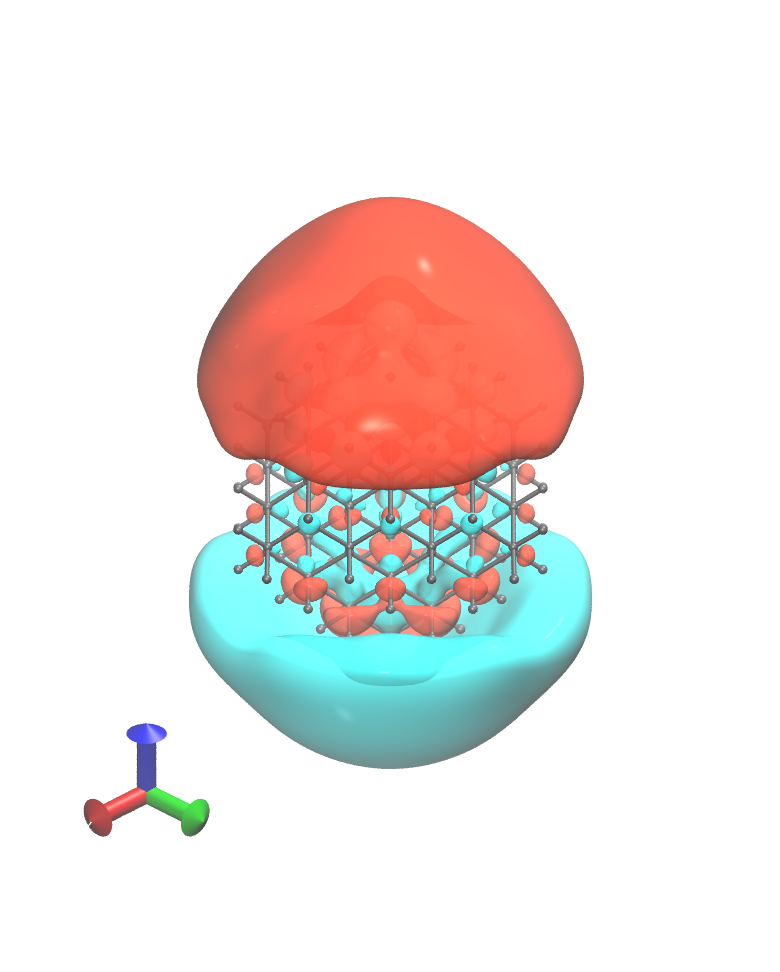}
          & $T_2$
          & $T_2$\\
          $\chi^{\alpha}_{287}$
          & \includegraphics[trim=0.0cm 0.27cm 0.0cm 0.40cm, clip, scale=1.3]{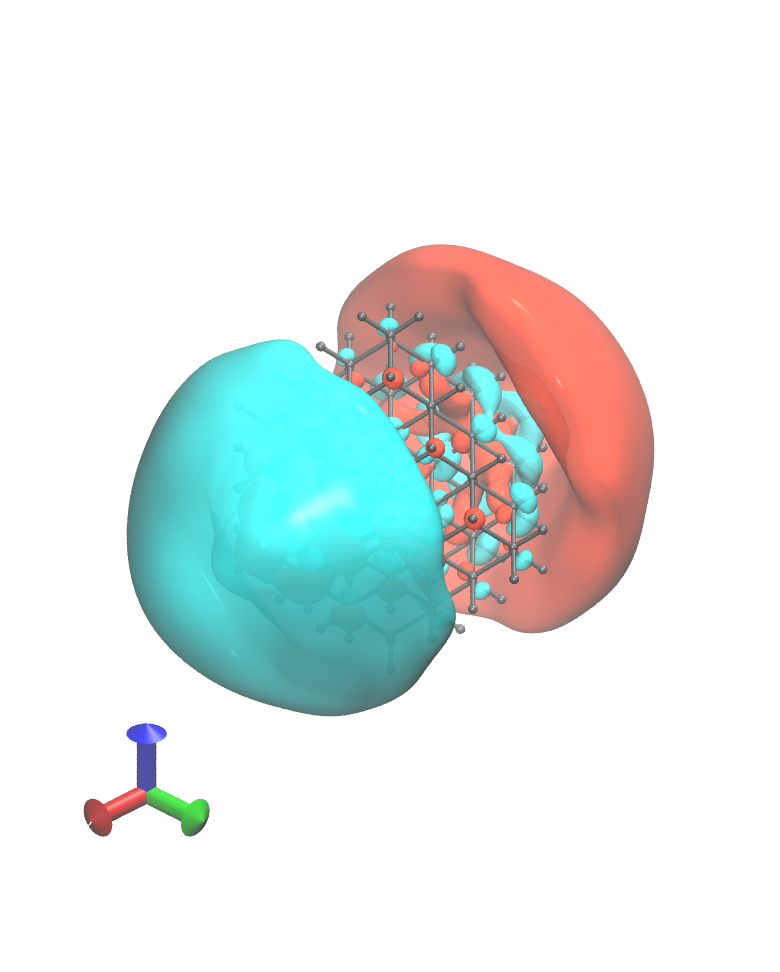}
          & $T_2$
          & $T_2$\\
          $\chi^{\alpha}_{288}$
          & \includegraphics[trim=0.0cm 0.27cm 0.0cm 0.40cm, clip, scale=1.3]{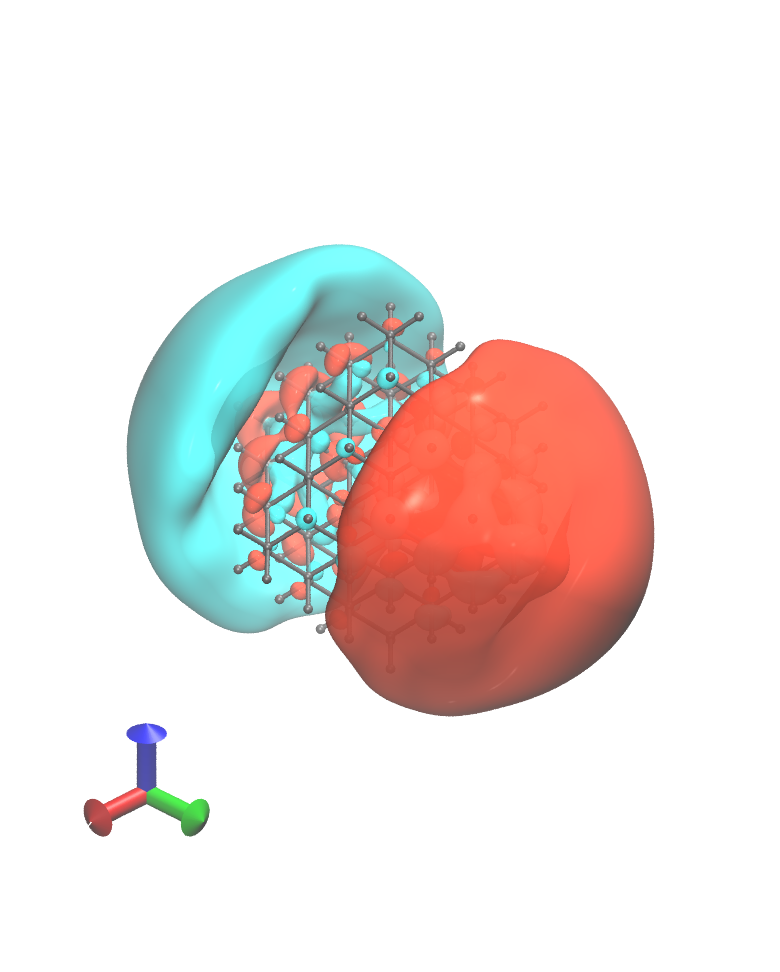}
          & $T_2$
          & $T_2$\\
          \bottomrule
          \end{tabular}
        \end{subtable}
      \end{table*}

      \begin{table*}
        \centering
        \caption{%
          Comparison of symmetry assignments of frontier occupied canonical MOs in \ce{C60} calculated at the CAM-B3LYP/6-31+G* level of theory using an $\mathcal{I}_h$-symmetrized geometry.
          For each MO $\chi(\mathbf{r})$, the isosurface is plotted at $\lvert \chi(\mathbf{r}) \rvert = 0.008$.
          The four- and five-dimensional irreducible representation labels follow the convention specified in Ref.~\citenum{book:Altmann2011}.
        }
        \label{tab:c60mos}
        \renewcommand*{\arraystretch}{0.5}
        \begin{tabular}{%
          M{0.80cm} M{2.0cm} M{1.70cm} M{1.70cm}
        }
        \toprule
        MO
        & Isosurface
        & \textsc{Q-Chem} ($\mathcal{C}_{2h}$)
        & \qsymsq{} ($\mathcal{I}_{h}$)\\
        \midrule
        $\chi^{\alpha}_{167}$
        & \includegraphics[trim=0.0cm 0.2cm 0.0cm 0.2cm, clip, scale=1.0]{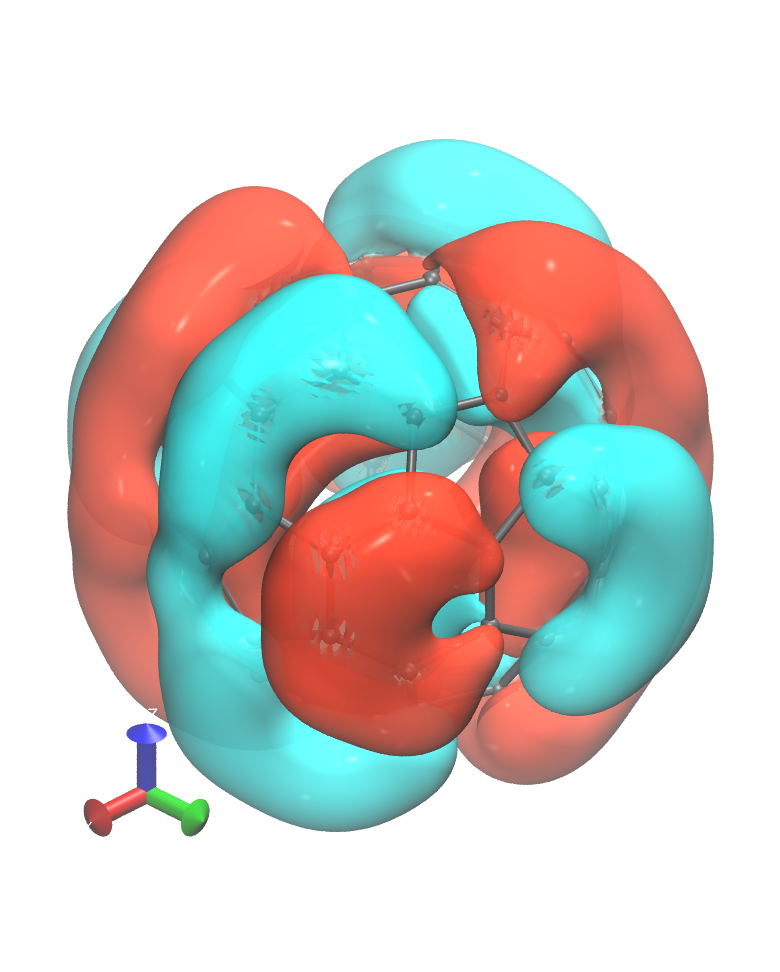}
        & $A_g$
        & $F_g$\\
        $\chi^{\alpha}_{168}$
        & \includegraphics[trim=0.0cm 0.2cm 0.0cm 0.2cm, clip, scale=1.0]{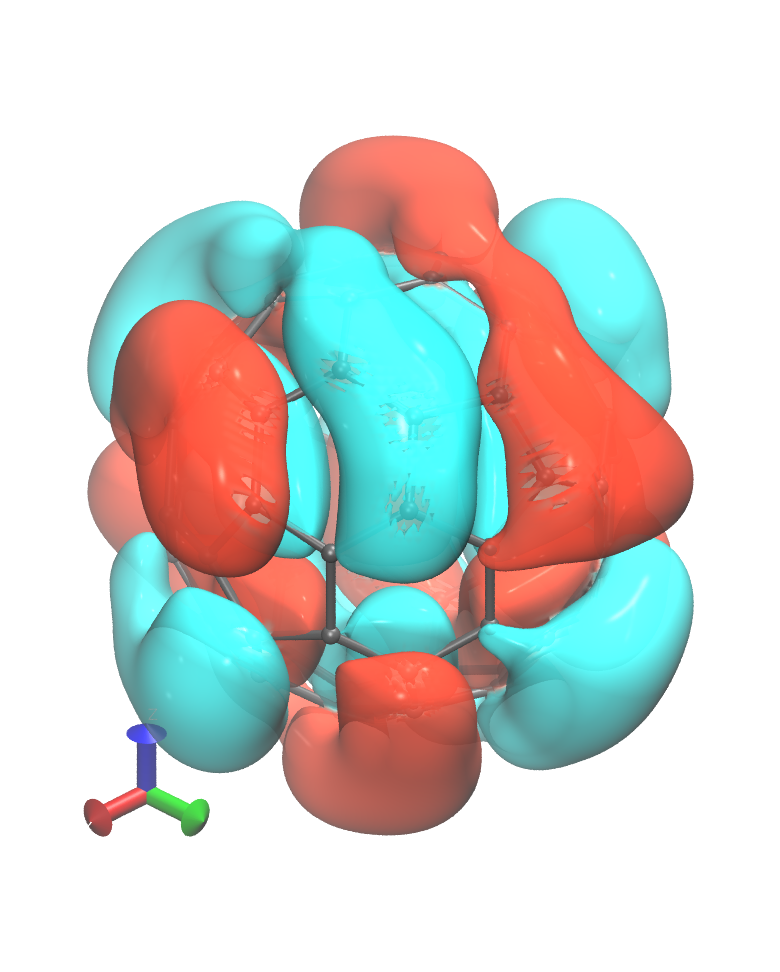}
        & $B_g$
        & $F_g$\\
        $\chi^{\alpha}_{169}$
        & \includegraphics[trim=0.0cm 0.2cm 0.0cm 0.2cm, clip, scale=1.0]{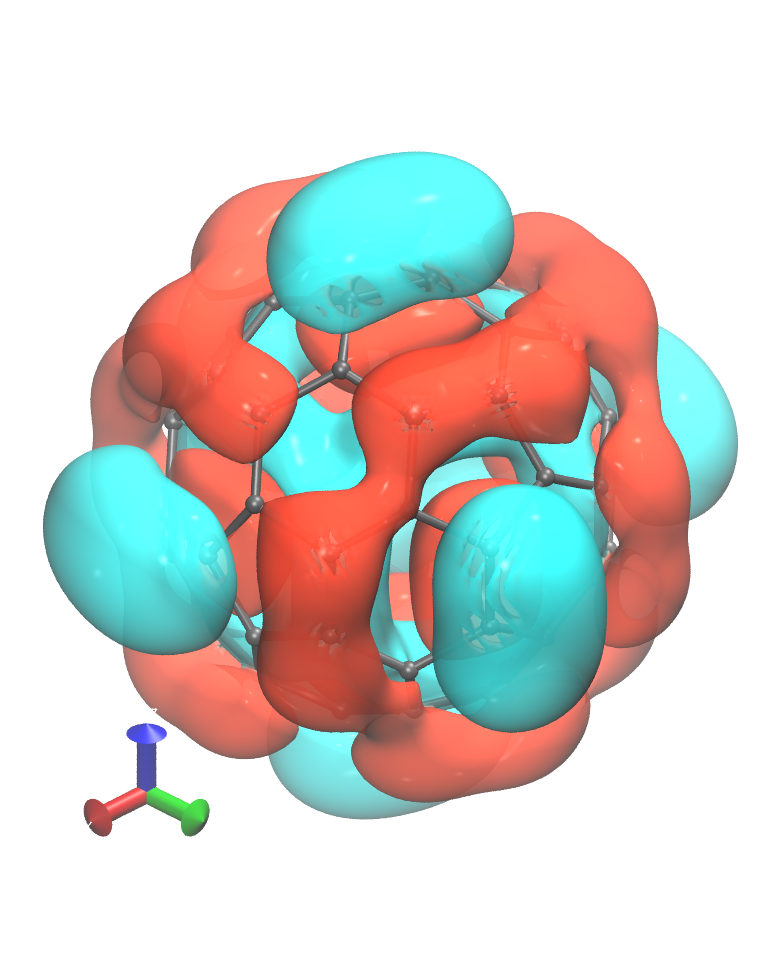}
        & $A_g$
        & $F_g$\\
        $\chi^{\alpha}_{170}$
        & \includegraphics[trim=0.0cm 0.2cm 0.0cm 0.2cm, clip, scale=1.0]{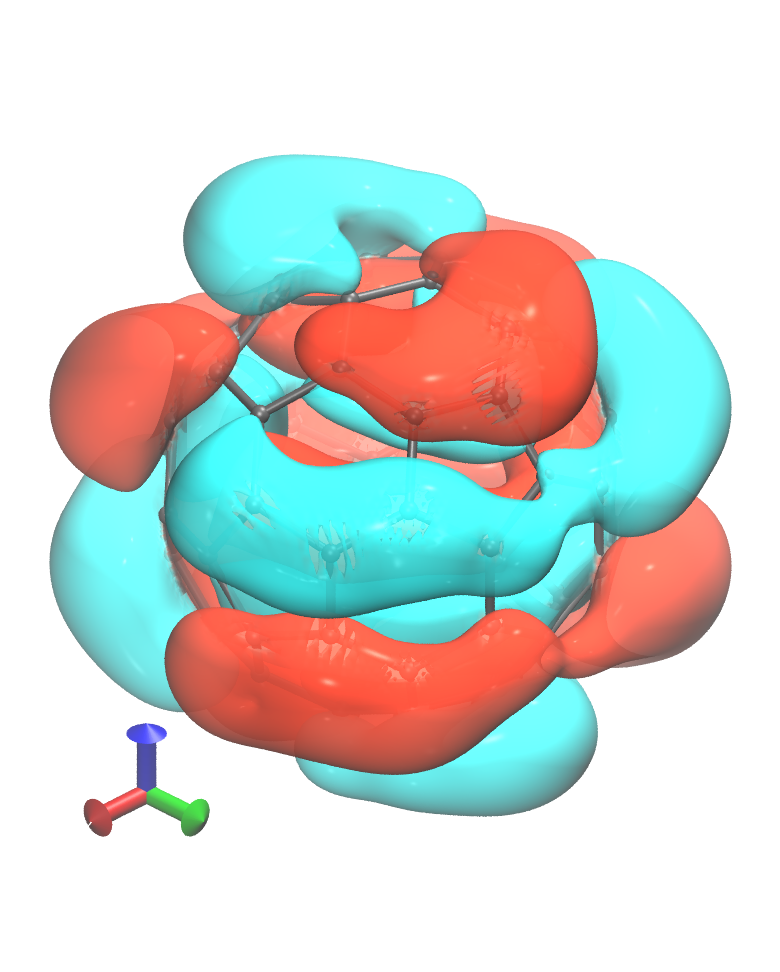}
        & $B_g$
        & $F_g$\\
        \bottomrule
        \end{tabular}

        \vspace{0.7cm}

        \begin{tabular}{%
          M{0.80cm} M{2.0cm} M{1.70cm} M{1.70cm}
        }
        \toprule
        MO
        & Isosurface
        & \textsc{Q-Chem} ($\mathcal{C}_{2h}$)
        & \qsymsq{} ($\mathcal{I}_{h}$)\\
        \midrule
        $\chi^{\alpha}_{171}$
        & \includegraphics[trim=0.0cm 0.2cm 0.0cm 0.2cm, clip, scale=1.0]{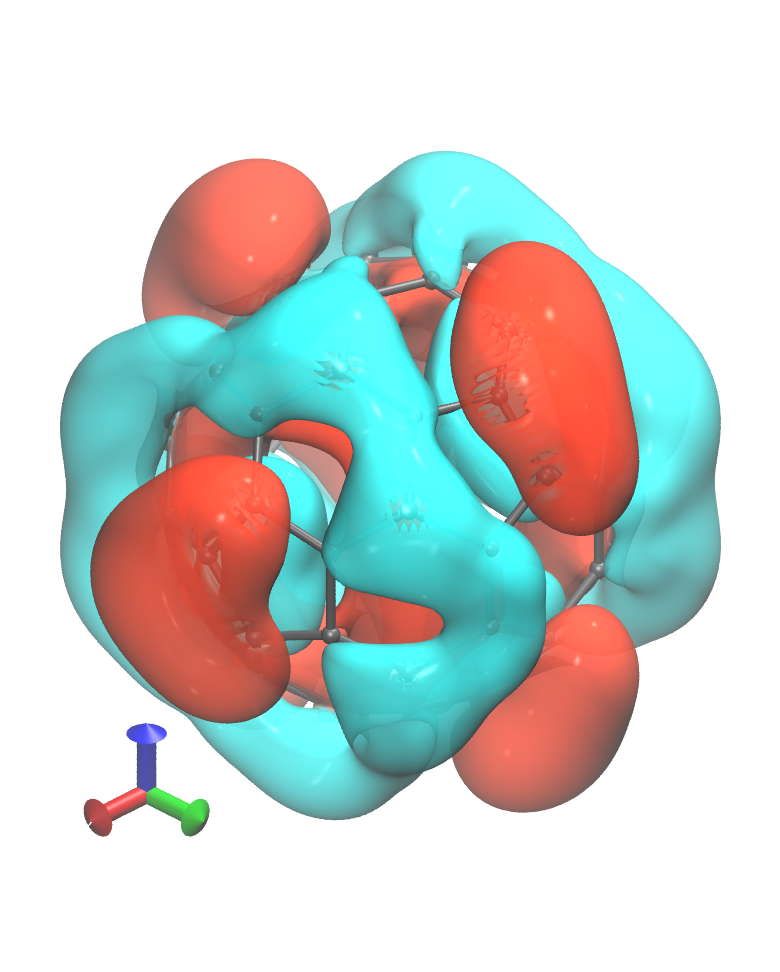}
        & $A_g$
        & $H_g$\\
        $\chi^{\alpha}_{172}$
        & \includegraphics[trim=0.0cm 0.2cm 0.0cm 0.2cm, clip, scale=1.0]{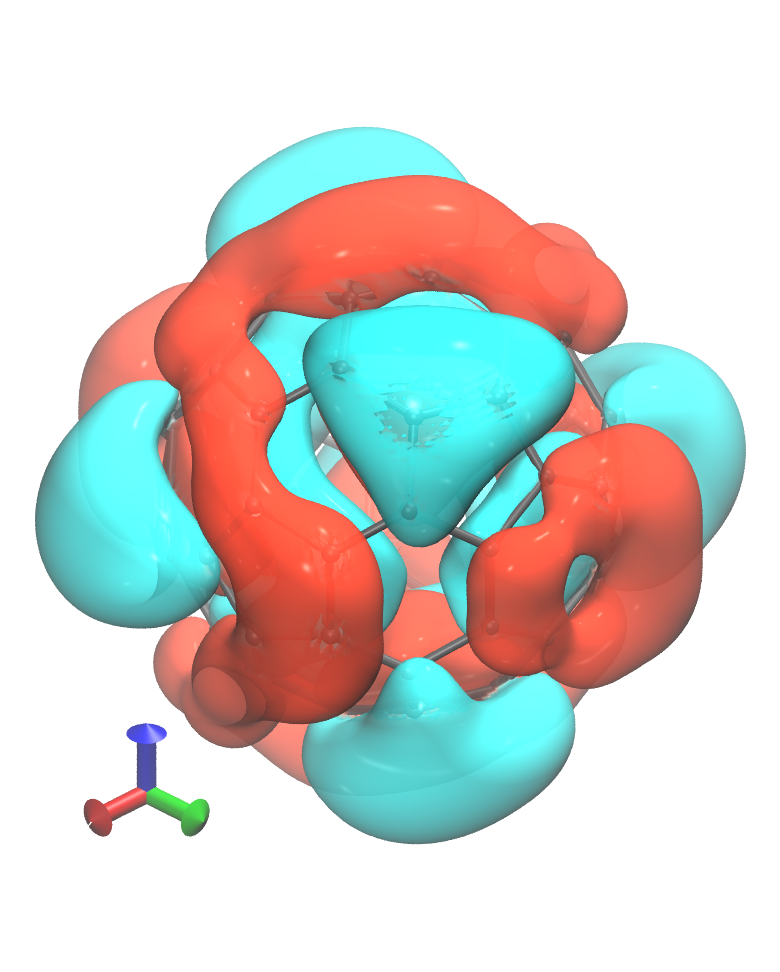}
        & $A_g$
        & $H_g$\\
        $\chi^{\alpha}_{173}$
        & \includegraphics[trim=0.0cm 0.2cm 0.0cm 0.2cm, clip, scale=1.0]{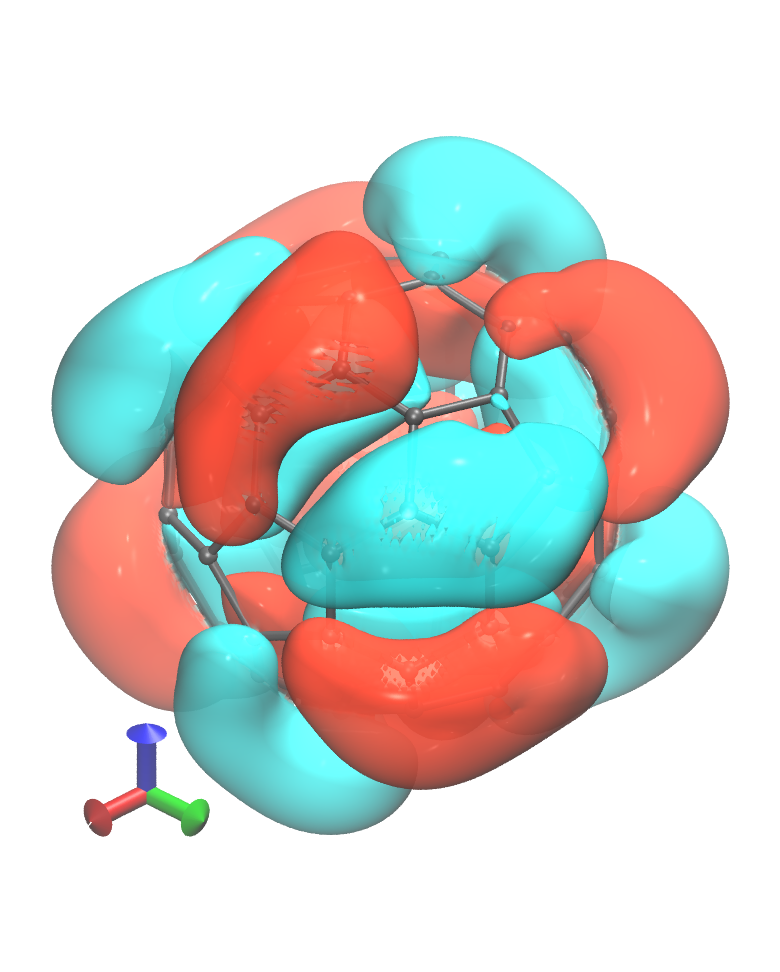}
        & $B_g$
        & $H_g$\\
        $\chi^{\alpha}_{174}$
        & \includegraphics[trim=0.0cm 0.2cm 0.0cm 0.2cm, clip, scale=1.0]{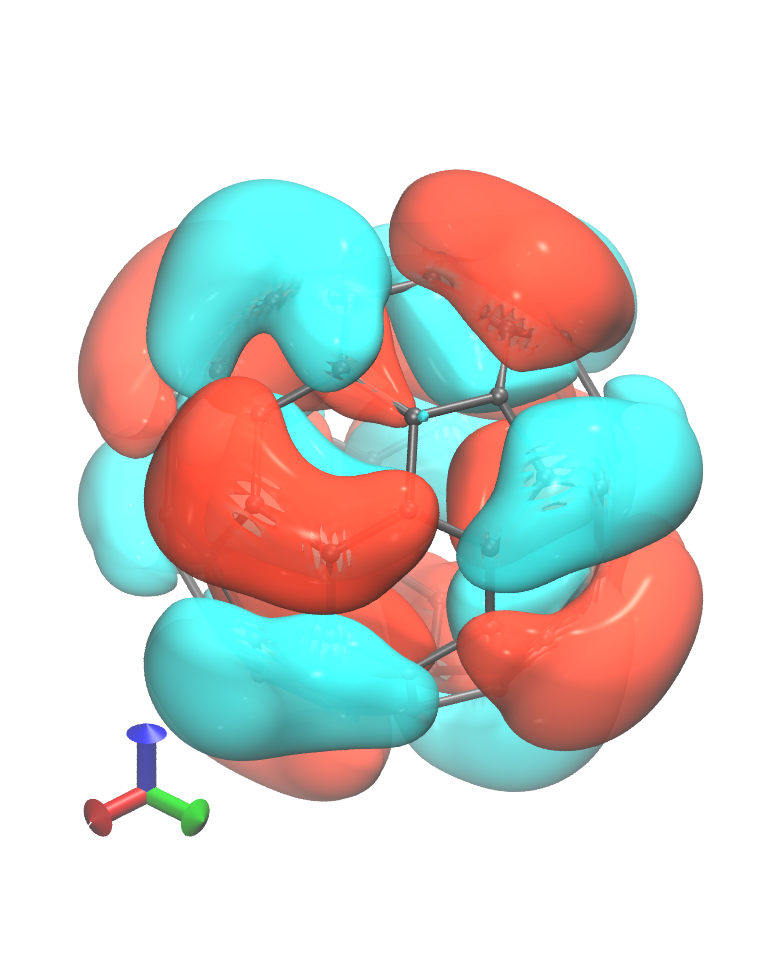}
        & $B_g$
        & $H_g$\\
        $\chi^{\alpha}_{175}$
        & \includegraphics[trim=0.0cm 0.2cm 0.0cm 0.2cm, clip, scale=1.0]{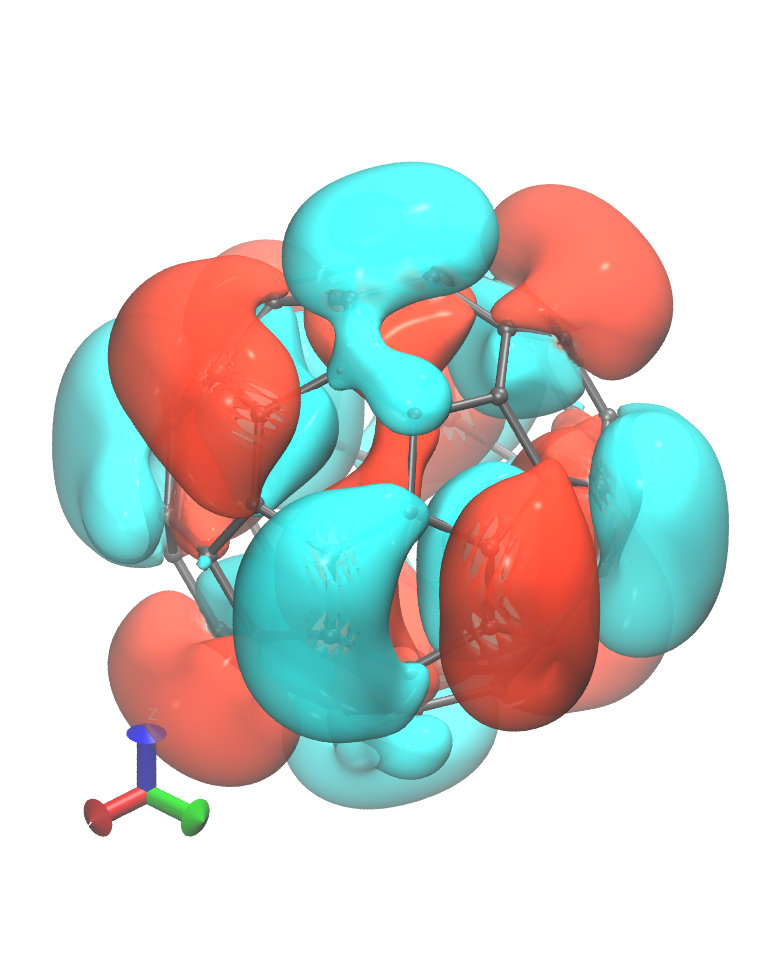}
        & $A_g$
        & $H_g$\\
        \bottomrule
        \end{tabular}
        \hspace{0.4cm}
        \begin{tabular}{%
          M{0.80cm} M{2.0cm} M{1.70cm} M{1.70cm}
        }
        \toprule
        MO
        & Isosurface
        & \textsc{Q-Chem} ($\mathcal{C}_{2h}$)
        & \qsymsq{} ($\mathcal{I}_{h}$)\\
        \midrule
        $\chi^{\alpha}_{176}$
        & \includegraphics[trim=0.0cm 0.2cm 0.0cm 0.2cm, clip, scale=1.0]{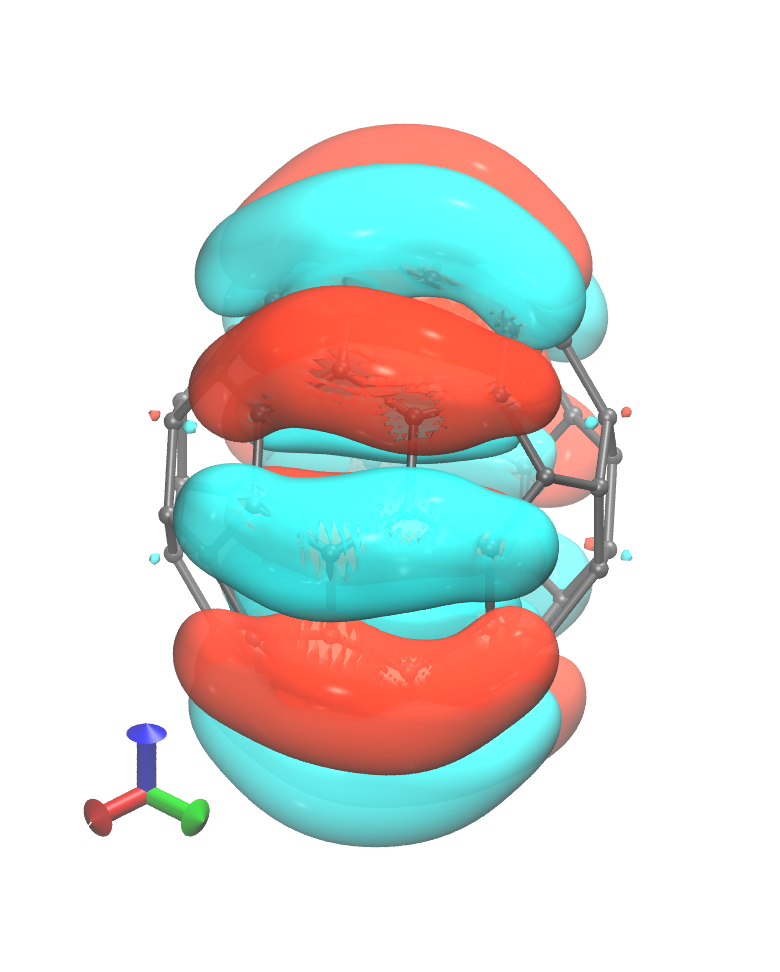}
        & $A_{u}$
        & $H_u$\\
        $\chi^{\alpha}_{177}$
        & \includegraphics[trim=0.0cm 0.2cm 0.0cm 0.2cm, clip, scale=1.0]{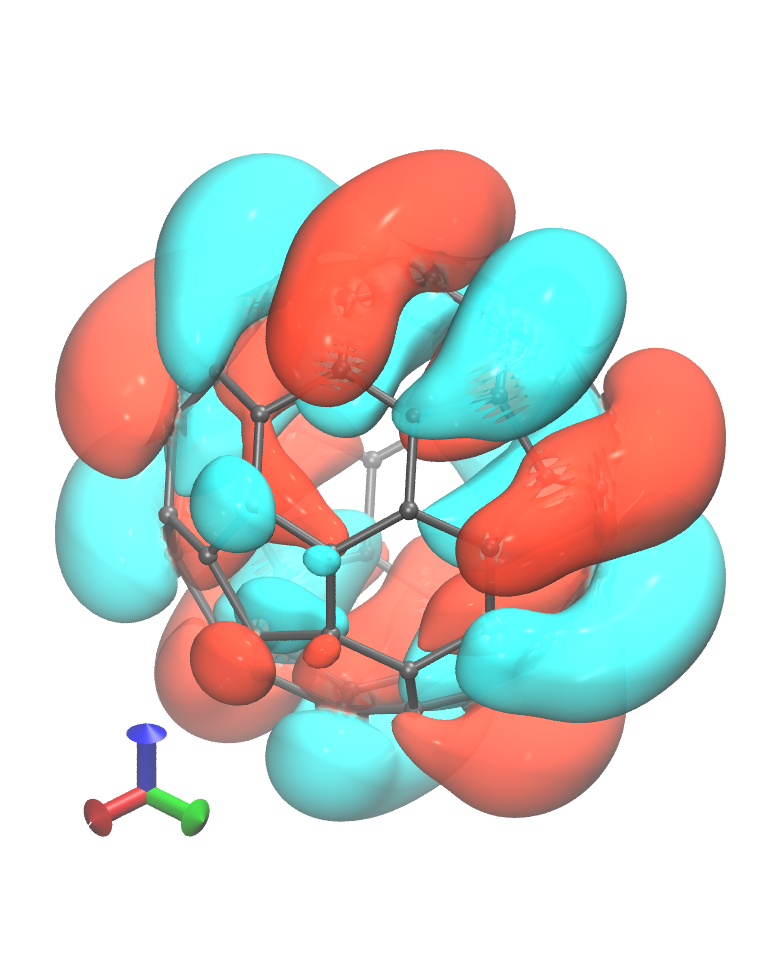}
        & $A_{u}$
        & $H_u$\\
        $\chi^{\alpha}_{178}$
        & \includegraphics[trim=0.0cm 0.2cm 0.0cm 0.2cm, clip, scale=1.0]{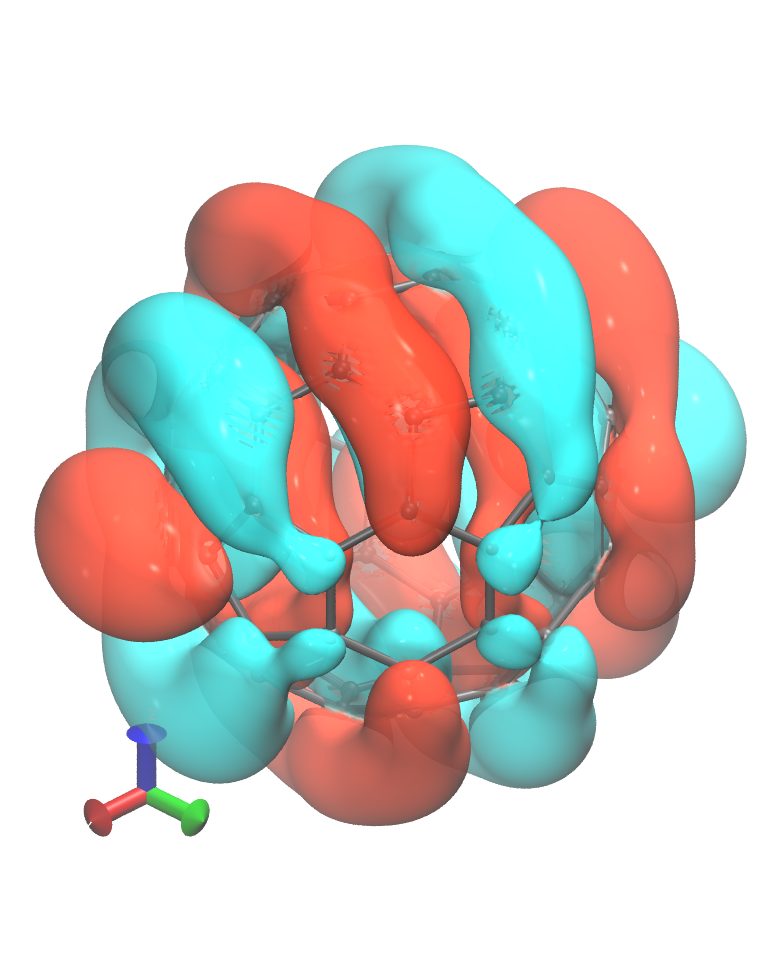}
        & $B_{u}$
        & $H_u$\\
        $\chi^{\alpha}_{179}$
        & \includegraphics[trim=0.0cm 0.2cm 0.0cm 0.2cm, clip, scale=1.0]{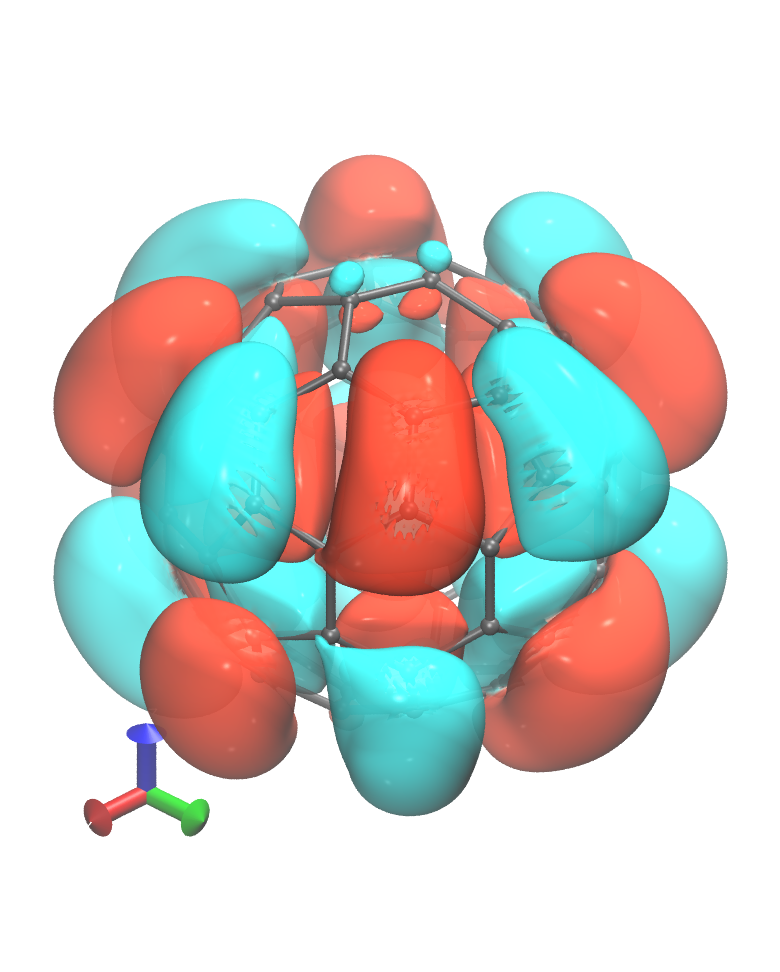}
        & $A_{u}$
        & $H_u$\\
        $\chi^{\alpha}_{180}$
        & \includegraphics[trim=0.0cm 0.2cm 0.0cm 0.2cm, clip, scale=1.0]{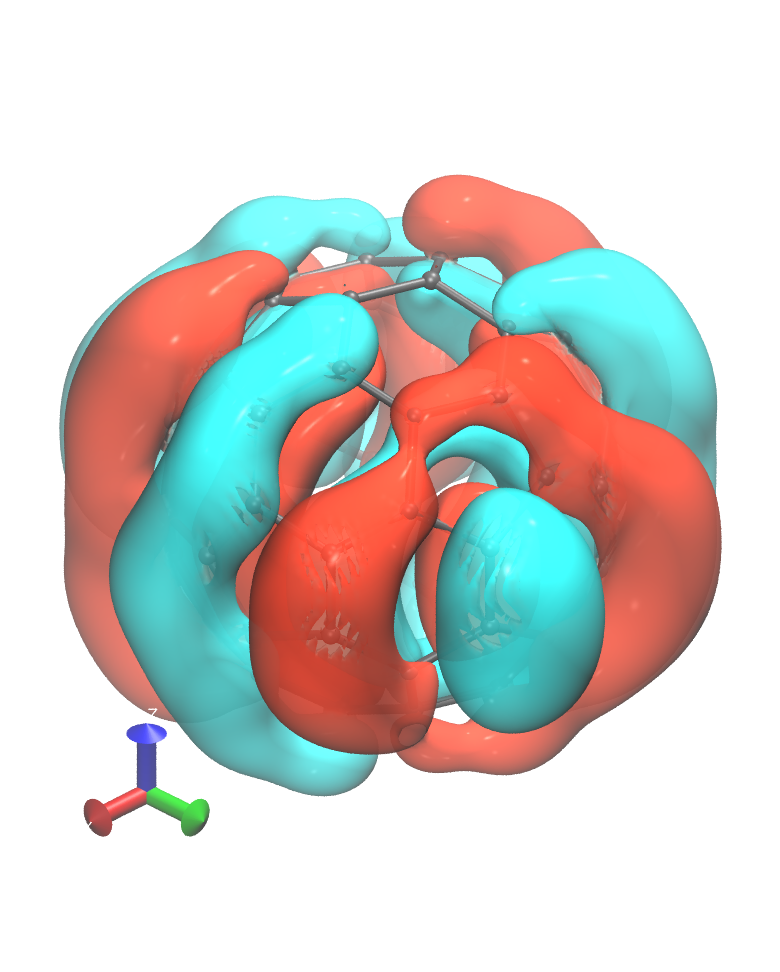}
        & $B_{u}$
        & $H_u$\\
        \bottomrule
        \end{tabular}
      \end{table*}

      We consider next icosahedral \ce{C60}, which has a higher symmetry than tetrahedral \ce{C84H64}, and for which an unrestricted CAM-B3LYP/6-31+G* calculation was also carried out to yield a set of KS MOs.
      It turns out that, even though \textsc{Q-Chem} is able to identify the symmetry group of this system as $\mathcal{I}_h$ when the symmetry tolerance value is set at \num{1e-4}, it seeks recourse to $\mathcal{C}_5$ (a subgroup of $\mathcal{I}_h$) to perform symmetry analysis, but then fails to produce any symmetry assignments for almost all MOs except those that are non-degenerate.
      Tightening the symmetry tolerance value to \num{1e-5} forces \textsc{Q-Chem} to identify the symmetry group as $\mathcal{C}_{2h}$ instead, but this then allows for a successful assignment of symmetry labels to MOs, albeit only under $\mathcal{C}_{2h}$.
      On the other hand, \qsymsq{} is able to both identify the symmetry group correctly as $\mathcal{I}_h$ and classify the symmetry of MOs using the irreducible representations of $\mathcal{I}_h$.
      Table~\ref{tab:c60mos} shows these symmetry assignments for the highest three degenerate sets of occupied MOs.

\begin{figure}
        \centering
        \includegraphics[trim=0.0cm 0.2cm 0.0cm 0.2cm, clip, scale=1.7]{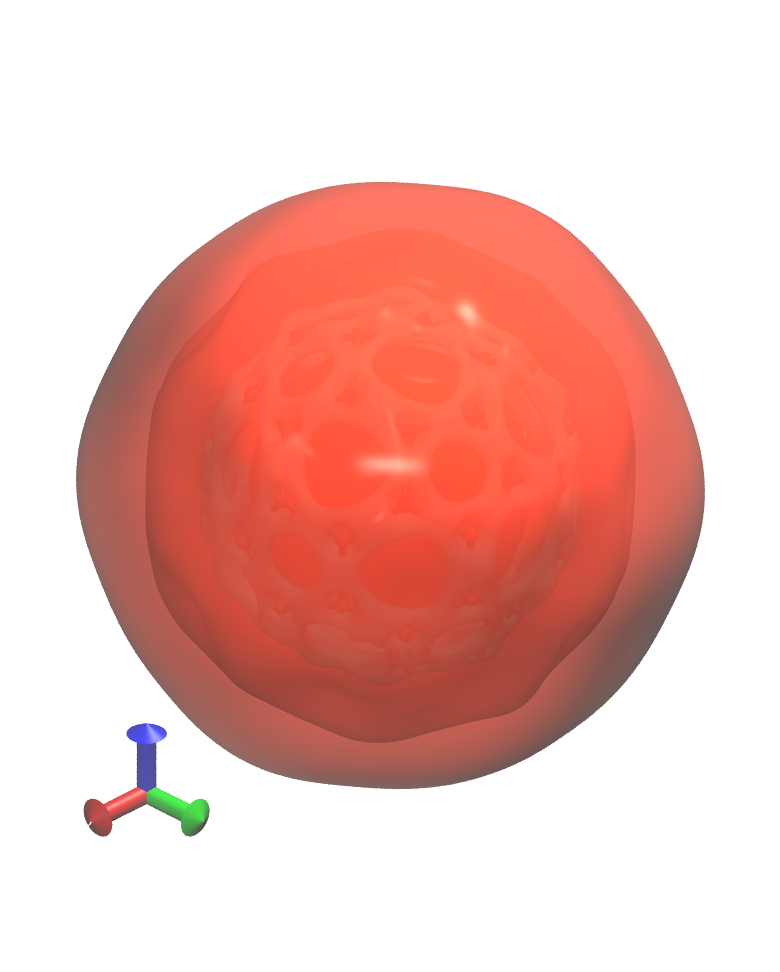}
        \caption{%
          Isosurface plot of the virtual canonical MO $\chi^{\alpha}_{195}$ at $\lvert \chi^{\alpha}_{195}(\mathbf{r}) \rvert = 0.009$ in \ce{C60} calculated at the CAM-B3LYP/6-31+G* level of theory using an $\mathcal{I}_h$-symmetrized geometry (see Table~\ref{tab:c60mos}).
          This MO has $A_g(\mathcal{I}_h)$ and $A_g(\mathcal{C}_{2h})$ symmetry.
        }
        \label{fig:c60-aplha195}
      \end{figure}

      An inspection of Table~\ref{tab:c60mos} makes clear the necessity of correct symmetry classifications of these MOs in the full group of the underlying molecule.
      Consider for instance the transition dipole moment integral $\braket{\chi^{\alpha}_{176} | \hat{\boldsymbol{\mu}} | \chi^{\alpha}_{195}}$, where $\hat{\boldsymbol{\mu}}$ is the dipole moment operator and $\chi^{\alpha}_{195}$ the first totally symmetric virtual $\alpha$-MO in the calculation shown in Table~\ref{tab:c60mos} (Figure~\ref{fig:c60-aplha195}).
      In $\mathcal{C}_{2h}$, the dipole moment operator transforms as $A_u \oplus 2B_u$, and so the integrand transforms as $A_u \otimes (A_u \oplus 2B_u) \otimes A_g \supset A_g$, indicating that the integral $\braket{\chi^{\alpha}_{176} | \hat{\boldsymbol{\mu}} | \chi^{\alpha}_{195}}$ contains up to one independent non-vanishing component.
      This would thus lead one to the incorrect expectation that the transition $\chi^{\alpha}_{195} \leftarrow \chi^{\alpha}_{176}$ is optically allowed.
      However, in $\mathcal{I}_{h}$, the dipole moment operator transforms as $T_{1u}$, and the above integral is part of a degenerate set $\braket{\chi^{\alpha}_{i} | \hat{\boldsymbol{\mu}} | \chi^{\alpha}_{195}}$, $i = 176, \ldots, 180$ whose integrands transform as $H_u \otimes T_{1u} \otimes A_g$ which does not contain $A_g$.
      The transitions $\chi^{\alpha}_{195} \leftarrow \chi^{\alpha}_{i}$, $i = 176, \ldots, 180$ are therefore all optically forbidden in $\mathcal{I}_{h}$.
      This example shows that, such MO symmetry considerations, when done accurately in the full symmetry group of the system, can improve the efficiency of algorithms that make use of symmetry to screen molecular integral evaluations\cite{article:Dacre1970,article:Elder1973,article:Pitzer1973,article:Haser1991} or sharpen the interpretation of spectroscopic results.

      \begin{table*}
        \centering
        \caption{%
          Comparison of symmetry assignments of valence canonical MOs in \ce{B9-} calculated at the B3LYP/def2-TZVP level of theory using the geometry optimized at the same level by \DJ{}or\dj{}evi\'{c} \textit{et al.}\cite{article:Dordevic2022}.
          For each MO $\chi(\mathbf{r})$, the isosurface is plotted at $\lvert \chi(\mathbf{r}) \rvert = 0.04$.
          In \qsymsq{}, the distance thresholds yielding $\mathcal{D}_{4h}$ and $\mathcal{D}_{8h}$ are $10^{-5}$ and $10^{-4}$, respectively (see Section~S1.4 of the Supporting Information for an explanation of this threshold).
        }
        \label{tab:b9mmos}
        \renewcommand*{\arraystretch}{0.5}
        \begin{tabular}{%
          l M{1.00cm} M{1.85cm} M{1.70cm} M{1.70cm} M{1.70cm}
        }
        \toprule
        & MO
        & Isosurface
        & \textsc{Q-Chem} ($\mathcal{D}_{4h}$)
        & \qsymsq{} ($\mathcal{D}_{4h}$)
        & \qsymsq{} ($\mathcal{D}_{8h}$)\\
        \midrule
        $\pi_1$
        & $\chi^{\alpha}_{18}$
        & \includegraphics[trim=0.0cm 0.30cm 0.0cm 0.35cm, clip, scale=0.85]{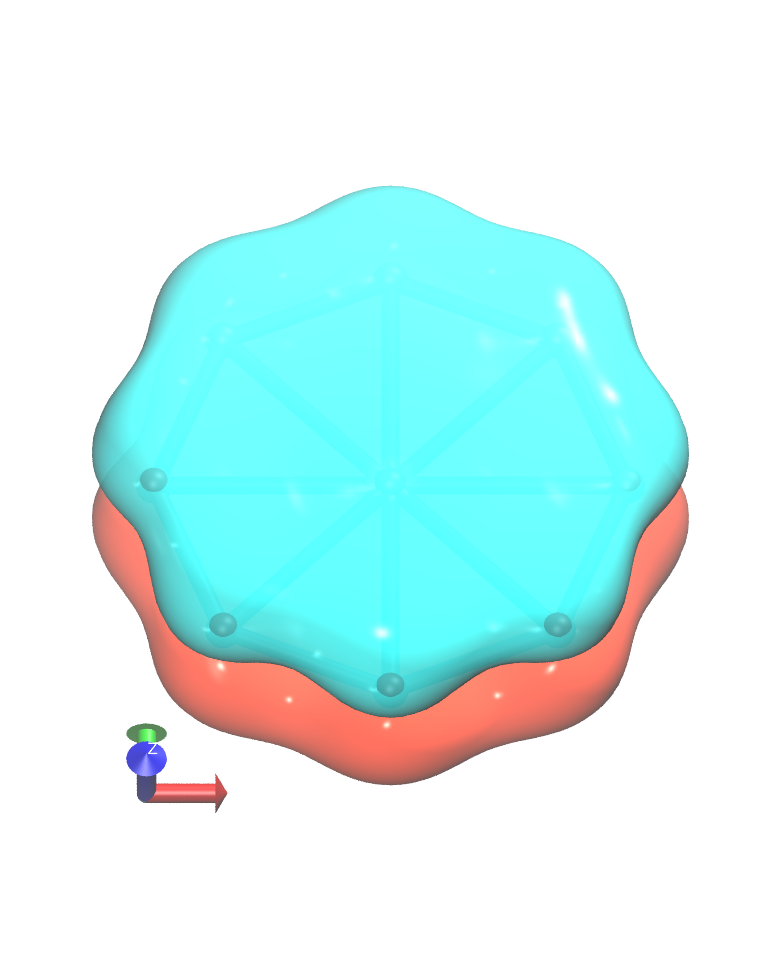}
        & $A_{2u}$
        & $A_{2u}$
        & $A_{2u}$\\
        & $\chi^{\alpha}_{22}$
        & \includegraphics[trim=0.0cm 0.30cm 0.0cm 0.35cm, clip, scale=0.85]{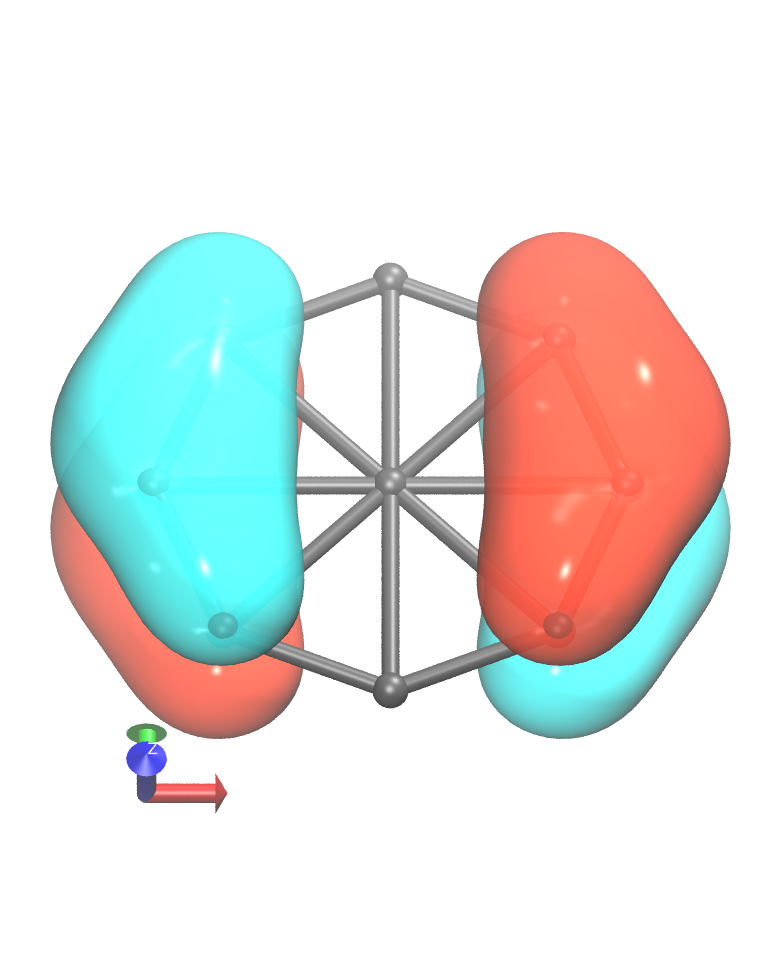}
        & $E_{g}$
        & $E_{g}$
        & $E_{1g}$\\
        & $\chi^{\alpha}_{23}$
        & \includegraphics[trim=0.0cm 0.30cm 0.0cm 0.35cm, clip, scale=0.85]{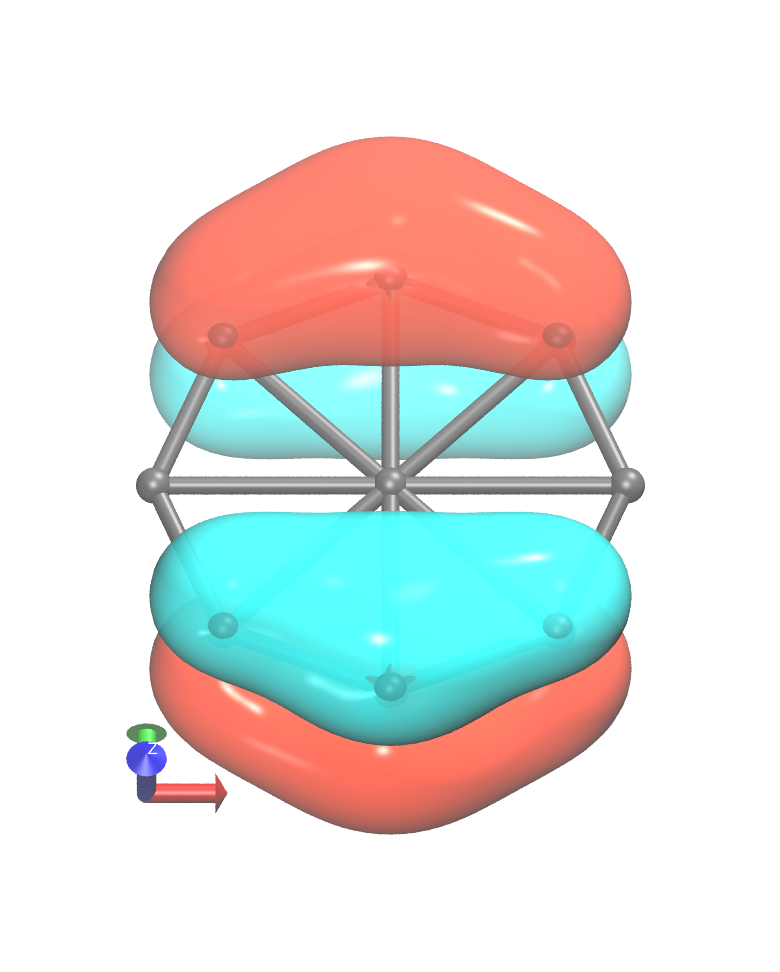}
        & $E_{g}$
        & $E_{g}$
        & $E_{1g}$\\
        \midrule
        $\sigma_1$
        & $\chi^{\alpha}_{10}$
        & \includegraphics[trim=0.0cm 0.30cm 0.0cm 0.35cm, clip, scale=0.85]{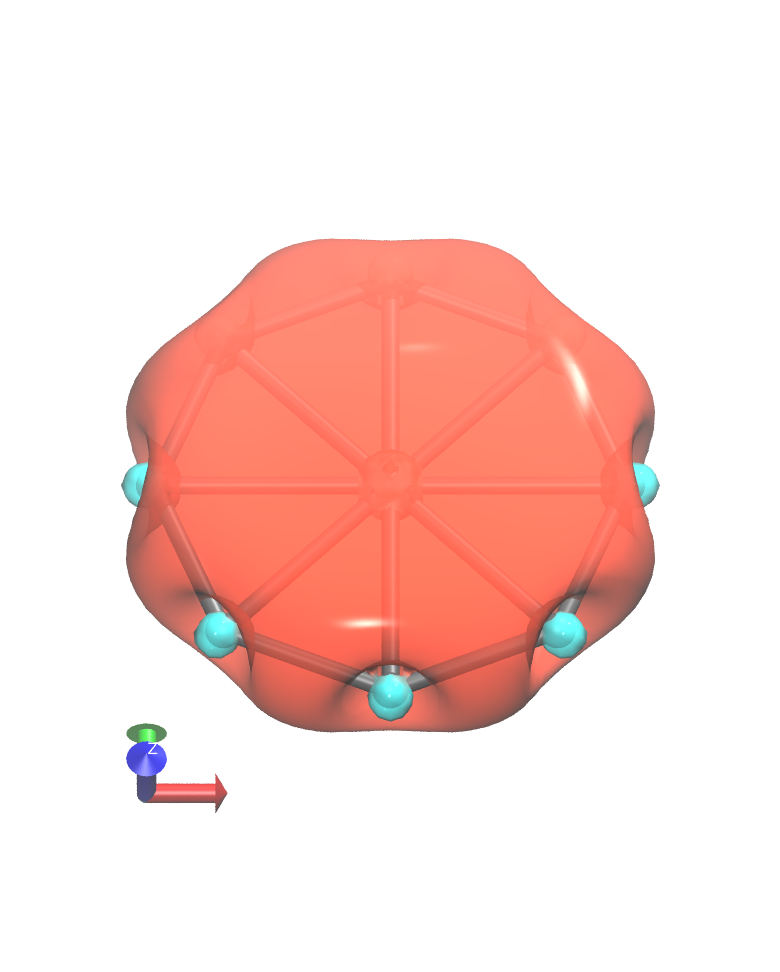}
        & $A_{1g}$
        & $A_{1g}$
        & $A_{1g}$\\
        & $\chi^{\alpha}_{11}$
        & \includegraphics[trim=0.0cm 0.30cm 0.0cm 0.35cm, clip, scale=0.85]{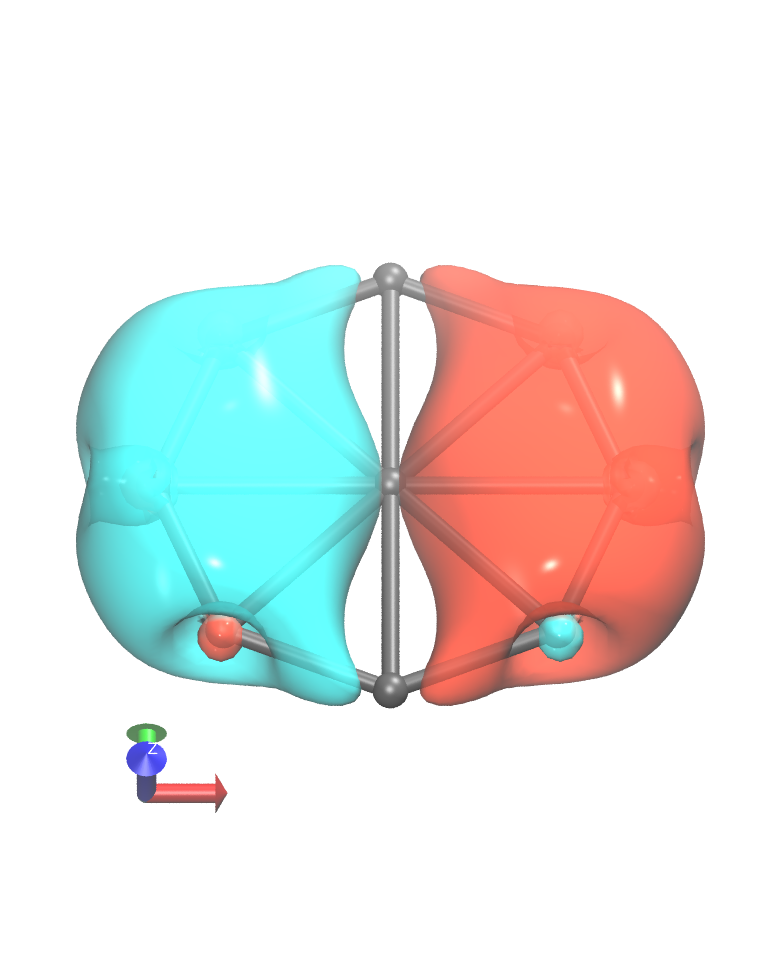}
        & $E_{u}$
        & $E_{u}$
        & $E_{1u}$\\
        & $\chi^{\alpha}_{12}$
        & \includegraphics[trim=0.0cm 0.30cm 0.0cm 0.35cm, clip, scale=0.85]{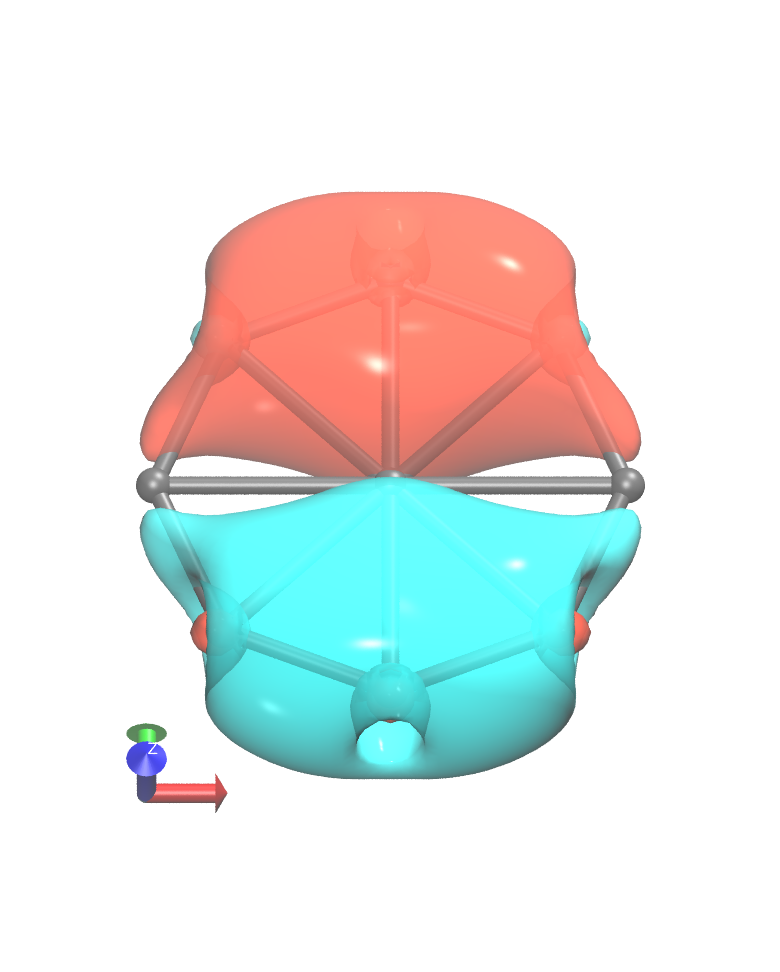}
        & $E_{u}$
        & $E_{u}$
        & $E_{1u}$\\
        & $\chi^{\alpha}_{13}$
        & \includegraphics[trim=0.0cm 0.30cm 0.0cm 0.35cm, clip, scale=0.85]{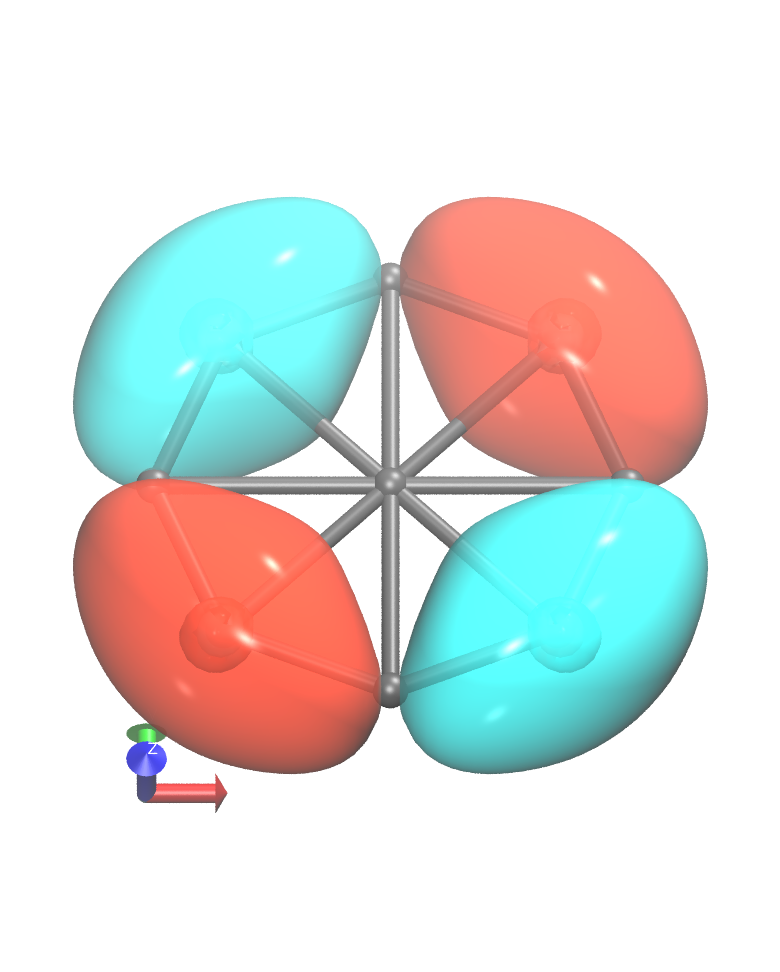}
        & $B_{2g}$
        & $B_{2g}$
        & $E_{2g}$\\
        & $\chi^{\alpha}_{14}$
        & \includegraphics[trim=0.0cm 0.30cm 0.0cm 0.35cm, clip, scale=0.85]{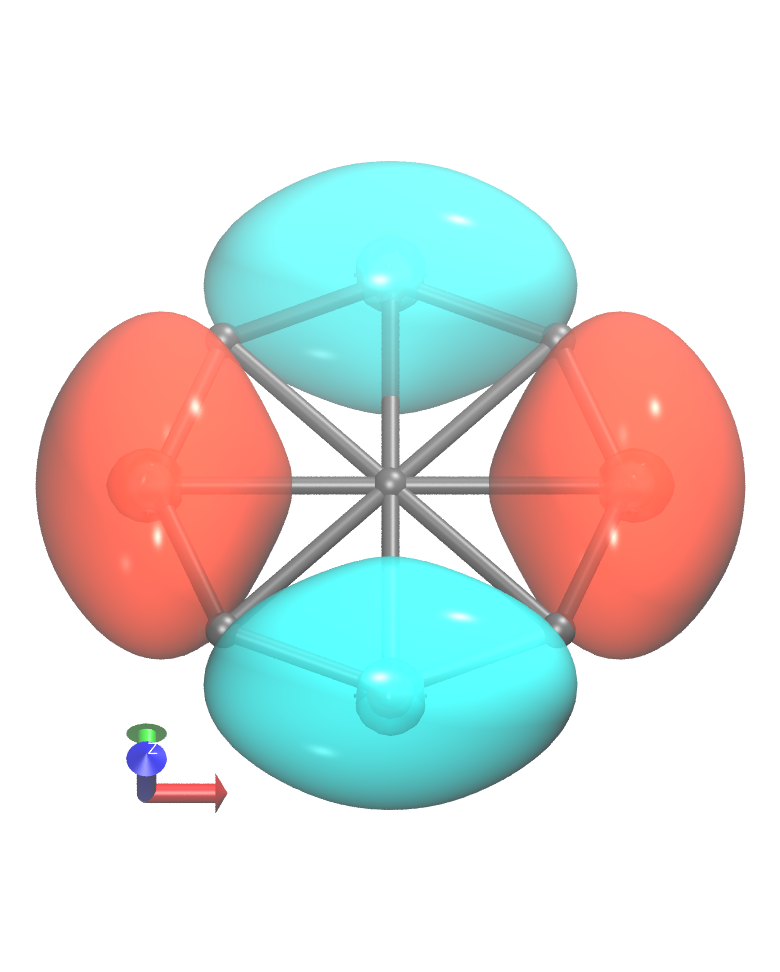}
        & $B_{1g}$
        & $B_{1g}$
        & $E_{2g}$\\
        & $\chi^{\alpha}_{16}$
        & \includegraphics[trim=0.0cm 0.30cm 0.0cm 0.35cm, clip, scale=0.85]{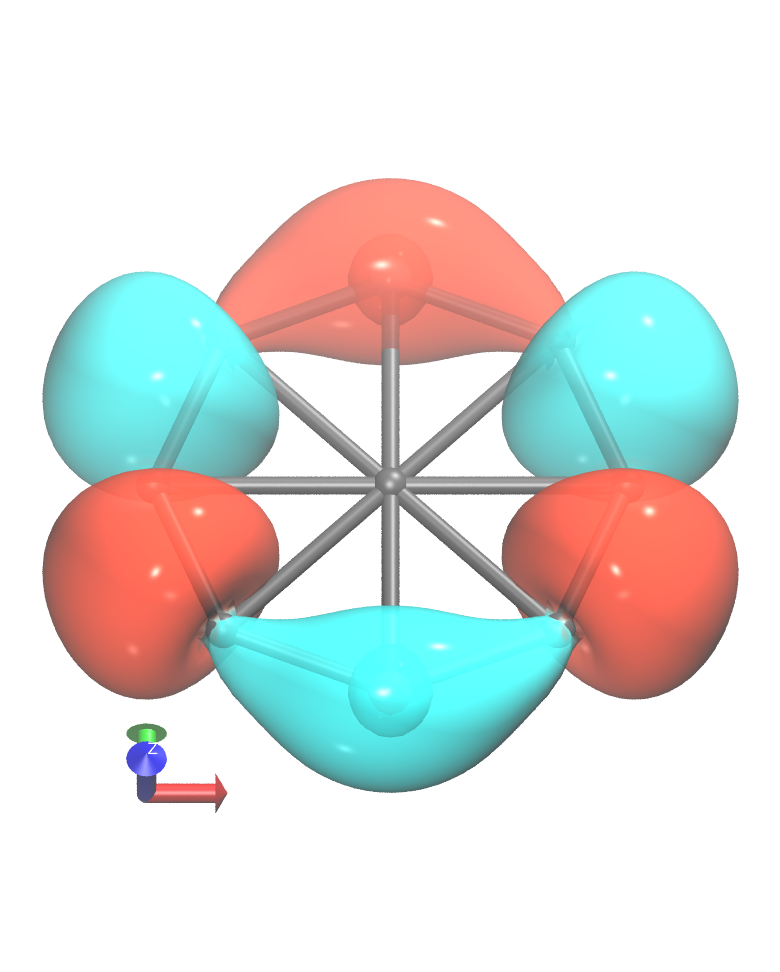}
        & $E_{u}$
        & $E_{u}$
        & $E_{3u}$\\
        & $\chi^{\alpha}_{17}$
        & \includegraphics[trim=0.0cm 0.30cm 0.0cm 0.35cm, clip, scale=0.85]{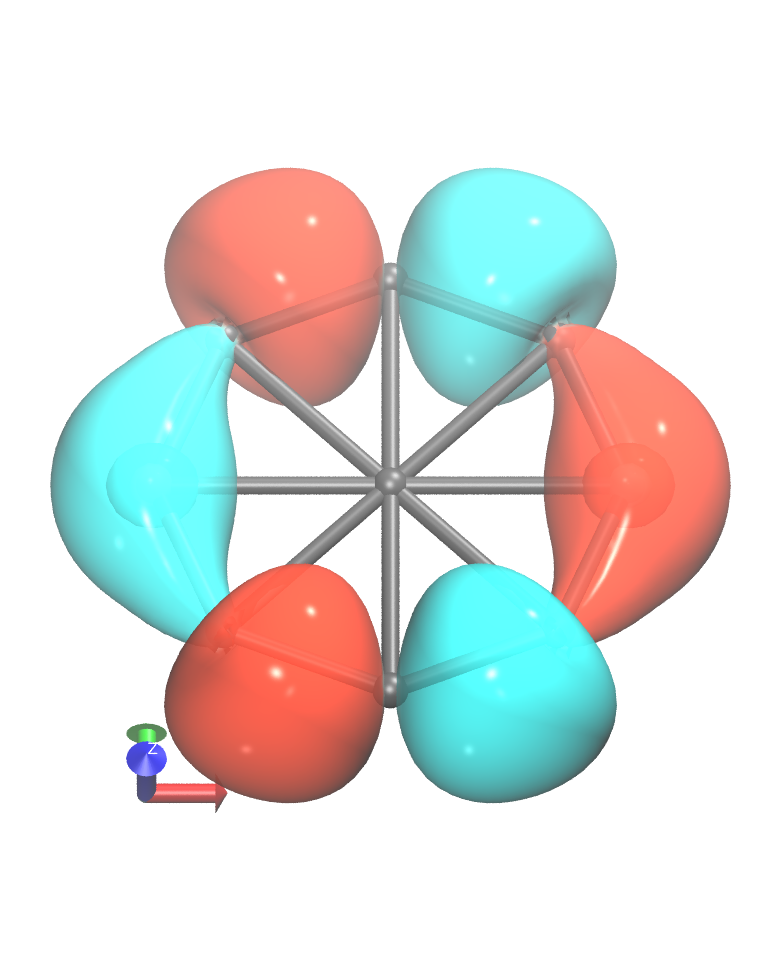}
        & $E_{u}$
        & $E_{u}$
        & $E_{3u}$\\
        & $\chi^{\alpha}_{19}$
        & \includegraphics[trim=0.0cm 0.30cm 0.0cm 0.35cm, clip, scale=0.85]{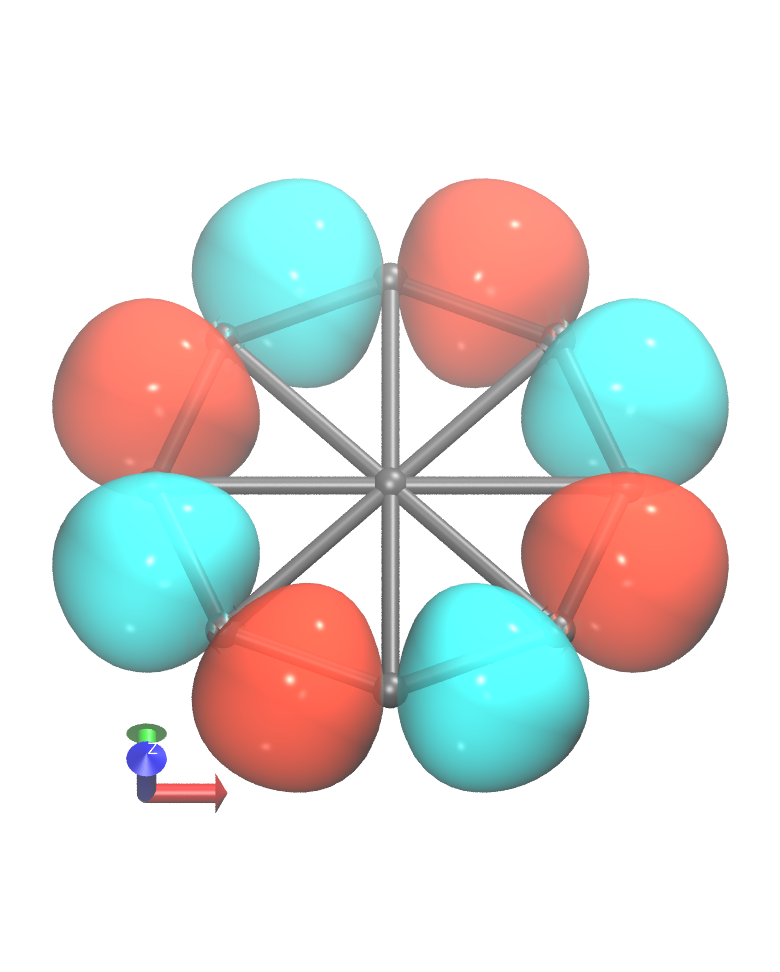}
        & $A_{2g}$
        & $A_{2g}$
        & $B_{2g}$\\
        \midrule
        $\sigma_2$
        & $\chi^{\alpha}_{15}$
        & \includegraphics[trim=0.0cm 0.30cm 0.0cm 0.35cm, clip, scale=0.85]{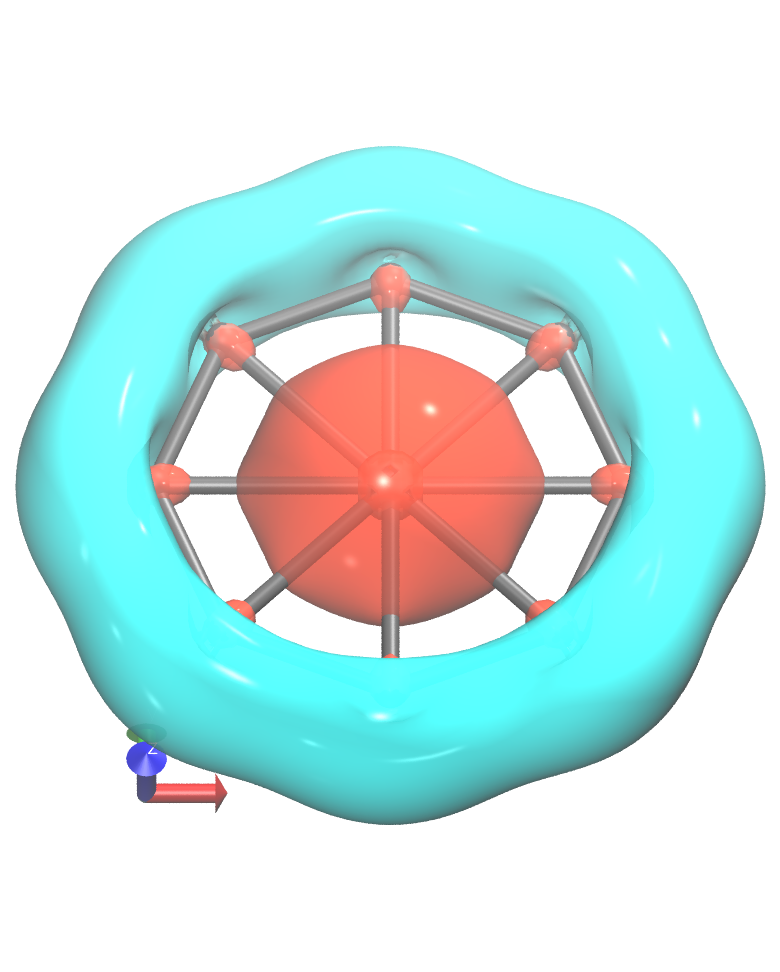}
        & $A_{1g}$
        & $A_{1g}$
        & $A_{1g}$\\
        & $\chi^{\alpha}_{20}$
        & \includegraphics[trim=0.0cm 0.30cm 0.0cm 0.35cm, clip, scale=0.85]{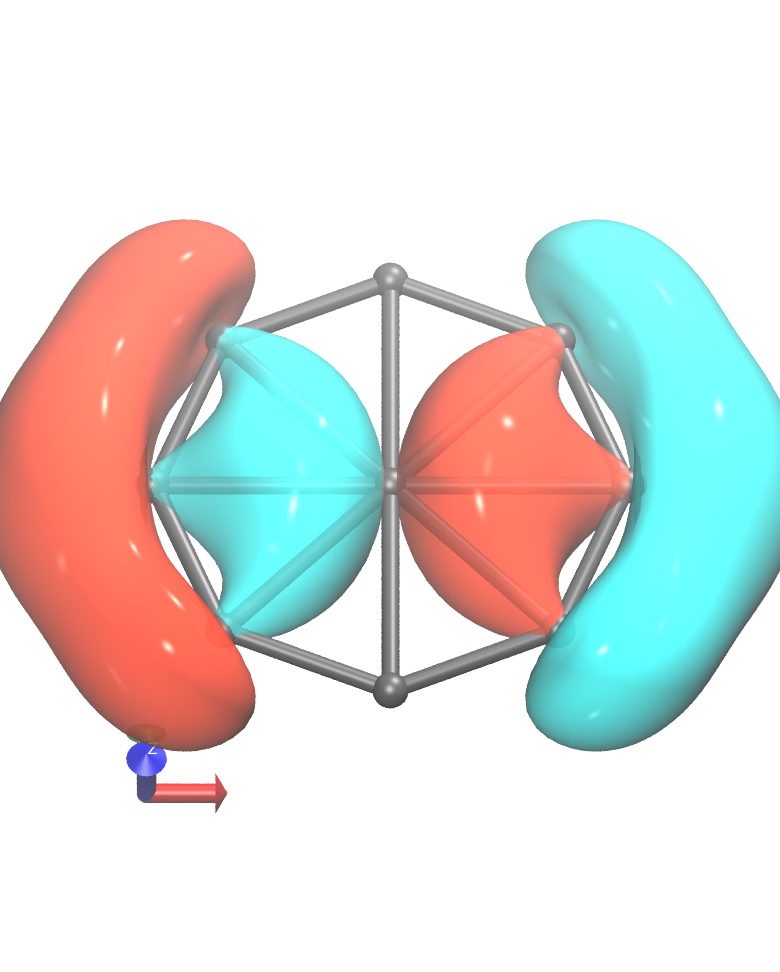}
        & $E_{u}$
        & $E_{u}$
        & $E_{1u}$\\
        & $\chi^{\alpha}_{21}$
        & \includegraphics[trim=0.0cm 0.15cm 0.0cm 0.15cm, clip, scale=0.85]{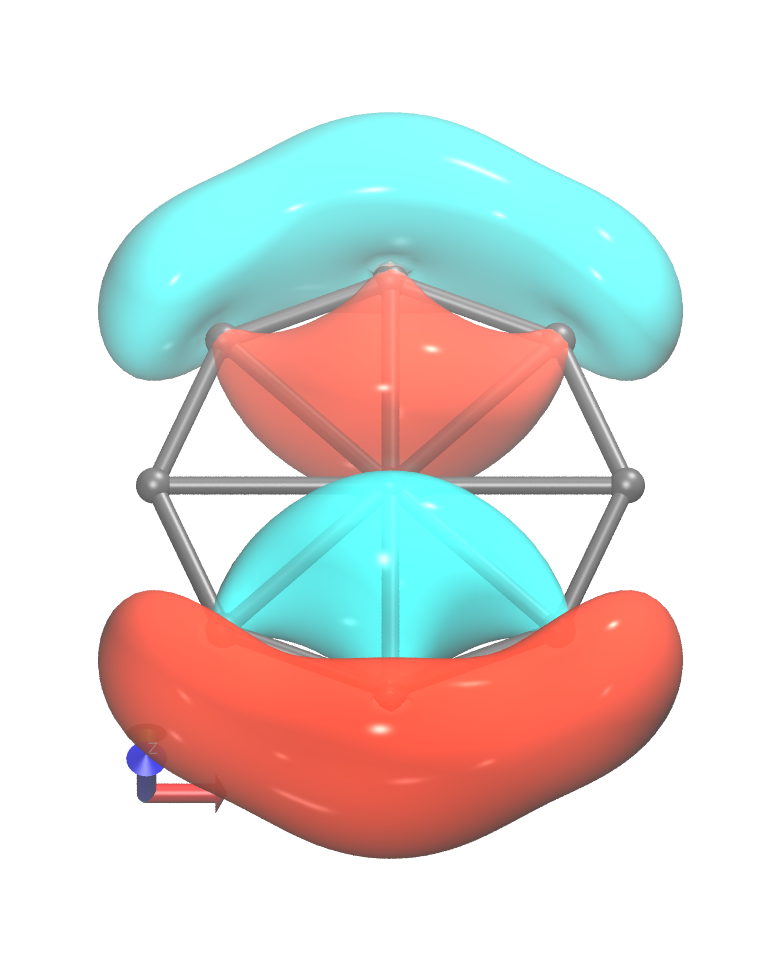}
        & $E_{u}$
        & $E_{u}$
        & $E_{1u}$\\
        \bottomrule
        \end{tabular}
      \end{table*}

      Finally, we consider an octagonal boron disc, \ce{B9-}, which has $\mathcal{D}_{8h}$ symmetry and should, in principle, be simpler than the $\mathcal{I}_h$ symmetry of \ce{C60}.
      Unfortunately, no matter the symmetry tolerance value, \textsc{Q-Chem} is only able to determine the symmetry group of this anion as $\mathcal{D}_{4h}$ and classify the MOs calculated at the B3LYP/def2-TZVP level of theory using the irreducible representations of this group.
      \qsymsq{}, on the other hand, is capable of identifying the symmetry group as either $\mathcal{D}_{4h}$ or $\mathcal{D}_{8h}$, depending on the choice of the distance threshold (Section~S1.4 of the Supporting Information), and then assigning MO symmetry labels using the irreducible representations of the corresponding groups.
      These are summarized in Table~\ref{tab:b9mmos} for the valence canonical MOs of \ce{B9-} as reported by \DJ{}or\dj{}evi\'{c} \textit{et al.}\cite{article:Dordevic2022}.
      It can be seen that, in $\mathcal{D}_{4h}$, the symmetry assignments from both \textsc{Q-Chem} and \qsymsq{} are in agreement, while those in $\mathcal{D}_{8h}$ computable only by \qsymsq{} provide a full symmetry classification of the MOs that is consistent with their nodal structures.

  \subsection{Symmetry breaking in SCF solutions and orbitals}
  \label{sec:symbreaking}

    The next case study seeks to demonstrate the ability of \qsymsq{} to detect and quantify symmetry breaking in SCF solutions and orbitals.
    For this purpose, an octahedral complex \ce{Fe(CN)6^{3-}} has been chosen.
    Having an unpaired electron due to the low-spin $d^5$ configuration on the \ce{Fe^{3+}} metal center surrounded by an octahedral ligand field, this complex is expected to give rise to multiple low-lying SCF UHF solutions, some of which break spatial symmetry\cite{article:Huynh2020}.

    \subsubsection{Computational details}
    \label{sec:symbreaking-compdetails}
      In all calculations, the structure of \ce{Fe(CN)6^{3-}} was held fixed at an $\mathcal{O}_h$ geometry with $\ce{Fe-C} = \SI{2.0256023}{\angstrom}$ and $\ce{C-N} = \SI{1.1570746}{\angstrom}$.
      Multiple SCF solutions at the UHF/def2-TZVP level of theory were found for this geometry  in \textsc{Q-Chem} 6.1.0 using metadynamics\cite{article:Thom2008} combined with the Direct Inversion in the Iterative Subspace (DIIS) algorithm\cite{article:Pulay1980}.
      SCF convergence was set at a DIIS error value of \num{1e-10} as implemented in \textsc{Q-Chem}.
      For each converged solution, its geometry information, basis set information, and Pipek--Mezey-localized\cite{article:Pipek1988,article:Pipek1989} MO coefficients were read in by \qsymsq{} from which symmetry assignments for the individual MOs, as well as for the overall wavefunction, were determined.
      Unfortunately, no benchmarking symmetry assignments were available because no existing programs were able to handle symmetry-broken quantities.

      For each quantity that is symmetry-analyzed in \qsymsq{}, a threshold $\lambda^{\mathrm{thresh}}_{\mathbf{S}}$ must be chosen to determine which eigenvalues of the orbit overlap matrix $\mathbf{S}$ [Equation~\eqref{eq:smat}] are non-zero so that the transformation matrix $\mathbf{X}$ can be constructed [Equation~\eqref{eq:xmat}].
      Whether or not a particular choice of threshold is reasonable depends on the gap between the eigenvalue of $\mathbf{S}$ that is immediately above the threshold, $\lambda^{>}_{\mathbf{S}}$, and the eigenvalue of $\mathbf{S}$ that is immediately below the threshold, $\lambda^{<}_{\mathbf{S}}$.
      In all cases for \ce{Fe(CN)6^{3-}}, the threshold was chosen such that $\log_{10} \lambda^{>}_{\mathbf{S}} - \log_{10} \lambda^{<}_{\mathbf{S}} \gtrsim 4$.

    \subsubsection{Symmetry breaking in \ce{Fe(CN)6^{3-}}}

      \begin{table*}
        \centering
        \caption{%
          Symmetry-broken SCF solutions of \ce{Fe(CN)6^{3-}} calculated at the UHF/def2-TZVP level of theory using an $\mathcal{O}_h$ geometry with $\ce{Fe-C} = \SI{2.0256023}{\angstrom}$ and $\ce{C-N} = \SI{1.1570746}{\angstrom}$.
          Each solution has $M_S = +1/2$.
          All symmetries were determined using \qsymsq{} with a linear independence cut-off $\lambda^{\mathrm{thresh}}_{\mathbf{S}} = \num{1e-7}$.
          See Section~\ref{sec:symbreaking-compdetails} in the main text for the description of $\lambda^{>}_{\mathbf{S}}$ and $\lambda^{<}_{\mathbf{S}}$.
        }
        \begin{subtable}{\textwidth}
          \captionsetup{justification=centering}
          \centering
          \caption{%
            Symmetries of four lowest-lying $M_S = +1/2$ UHF solutions located in \textsc{Q-Chem} using SCF metadynamics and DIIS with a convergence threshold of \num{1e-10}.
            Solutions are labeled alphabetically in ascending order of energy.
          }
          \label{tab:fecn3mscf}
          \renewcommand*{\arraystretch}{1.3}
          \begin{tabular}{%
            M{2.0cm}
            S[table-format=+4.9, table-alignment-mode=format]
            M{5.0cm}
            S[table-format=1.2, table-number-alignment=center]
            S[table-format=1.2e+2, table-number-alignment=center]
          }
            \toprule
            \shortstack{SCF\\solution}
            & {Energy}
            & Symmetry
            & $\lambda^{>}_{\mathbf{S}}$
            & $\lambda^{<}_{\mathbf{S}}$\\
            \midrule
            A
            & -1816.083884703
            & $T_{1u} \oplus T_{2u}$
            & +8.00e0
            & +2.54e-12\\
            B
            & -1816.047307148
            & $T_{1u} \oplus T_{2u}$
            & +8.00e0
            & +3.82e-9\\
            C
            & -1815.986542495
            & $T_{1g} \oplus T_{2g}$
            & +8.00e0
            & +3.92e-15\\
            D
            & -1815.917430101
            & $A_{1g} \oplus A_{2g} \oplus 2E_{g} \oplus T_{1g} \oplus T_{2g}$
            & +2.03e0
            & +2.18e-9\\
            \bottomrule
          \end{tabular}
        \end{subtable}

        \vspace{0.7cm}

        \renewcommand*{\arraystretch}{0.5}
        \begin{subtable}[t]{.49\textwidth}
          \captionsetup{justification=centering}
          \centering
          \caption{%
              Pipek--Mezey-localized $d$-MOs of solution A.
              All isosurfaces are plotted at $\lvert \chi(\mathbf{r}) \rvert = 0.014$.
          }
          \label{tab:dMOsA}
          \scalebox{.65}{
            \begin{tabular}{%
              l
              M{2.20cm}
              M{2.90cm}
              S[table-format=1.2e+1, table-alignment-mode=format, table-number-alignment=center]
              S[table-format=1.2e+2, table-alignment-mode=format, table-number-alignment=center]
            }
              \toprule
              MO
              & Isosurface
              & Symmetry
              & $\lambda^{>}_{\mathbf{S}}$
              & $\lambda^{<}_{\mathbf{S}}$\\
              \midrule
              $\chi^{\alpha}_{38}$
              & \includegraphics[trim=0.0cm 0.10cm 0.0cm 0.10cm, clip, scale=1.3]{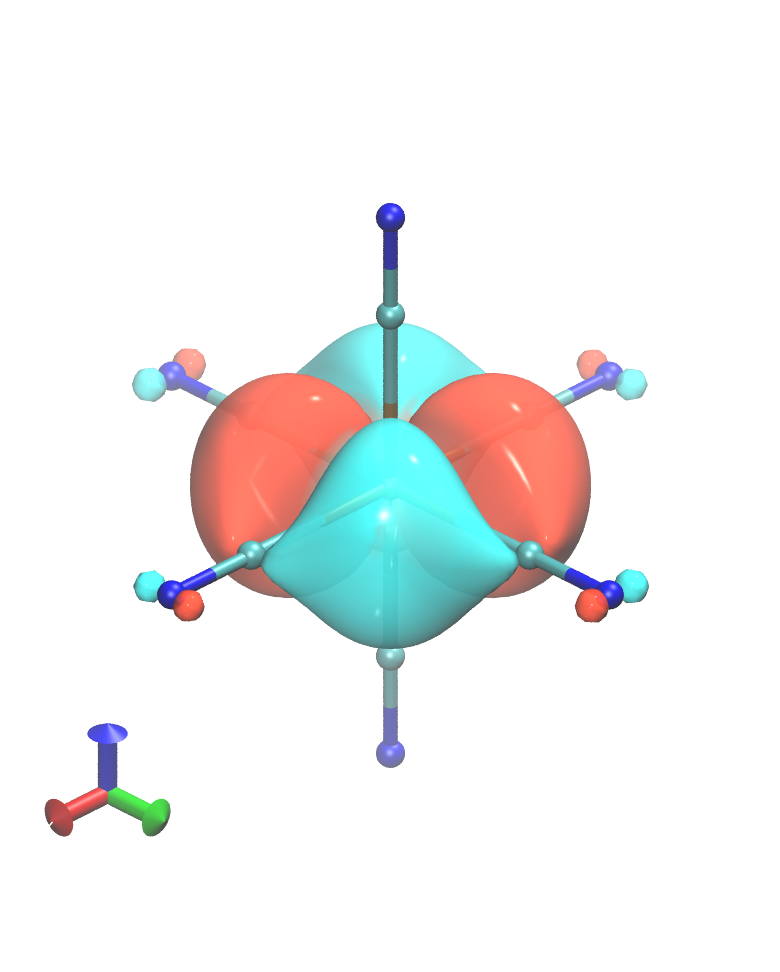}
              & $T_{1g} \oplus T_{2g}$
              & 4.17e-5
              & 4.00e-12\\
              $\chi^{\alpha}_{52}$
              & \includegraphics[trim=0.0cm 0.10cm 0.0cm 0.10cm, clip, scale=1.3]{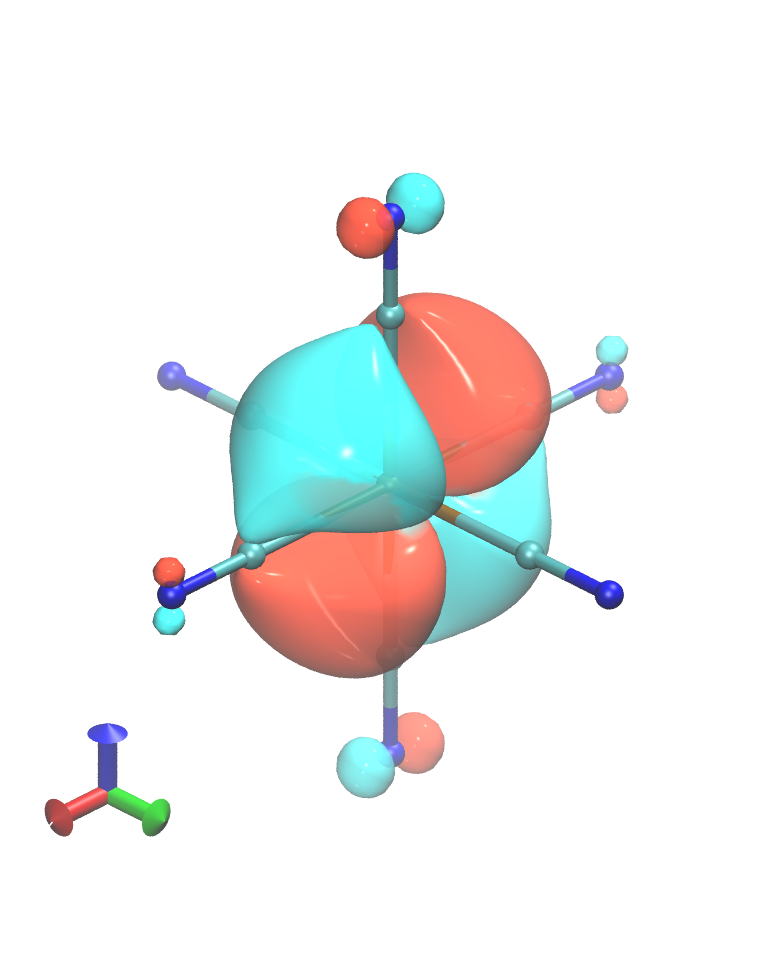}
              & $T_{1g} \oplus T_{2g}$
              & 5.70e-3
              & 3.06e-13\\
              $\chi^{\alpha}_{54}$
              & \includegraphics[trim=0.0cm 0.10cm 0.0cm 0.10cm, clip, scale=1.3]{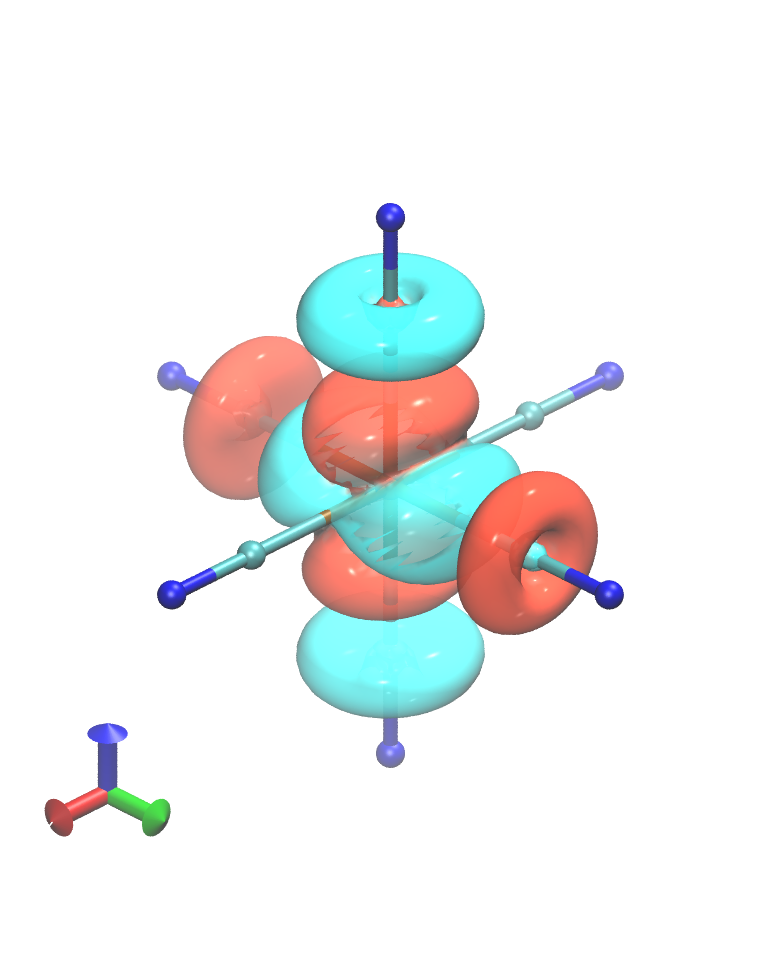}
              & $A_{1g} \oplus A_{2g} \oplus 2E_{g}$
              & 1.84e-6
              & 3.96e-14\\
              \midrule
              $\chi^{\beta}_{34}$
              & \includegraphics[trim=0.0cm 0.10cm 0.0cm 0.10cm, clip, scale=1.3]{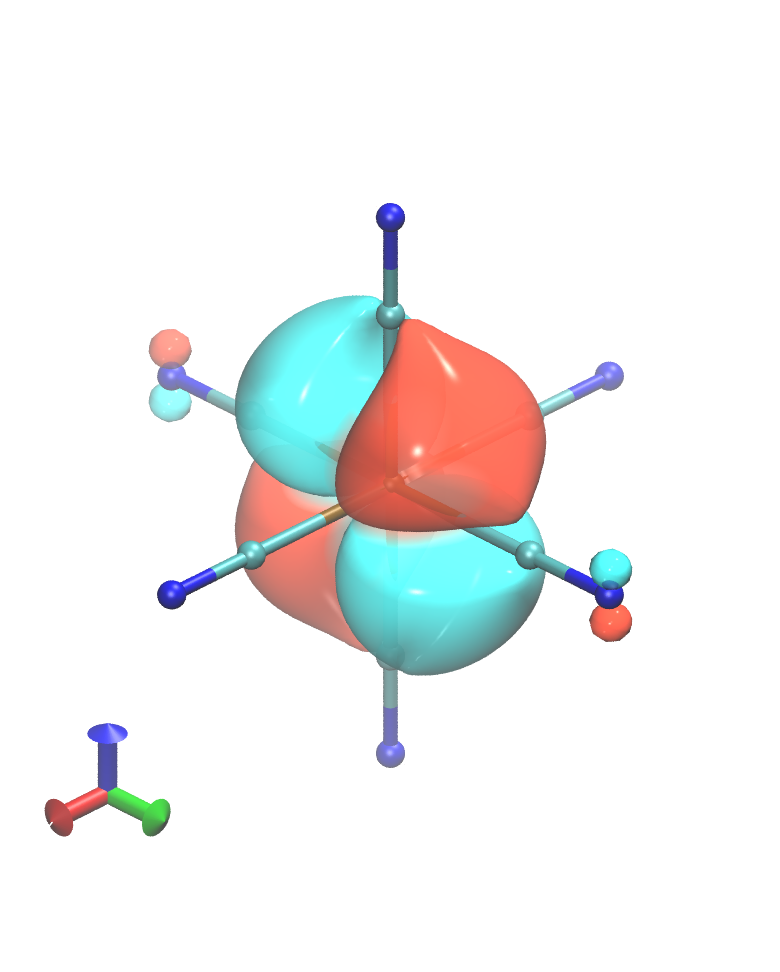}
              & $T_{1g} \oplus T_{2g}$
              & 5.53e-3
              & 2.33e-13\\
              $\chi^{\beta}_{37}$
              & \includegraphics[trim=0.0cm 0.10cm 0.0cm 0.10cm, clip, scale=1.3]{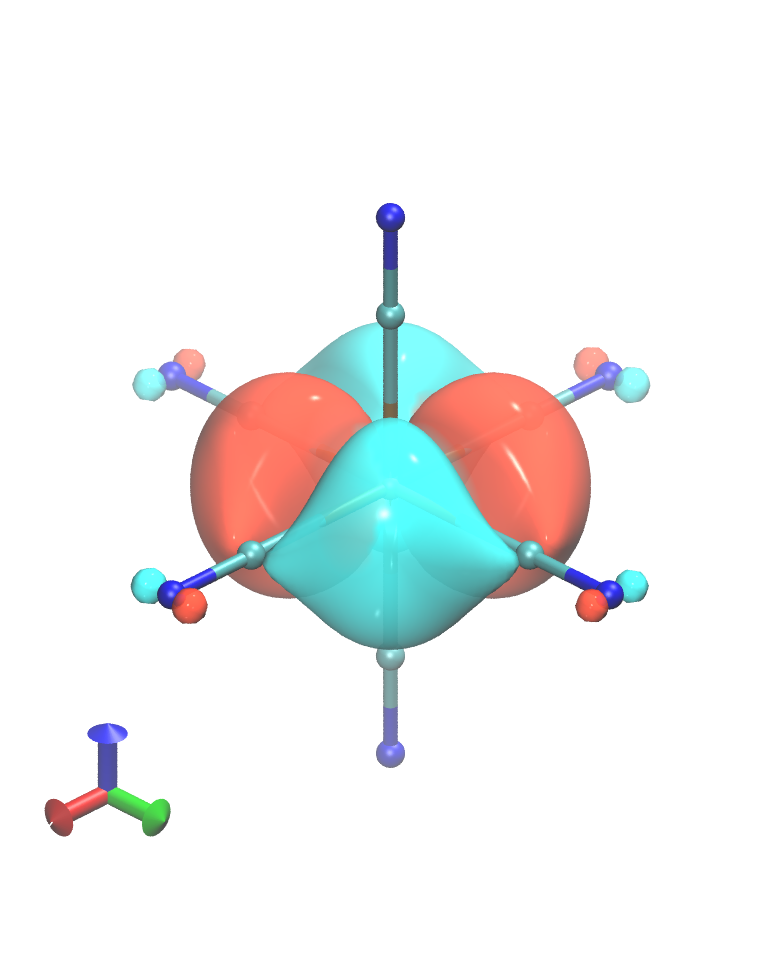}
              & $T_{1g} \oplus T_{2g}$
              & 4.86e-5
              & 3.35e-12\\
              $\chi^{\beta}_{44}$
              & \includegraphics[trim=0.0cm 0.10cm 0.0cm 0.10cm, clip, scale=1.3]{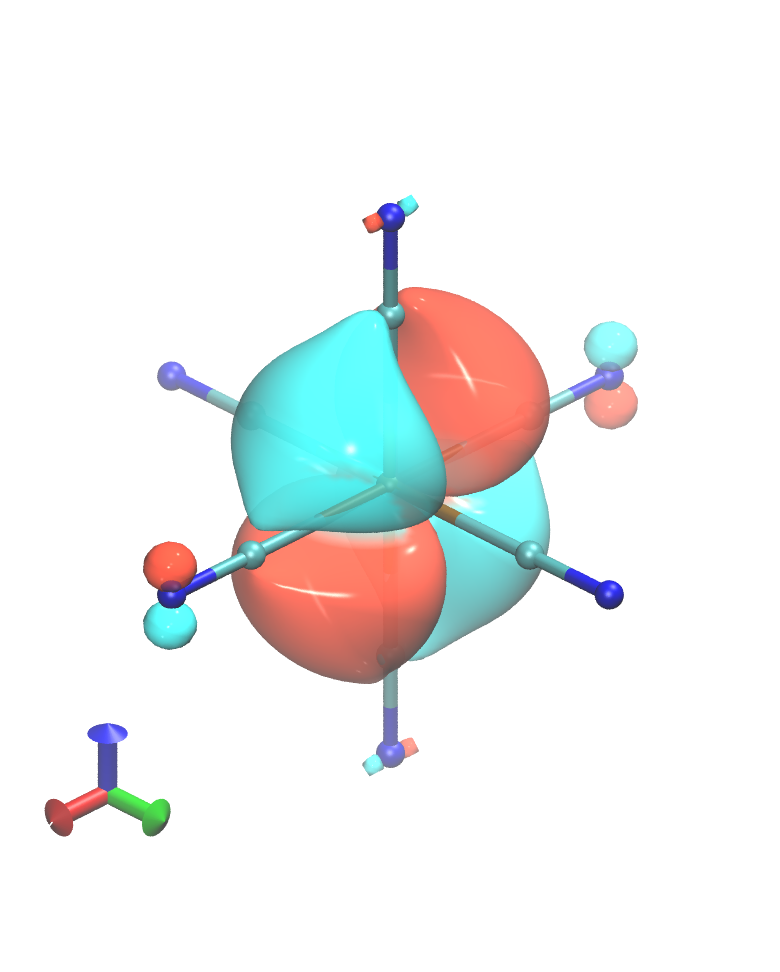}
              & $T_{1g} \oplus T_{2g}$
              & 7.84e-3
              & 1.95e-13\\
              \bottomrule
            \end{tabular}
          }
        \end{subtable}
        \hfill
        \begin{subtable}[t]{.49\textwidth}
          \captionsetup{justification=centering}
          \centering
          \caption{%
            Pipek--Mezey-localized $d$-MOs of solution C.
            All isosurfaces are plotted at $\lvert \chi(\mathbf{r}) \rvert = 0.008$.
          }
          \label{tab:dMOsC}
          \scalebox{.65}{
            \begin{tabular}{%
              l
              M{2.20cm}
              M{2.90cm}
              S[table-format=1.2e+1, table-alignment-mode=format, table-number-alignment=center]
              S[table-format=1.2e+2, table-alignment-mode=format, table-number-alignment=center]
            }
              \toprule
              MO
              & Isosurface
              & Symmetry
              & $\lambda^{>}_{\mathbf{S}}$
              & $\lambda^{<}_{\mathbf{S}}$\\
              \midrule
              $\chi^{\alpha}_{29}$
              & \includegraphics[trim=0.0cm 0.10cm 0.0cm 0.10cm, clip, scale=1.3]{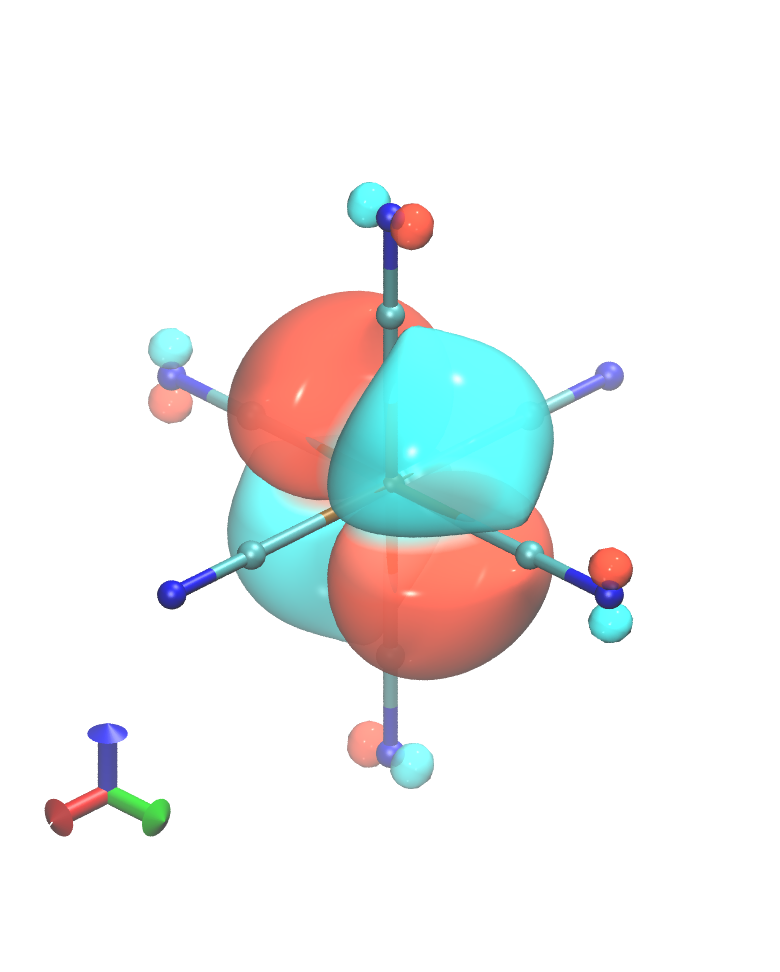}
              & $T_{1g} \oplus T_{2g}$
              & 6.33e-4
              & 4.02e-15\\
              $\chi^{\alpha}_{30}$
              & \includegraphics[trim=0.0cm 0.10cm 0.0cm 0.10cm, clip, scale=1.3]{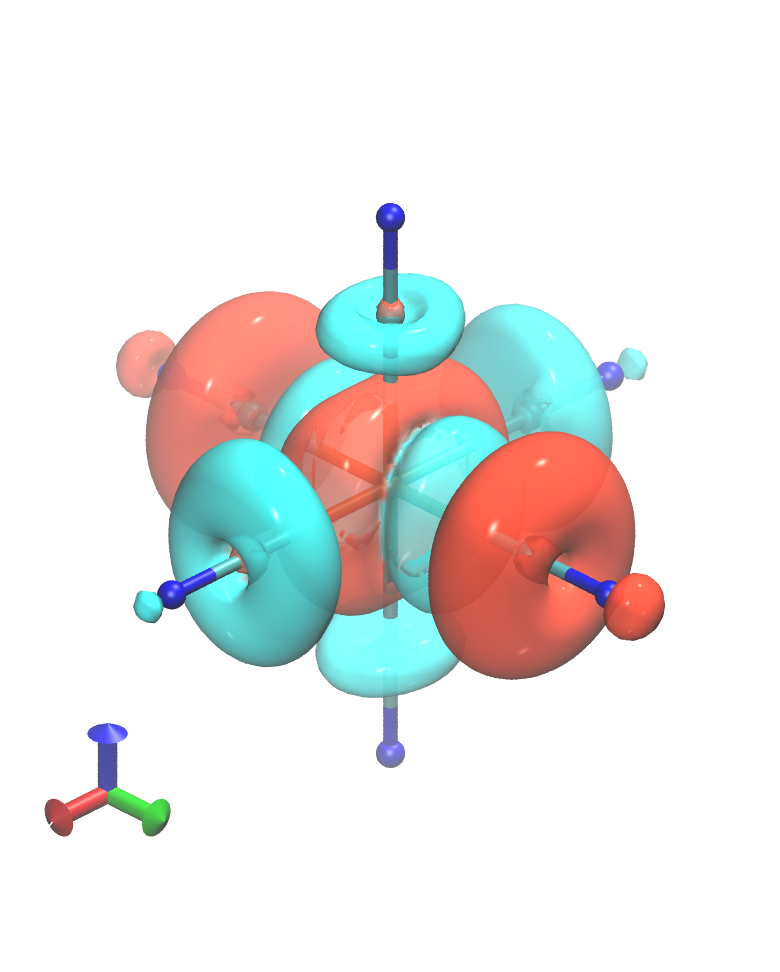}
              & $A_{1g} \oplus A_{2g} \oplus 2E_{g}$
              & 1.35e-7
              & 4.55e-15\\
              $\chi^{\alpha}_{34}$
              & \includegraphics[trim=0.0cm 0.10cm 0.0cm 0.10cm, clip, scale=1.3]{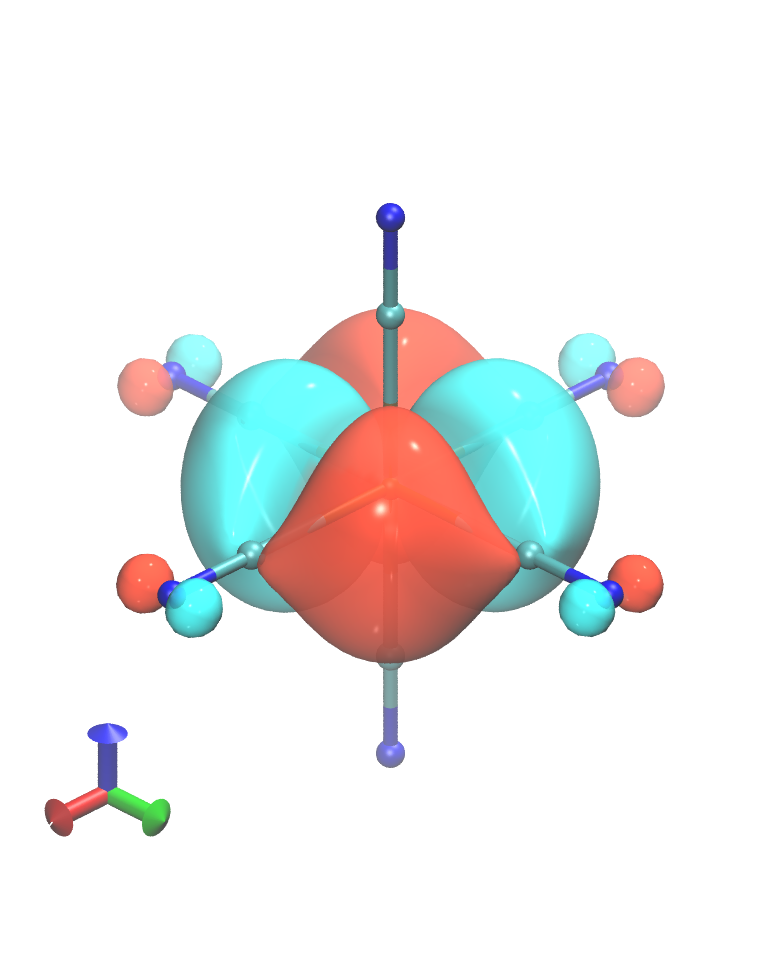}
              & $T_{1g} \oplus T_{2g}$
              & 2.45e-4
              & 4.67e-15\\
              \midrule
              $\chi^{\beta}_{35}$
              & \includegraphics[trim=0.0cm 0.10cm 0.0cm 0.10cm, clip, scale=1.3]{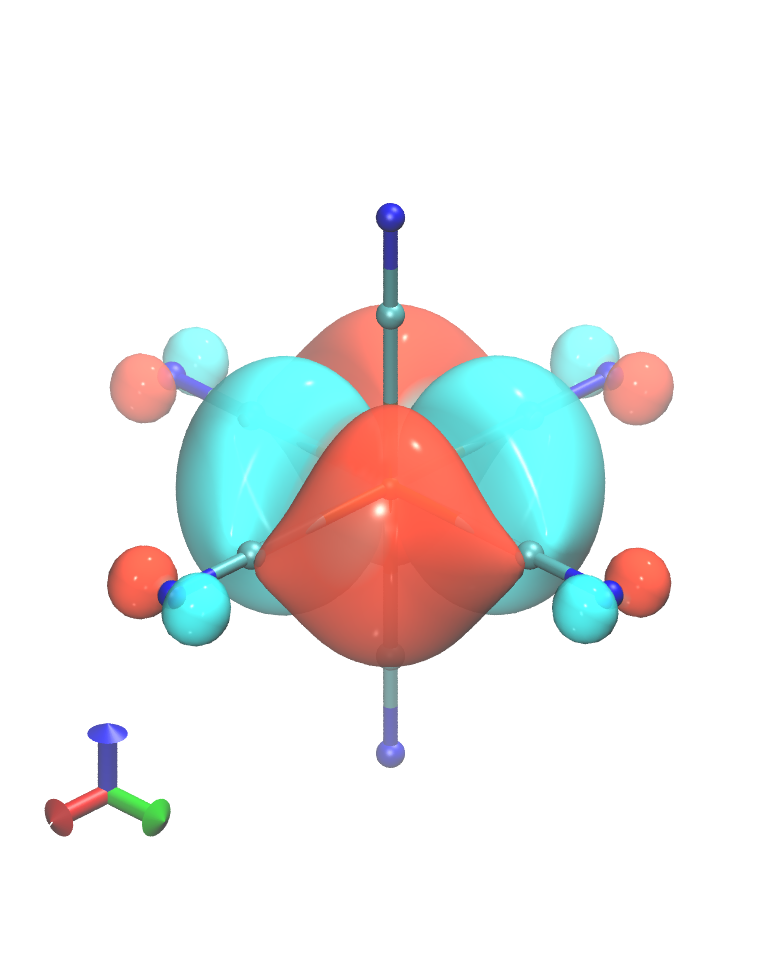}
              & $T_{1g} \oplus T_{2g}$
              & 3.40e-4
              & 6.60e-15\\
              $\chi^{\beta}_{53}$
              & \includegraphics[trim=0.0cm 0.10cm 0.0cm 0.10cm, clip, scale=1.3]{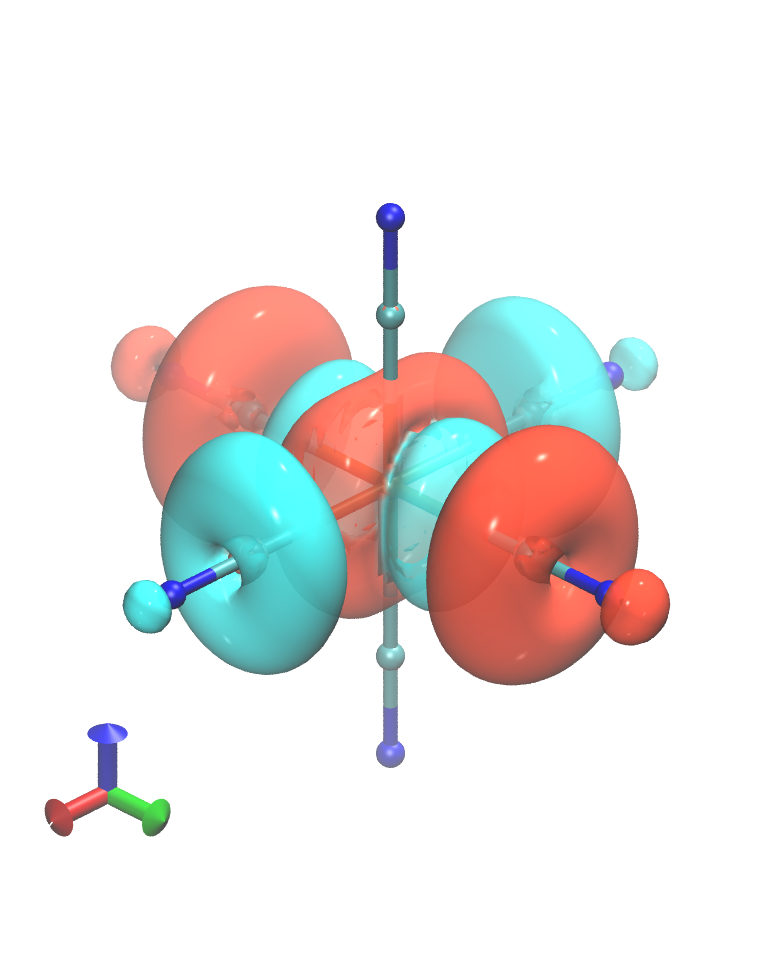}
              & $A_{1g} \oplus A_{2g} \oplus 2E_{g}$
              & 2.56e-6
              & 6.39e-15\\
              \bottomrule
            \end{tabular}
          }
        \end{subtable}
      \end{table*}

      Table~\ref{tab:fecn3mscf} presents the energies of four lowest-lying $M_S = +1/2$ UHF solutions of \ce{Fe(CN)6^{3-}} alongside their symmetry assignments from \qsymsq{} determined at the linear independence threshold $\lambda^{\mathrm{thresh}}_{\mathbf{S}} = \num{1e-7}$.
      All four solutions are found to be symmetry-broken (\textit{i.e.}, they each span a reducible representation space of $\mathcal{O}_h$), and interestingly, the lowest two solutions, A and B, do not contain the $T_{2g}$ term na\"{\i}vely expected of the ground state for a $d^5$ configuration in an octahedral strong-field environment\cite{article:Tanabe1954b}, whereas the other two solutions, C and D, do.
      In addition, the A and B solutions have a different inversion symmetry (\textit{ungerade}) compared to that of the C and D solutions (\textit{gerade}).
      This strongly suggests that there is a qualitative difference between these two pairs of solutions.

      For illustrative purposes, we focus only on the lower-energy solution in each of the two pairs, namely the A and C solutions.
      To acquire a crude understanding of the origin of this qualitative difference, we turn to the Pipek--Mezey-localized MOs\cite{article:Pipek1988,article:Pipek1989} obtainable from the canonical MOs of these two solutions, mainly because localized MOs have been known to provide a useful link between detailed quantum-chemical calculations and classical chemical concepts such as non-bonding orbitals, lone pairs, and multiple bonds with which most chemists have gained great familiarity and intuition\cite{booksection:England1971,article:Thom2009}.
      In the particular case of \ce{Fe(CN)6^{3-}}, localized orbitals help quantify the number of $d$-electrons on the iron center and allow for a discussion on the nature of the UHF solutions in terms of the metal $d^n$-electronic-configuration and oxidation-state descriptors that are common in coordination chemistry.

      In Tables~\ref{tab:dMOsA}~and~\ref{tab:dMOsC}, the Pipek--Mezey-localized $d$-MOs for the A and C solutions are listed, respectively.
      Each of these MOs has a $d$-shell Mulliken population of at least \num{0.980} and can therefore be regarded as being predominantly contributed to by a $d$-electron on the iron center.
      Clearly, the iron center in solution A admits a $d^6$ configuration, whereas that in solution C admits a $d^5$ configuration.
      This is also confirmed by a LOBA oxidation-state analysis formulated by Thom \textit{et al.}\cite{article:Thom2009}: the iron center in solution A has an oxidation state of \num[explicit-sign=+]{+2}, whereas that in solution C has an oxidation state of \num[explicit-sign=+]{+3}.
      Noting that all $d$-orbitals on the iron center must have \textit{gerade} inversion symmetry, and also that all $p$-orbitals on the iron center have been confirmed to be paired, we conclude that the \textit{ungerade} inversion symmetry in solution A must arise from an unmatched \textit{ungerade} ligand orbital between the two spin spaces.
      The seemingly innocent \textit{ungerade} inversion symmetry found in solution A turns out to be a manifestation of a ligand-to-metal charge transfer process.

      The fact that the four UHF solutions A--D are symmetry-broken means that none of them is able to provide a physical description of the ground state of the system\cite{article:Lykos1963}.
      However, it has been demonstrated elsewhere\cite{article:Huynh2020} that post-HF methods such as NOCI can yield multi-determinantal wavefunctions that conserve symmetry and thus provide more appropriate approximations of the ground state.
      For this to be viable, either a basis $\mathcal{B}$ spanning a complete representation space $W$ [Equation~\eqref{eq:wtransformed}] or a full symmetry-equivalent orbit spanning the same space [Equation~\eqref{eq:orbitGw}] must be provided as the basis for NOCI --- both of which can be generated by \qsymsq{}.

\begin{figure*}
        \centering
        \ifdefined\twocolumnmode
          \includegraphics[scale=.8]{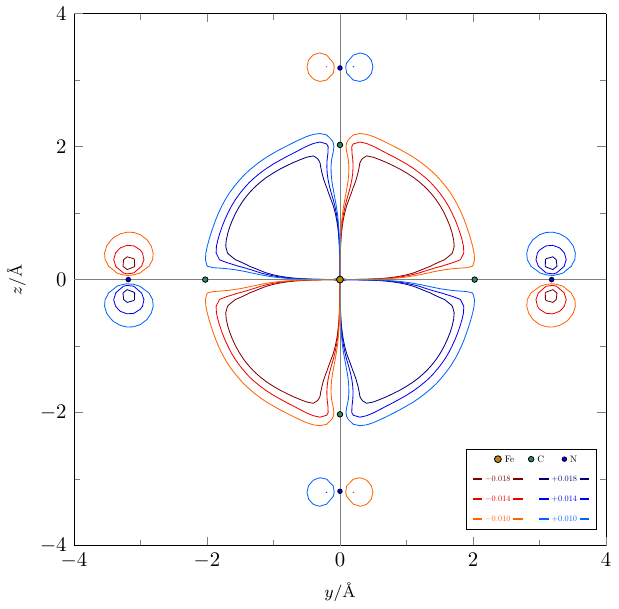}
        \else
          \includegraphics[scale=.8]{figures/fig5-symbrokencontour/symbrokencontour.tikz.pdf}
        \fi
        \caption{%
          Contours of Pipek--Mezey-localized MO $\chi^{\beta}_{34}$ of solution A in the $yz$-plane.
          The inequivalence between the interactions of the \ce{(CN)-} $\pi$ orbitals in the $y$- and $z$-directions with the \ce{Fe}-based $d_{yz}$ orbital accounts for the $T_{1g} \oplus T_{2g}$ symmetry breaking of this MO.
        }
        \label{fig:Achib34contour}
      \end{figure*}

      We conclude this case study with a remark that, as expected, the symmetry breaking of the overall determinants shown in Table~\ref{tab:fecn3mscf} can be traced back to the symmetry breaking of the constituting orbitals.
      This is in fact demonstrated by the symmetry assignments for the Pipek--Mezey-localized $d$-MOs of the A and C solutions in Tables~\ref{tab:dMOsA}~and~\ref{tab:dMOsC}, respectively.
      It should be noted, however, that symmetry breaking effects can sometimes be subtle and difficult to discern from a mere visual inspection of isosurface plots.
      A detailed analysis based on the formulation in Section~\ref{sec:repanalysis} should therefore be preferred to obtain reliable symmetry information.
      For instance, consider the MO $\chi^{\beta}_{34}$ of solution A whose isosurface at \num{0.014} is shown in Table~\ref{tab:dMOsA}.
      At first glance, this orbital appears just like a typical $d_{yz}$ orbital (with some distortions due to interactions with the ligands) and should just have $T_{2g}$ symmetry.
      However, a close inspection of this isosurface, with the aid of the contour plot in the $yz$-plane shown in Figure~\ref{fig:Achib34contour}, reveals that the interactions with the ligands on the $y$-axis are not equivalent to those on the $z$-axis, thus causing the relation $\hat{C}_4^{x} \chi^{\beta}_{34} = -\chi^{\beta}_{34}$ to fail to hold.
      In other words, $\chi^{\beta}_{34}$ and $\hat{C}_4^{x} \chi^{\beta}_{34}$ are linearly independent, which gives rise to the $T_{1g} \oplus T_{2g}$ symmetry breaking as observed.
      The large gap between the boundary orbit overlap eigenvalues $\lambda^{>}_{\mathbf{S}}$ and $\lambda^{<}_{\mathbf{S}}$ (\textit{ca.} 10 orders of magnitude) indicates that this symmetry breaking is in fact not just a numerical artifact of the analysis, but rather an intrinsic feature of this MO.

  \subsection{Symmetry in external fields}
    Thus far, we have demonstrated the symmetry analysis capability of \qsymsq{} for real orbitals and determinants in the absence of any external fields.
    In this next case study, we illustrate how \qsymsq{} can be used to understand quantum-chemical behaviors when external fields are introduced.
    In particular, we shall show that, for a hydrogen fluoride molecule in a uniform magnetic field, a knowledge of the symmetries of the complex-valued MOs helps rationalize the reversal of the electric dipole moment along the inter-nuclear axis observed by Irons \textit{et al.}\cite{article:Irons2022} at strong fields perpendicular to the molecule but not, curiously, at parallel fields.

    \subsubsection{Computational details}
      We followed Irons \textit{et al.}\cite{article:Irons2022} and performed current-DFT calculations with the cTPSS functional in the uncontracted aug-cc-pVQZ basis set\cite{article:Dunning1989,article:Wilson1999} employing the resolution-of-the-identity approximation with the \textsc{AutoAux} auxiliary basis\cite{article:Stoychev2017} in \textsc{QUEST}\cite{software:Quest2022}.
      The obtained KS MOs were then passed to \qsymsq{} for symmetry analysis in the appropriate unitary symmetry group $\mathcal{G}$ of the molecule-plus-field system (\textit{cf.} Section~\ref{sec:magsym}).
      In all cases, the gauge origin of the magnetic field and the center of mass of the molecule were placed at the origin of the Cartesian coordinate system so that the gauge origin would always be left invariant by the applications of all symmetry operations during the orbit generation [Equation~\eqref{eq:orbitGw}].\bibnote{This choice is merely for computational convenience because symmetry properties of MOs must be gauge-origin-invariant.}
      For the parallel and perpendicular field orientations, complex MO isosurfaces were also plotted in \textsc{VMD}\cite{article:Humphrey1996} using the method described by Al-Saadon \textit{et al.}\cite{article:Al-Saadon2019}.

\begin{figure}
        \centering
        \includegraphics[width=.35\textwidth]{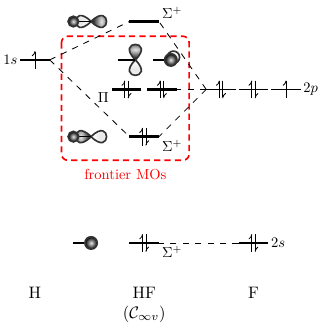}
        \caption{A simplistic depiction of the MOs in hydrogen fluoride at zero field.}
        \label{fig:hffmos}
      \end{figure}

      In the cases where $\mathcal{G}$ is an infinite group (\textit{i.e.}, $\mathcal{C}_{\infty v}$ at zero field or $\mathcal{C}_{\infty}$ at parallel-field orientations), a finite integer order $n \ge 2$ is chosen for the infinite-order rotation axis $C_{\infty}$ so that $\mathcal{G}$ is restricted to a suitable finite subgroup $\mathcal{G}_n$ (\textit{i.e.}, $\mathcal{C}_{nv}$ or $\mathcal{C}_n$, respectively) in which the representation analysis of Section~\ref{sec:repanalysis} is carried out by \qsymsq{}.
      The actual representations in $\mathcal{G}$ can then be deduced by the representations in $\mathcal{G}_n$ produced by \qsymsq{} according to the following subduction rules:
      \ifdefined\twocolumnmode
        \begin{scriptsize}
          \begin{alignat*}{6}
            \Sigma^+(\mathcal{C}_{\infty v})
              & \downarrow \mathcal{C}_{nv}
              && = A_1
              &\qquad
              \Sigma(\mathcal{C}_{\infty})
                & \downarrow \mathcal{C}_{n}
                && = A \\
            \Sigma^-(\mathcal{C}_{\infty v})
              & \downarrow \mathcal{C}_{nv}
              && = A_2 \\
            \Pi(\mathcal{C}_{\infty v})
              & \downarrow \mathcal{C}_{nv}
              && = E_1
              &\qquad
              \Gamma_1(\mathcal{C}_{\infty})
                & \downarrow \mathcal{C}_{n}
                && = \Gamma_1\\[-3pt]
            & && &\qquad
              \bar{\Gamma}_1(\mathcal{C}_{\infty})
                & \downarrow \mathcal{C}_{n}
                && = \bar{\Gamma}_1\\
            \Delta(\mathcal{C}_{\infty v})
              & \downarrow \mathcal{C}_{nv}
              && = E_2
              &\qquad
              \Gamma_2(\mathcal{C}_{\infty})
                & \downarrow \mathcal{C}_{n}
                && = \Gamma_2\\[-3pt]
            & && &\qquad
              \bar{\Gamma}_2(\mathcal{C}_{\infty})
                & \downarrow \mathcal{C}_{n}
                && = \bar{\Gamma}_2\\
              & && \vdotswithin{=} & & && \vdotswithin{=} \\
            E_{\lceil n/2 \rceil - 1}(\mathcal{C}_{\infty v})
              & \downarrow \mathcal{C}_{nv}
              && = E_{\lceil n/2 \rceil - 1}
              &\qquad
              \Gamma_{\lceil n/2 \rceil - 1}(\mathcal{C}_{\infty})
                & \downarrow \mathcal{C}_{n}
                && = \Gamma_{\lceil n/2 \rceil - 1}\\[-3pt]
            & && &\qquad
              \bar{\Gamma}_{\lceil n/2 \rceil - 1}(\mathcal{C}_{\infty})
                & \downarrow \mathcal{C}_{n}
                && = \bar{\Gamma}_{\lceil n/2 \rceil - 1}
          \end{alignat*}
        \end{scriptsize}
      \else
        \begin{alignat*}{6}
          \Sigma^+(\mathcal{C}_{\infty v})
            & \downarrow \mathcal{C}_{nv}
            && = A_1
            &\qquad\qquad
            \Sigma(\mathcal{C}_{\infty})
              & \downarrow \mathcal{C}_{n}
              && = A \\
          \Sigma^-(\mathcal{C}_{\infty v})
            & \downarrow \mathcal{C}_{nv}
            && = A_2 \\
          \Pi(\mathcal{C}_{\infty v})
            & \downarrow \mathcal{C}_{nv}
            && = E_1
            &\qquad\qquad
            \Gamma_1(\mathcal{C}_{\infty})
              & \downarrow \mathcal{C}_{n}
              && = \Gamma_1\\[-9pt]
          & && &\qquad\qquad
            \bar{\Gamma}_1(\mathcal{C}_{\infty})
              & \downarrow \mathcal{C}_{n}
              && = \bar{\Gamma}_1\\
          \Delta(\mathcal{C}_{\infty v})
            & \downarrow \mathcal{C}_{nv}
            && = E_2
            &\qquad\qquad
            \Gamma_2(\mathcal{C}_{\infty})
              & \downarrow \mathcal{C}_{n}
              && = \Gamma_2\\[-9pt]
          & && &\qquad\qquad
            \bar{\Gamma}_2(\mathcal{C}_{\infty})
              & \downarrow \mathcal{C}_{n}
              && = \bar{\Gamma}_2\\
            & && \vdotswithin{=} & & && \vdotswithin{=} \\
          E_{\lceil n/2 \rceil - 1}(\mathcal{C}_{\infty v})
            & \downarrow \mathcal{C}_{nv}
            && = E_{\lceil n/2 \rceil - 1}
            &\qquad\qquad
            \Gamma_{\lceil n/2 \rceil - 1}(\mathcal{C}_{\infty})
              & \downarrow \mathcal{C}_{n}
              && = \Gamma_{\lceil n/2 \rceil - 1}\\[-9pt]
          & && &\qquad\qquad
            \bar{\Gamma}_{\lceil n/2 \rceil - 1}(\mathcal{C}_{\infty})
              & \downarrow \mathcal{C}_{n}
              && = \bar{\Gamma}_{\lceil n/2 \rceil - 1}
        \end{alignat*}
      \fi
      where $\Gamma_k$ and $\bar{\Gamma}_k$ in $\mathcal{C}_{\infty}$ and $\mathcal{C}_{n}$ are complex-conjugate one-dimensional irreducible representations with character functions
      \begin{align*}
        \chi^{\Gamma_k}[\hat{C}(\phi)] &= \exp(ik \phi),\\
        \chi^{\bar{\Gamma}_k}[\hat{C}(\phi)] &= \exp(-ik \phi).
      \end{align*}
      For each choice of integer $n \ge 2$, irreducible representations up to and including $E_{\lceil n/2 \rceil - 1}$ in $\mathcal{G} = \mathcal{C}_{\infty v}$ and $\Gamma_{\lceil n/2 \rceil - 1}$ and $\bar{\Gamma}_{\lceil n/2 \rceil - 1}$ in $\mathcal{G} = \mathcal{C}_{\infty}$ remain irreducible in the respective subgroups $\mathcal{G}_n = \mathcal{C}_{nv}$ and $\mathcal{C}_{n}$.
      These irreducible representations in $\mathcal{G}$ can therefore be deduced unambiguously from those in $\mathcal{G}_n$.
      In the current case study of hydrogen fluoride, it is known from basic MO theory (Figure~\ref{fig:hffmos}) that MOs of up to $\Pi$ symmetry at zero field are occupied in the ground state, and so we require $n \ge 3$ so that $\mathcal{C}_{nv}$ and $\mathcal{C}_{n}$ have enough irreducible representations to describe $\Pi(\mathcal{C}_{\infty v})$, $\Gamma_1(\mathcal{C}_{\infty})$, and $\bar{\Gamma}_1(\mathcal{C}_{\infty})$ symmetries unequivocally.
      In fact, for good measure, we chose $n = 8$ in all infinite-group symmetry analyses for hydrogen fluoride in \qsymsq{}.

    \subsubsection{MO description of electric dipole reversal in a magnetic field}

      \begin{figure*}
        \centering
        \begin{subfigure}[t]{.45\textwidth}
          \centering
          \includegraphics[width=\textwidth]{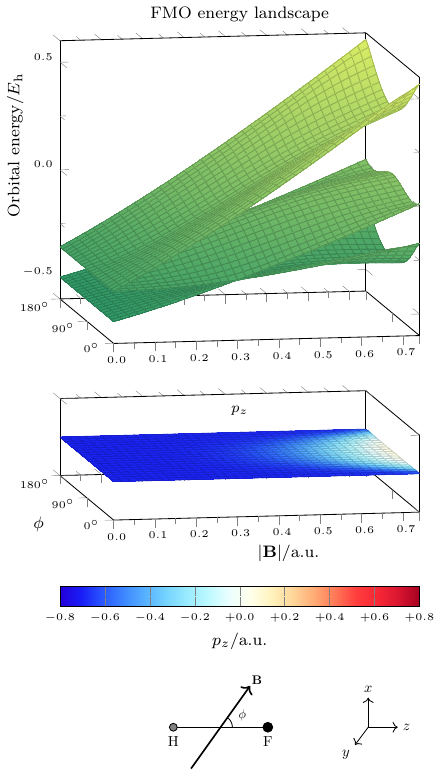}
          \caption{}
          \label{fig:hfbfieldlandscape}
        \end{subfigure}
        \hfill
        \begin{subfigure}[t]{.523\textwidth}
          \centering
          \includegraphics[width=\textwidth]{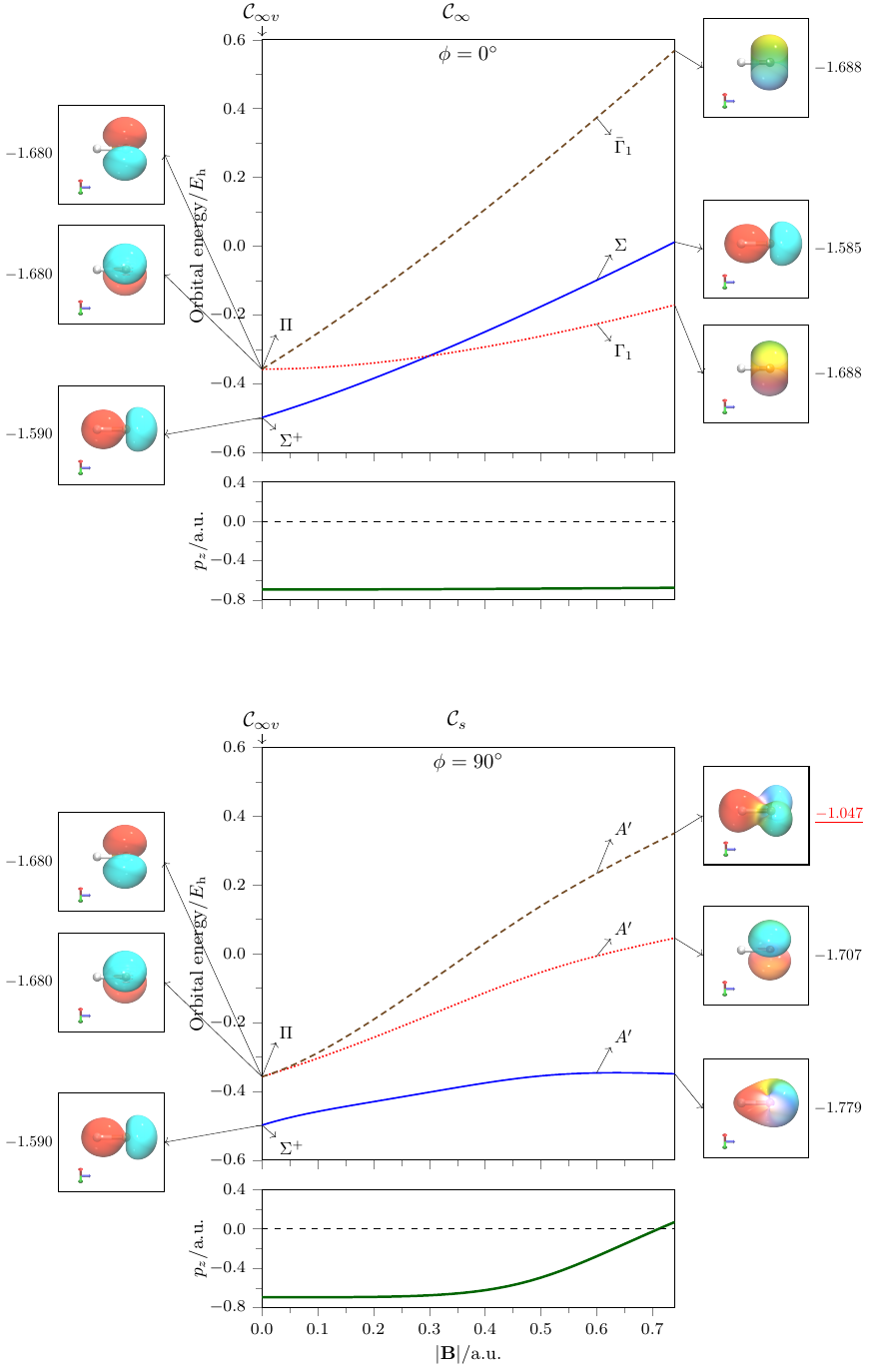}
          \caption{}
          \label{fig:hfbfieldcrosssections}
        \end{subfigure}

        \begin{subfigure}[t]{\textwidth}
          \centering
          \includegraphics[scale=0.8]{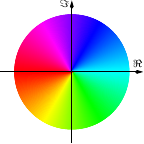}
        \end{subfigure}
        \caption{%
          (a) Energy landscapes of the frontier $m_s = +1/2$ MOs in hydrogen fluoride at various magnetic field strengths and angles.
          (b) The cross-sections through these landscapes at parallel (top) and perpendicular (bottom) field orientations.
          Annotated on these cross-sections are the MO symmetries and isosurfaces plotted at $\lvert \chi(\mathbf{r}) \rvert = 0.100$.
          The color at each point $\mathbf{r}$ on an isosurface indicates the value of $\arg \chi(\mathbf{r})$ at that point according to the accompanying color wheel.
          The numerical value next to each isosurface gives the value of the orbital electronic dipole moment $\braket{\chi | \hat{\mu}_z | \chi}$ for the associated MO.
          In $\mathcal{C}_{\infty}$, the one-dimensional irreducible representation $\Gamma_1$ has character function $\chi^{\Gamma_1}[\hat{C}_{\infty}(\phi)] = \exp(i\phi)$, and the corresponding complex-conjugate one-dimensional irreducible representation $\bar{\Gamma}_1$ has character function $\chi^{\bar{\Gamma}_1}[\hat{C}_{\infty}(\phi)] = \exp(-i\phi)$, where $\hat{C}_{\infty}(\phi)$ denotes an anticlockwise rotation through an angle $\phi$ as viewed down the $z$-axis.
        }
        \label{fig:hfbfield}
      \end{figure*}

      Figure~\ref{fig:hfbfieldlandscape} shows the landscapes of the $m_s = +1/2$ frontier MOs in hydrogen fluoride (\textit{cf.} Figure~\ref{fig:hffmos}) at various strengths and orientations of the external magnetic field together with the values of the electric dipole moment component along the inter-nuclear axis, $p_z$.
      The \ce{H-F} bond length is kept fixed at its zero-field equilibrium value, \SI{0.92897}{\angstrom}, for all field strengths and orientations.
      It can be seen that, in the region where $p_z$ becomes less negative and approaches zero ($\lvert \mathbf{B} \rvert \ge 0.5 B_0 \approx \SI{1.18e5}{\tesla}$ and $\phi \approx 90^{\circ}$), the frontier MO landscapes display significant curvature.
      This suggests that these MOs interact with one another strongly in this region, and these interactions might be responsible for the observed electric dipole reversal.
      However, in parallel fields, even at very high field strengths ($\lvert \mathbf{B} \rvert \ge 0.7 B_0 \approx \SI{1.65e5}{\tesla}$), $p_z$ remains at \textit{ca.} \SI{-0.7}{\atomicunit} which is approximately the same value as that at zero field.
      The energy landscapes of the frontier MOs also show very little curvature in the vicinity of $\phi = 0^{\circ}$ or $180^{\circ}$, thus implying a lack of interaction between these MOs and further strengthening the conjecture that these MOs must interact in some way to result in a reversal of the electric dipole moment.

      In Figure~\ref{fig:hfbfieldcrosssections}, cross-sections through the frontier MO energy landscapes and $p_z$ plots at $\phi = 0^{\circ}$ and $90^{\circ}$ are shown together with MO symmetry assignments from \qsymsq{} and complex isosurface plots as described earlier.
      It becomes immediately obvious that, at $\phi = 0^{\circ}$, the three frontier MOs have different symmetries at all values of $\lvert \mathbf{B} \rvert$ and are therefore unable to interact via the KS operator.
      In fact, the $\Sigma$ and $\Gamma_1$ energy curves are able to cross because of their different symmetries.
      Even though their energies vary quite significantly as $\lvert \mathbf{B} \rvert$ increases, this variation is due only to the interactions of these MOs with the applied field.
      Although these interactions do lead to qualitative changes in the shapes of the MOs, most notably the disappearance of nodal planes in the two $\Pi$ MOs at zero field that become $\Gamma_1$/$\bar{\Gamma}_1$ MOs at $\lvert \mathbf{B} \rvert > 0$, these changes do not actually affect the distribution of electrons along the inter-nuclear axis in any significant way.
      This is indeed confirmed by the near equality of the $\braket{\chi | \hat{\mu}_z | \chi}$ values of these MOs at $\lvert \mathbf{B} \rvert = 0$ and $0.74 B_0$.
      Consequently, the electric dipole moment along the inter-nuclear axis remains almost unchanged.

      The situation is markedly different at $\phi = 90^{\circ}$.
      The three frontier MOs now have the same symmetry in $\mathcal{C}_s$ and are thus permitted to interact via the KS operator.
      Indeed they do, as is evident from the distortions in their energy curves for $\lvert \mathbf{B} \rvert \ge 0.5 B_0$ and also in the shapes of their isosurfaces.
      Most significantly, the highest occupied MO (HOMO) shows the most drastic change from a $2p$ orbital localized entirely on the fluorine atom to a laterally delocalized MO with a pronounced lobe on the hydrogen atom.
      Associated with this change is the large increase in the value of $\braket{\chi | \hat{\mu}_z | \chi}$ for this MO from \SI{-1.680}{\atomicunit} at zero field to \SI{-1.047}{\atomicunit} at $\lvert \mathbf{B} \rvert = 0.74 B_0$, which more than outweighs the decreases in the values of $\braket{\chi | \hat{\mu}_z | \chi}$ for the other two frontier MOs.
      There is thus a partial charge transfer from the fluorine atom to the hydrogen atom in the HOMO induced by the perpendicular magnetic field, which is responsible for the observed dipole reversal.

  \subsection{Symmetry of electron densities}

      In the final set of case studies, we demonstrate the ability of \qsymsq{} to perform symmetry analysis for electron densities, as formulated in Section~\ref{sec:exampleslinearspacequantities}.
      In particular, we show how the symmetry of the electron density is intimately related to that of the underlying electronic wavefunction, both at zero field and in the presence of external electric and magnetic fields.

    \subsubsection{Computational details}
      For the above purpose, we chose the equilateral geometry of \ce{H3^{+}} that has been found in Ref.~\citenum{article:Wibowo2023} to be the optimal geometry for the lowest $M_S = -1$ electronic state when a uniform magnetic field of strength $\lvert \mathbf{B} \rvert = 1.0 B_0$ is applied perpendicular to the plane of the molecule.
      For this geometry, the lowest $M_S = -1$ wavefunctions and densities were computed in \textsc{QUEST}\cite{software:Quest2022} at the UHF/6-311++(2+,2+)G** level of theory in three cases: at zero field, in the presence of a perpendicular uniform electric field with strength $\lvert \boldsymbol{\mathcal{E}} \rvert = \SI{0.1}{\atomicunit}$, and in the presence of a perpendicular uniform magnetic field with strength $\lvert \mathbf{B} \rvert = 1.0 B_0$.
      The symmetry assignments for the resulting determinantal wavefunctions and the corresponding electron densities were then determined by \qsymsq{}.
      In all cases, the \ce{H3^{+}} structure was placed in the $yz$-plane so that any external field applied perpendicular to the molecule would be along the $x$-direction.

    \subsubsection{Density symmetries in \ce{H3+}}
      Table~\ref{tab:densym} shows the wavefunction and density symmetries of the lowest $M_S = -1$ UHF wavefunction in the three cases described above.
      We examine first the perpendicular magnetic field case (labeled $B_x$ in Table~\ref{tab:densym}) where the unitary symmetry group of the molecule-plus-field system is $\mathcal{C}_{3h}$.
      The lowest $M_S = -1$ UHF wavefunction has already been reported in Ref.~\citenum{article:Wibowo2023} to have $\Gamma'(\mathcal{C}_{3h})$ symmetry, which is a one-dimensional irreducible representation in $\mathcal{C}_{3h}$ whose character function satisfies $\chi^{\Gamma'}(\hat{C}_3) = \exp(2i\pi/3)$ and $\chi^{\Gamma'}(\hat{\sigma}_h) = 1$.
      As this is a non-degenerate wavefunction, the corresponding density must be totally symmetric in $\mathcal{C}_{3h}$, which is indeed the case as verified by the density symmetry assignment and also by the density isosurface and contours in the $yz$-plane.
      Here, the electron cloud can be seen to be equi-distributed over the three symmetry-equivalent hydrogen nuclei.

      \begin{table*}
        \centering
        \caption{%
          Electronic wavefunction and total density symmetry of the lowest $M_S = -1$ state of \ce{H3^{+}} in the presence of external electric and magnetic fields.
          Calculations were performed at the UHF/6-311++(2+,2+)G** level of theory.
          The magnitude of the applied electric field is \SI{0.1}{\atomicunit} and that of the applied magnetic field is $1.0 B_0$.
          Isosurfaces of total densities are plotted at $\lvert \rho(\mathbf{r}) \rvert = 0.050$.
        }
        \label{tab:densym}
        \renewcommand*{\arraystretch}{1.5}
        \begin{tabular}{%
            R{3.1cm} M{3.9cm} M{3.9cm} M{3.9cm}
          }
          \toprule
              Field
          & $\mathbf{0}$
          & $\mathcal{E}_{x}$
          & $B_{x}$\\
          \midrule
          Symmetry group
            & $\mathcal{D}_{3h}$
            & $\mathcal{C}_{3v}$
            & $\mathcal{C}_{3h}$\\
          Wavefunction symmetry
            & $A_{1}' \oplus E'$
            & $A_{1} \oplus E$
            & $\Gamma'$\\
          Density symmetry
            & $A_{1}' \oplus E'$
            & $A_{1} \oplus E$
            & $A'$\\
          Density isosurface
            & \includegraphics[trim=0.0cm 0.10cm 0.0cm 0.15cm, clip, scale=1.6]{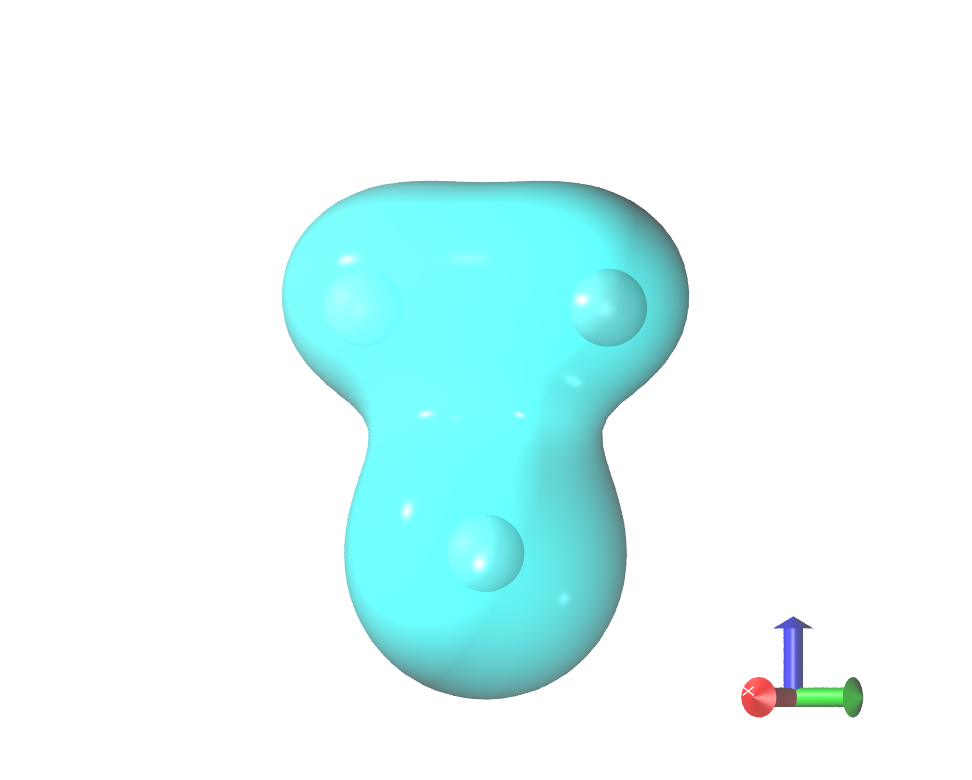}
            & \includegraphics[trim=0.0cm 0.10cm 0.0cm 0.15cm, clip, scale=1.6]{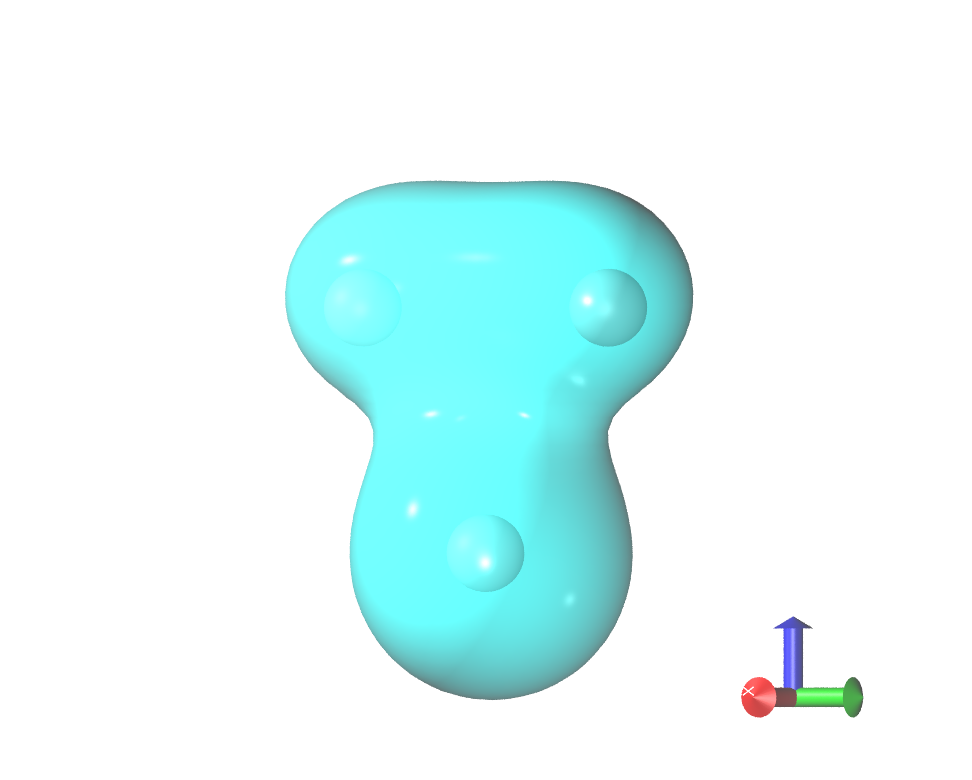}
            & \includegraphics[trim=0.0cm 0.10cm 0.0cm 0.15cm, clip, scale=1.6]{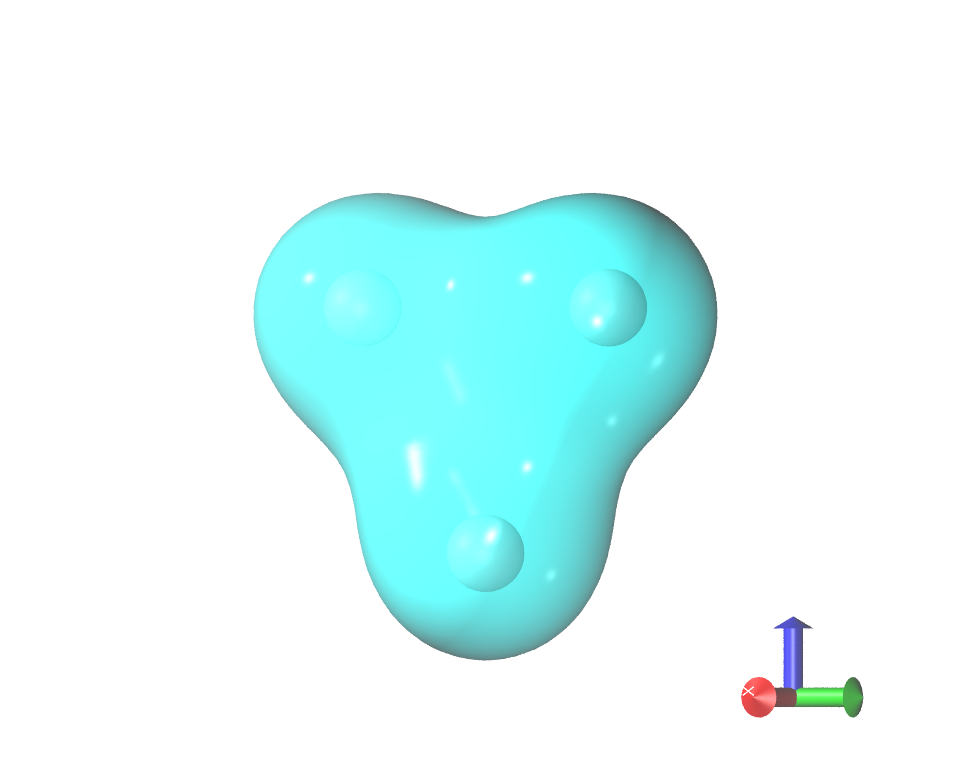}\\
          Density contours in $yz$-plane
            & \includegraphics[width=\linewidth]{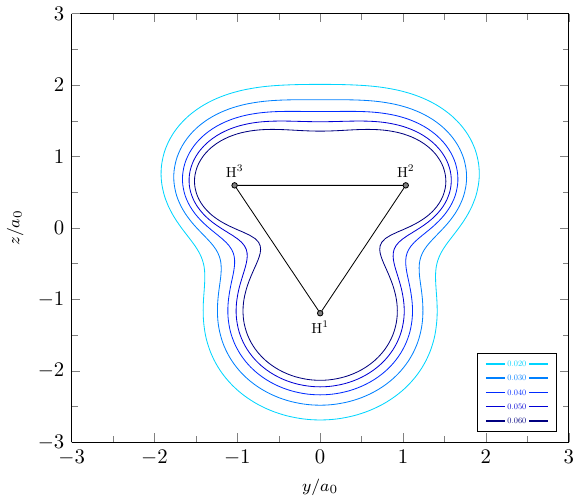}
            & \includegraphics[width=\linewidth]{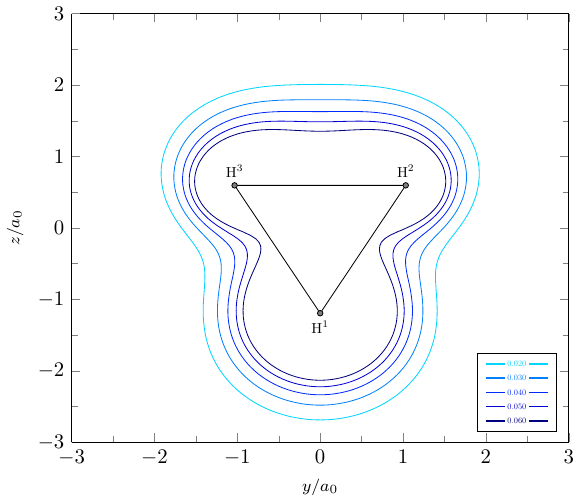}
            & \includegraphics[width=\linewidth]{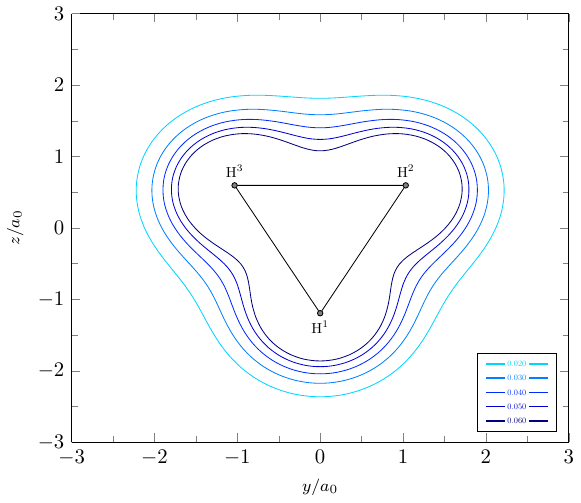}\\
          \bottomrule
        \end{tabular}
      \end{table*}

      The situation is rather different in the other two cases.
      In the absence of any external fields (labeled $\mathbf{0}$ in Table~\ref{tab:densym}), for which the unitary symmetry group is $\mathcal{D}_{3h}$, the lowest $M_S = -1$ UHF wavefunction at the same geometry turns out to exhibit symmetry breaking due to its $A_{1}' \oplus E'(\mathcal{D}_{3h})$ symmetry.
      The symmetry analysis of the corresponding total density shows that the density also exhibits an $A_{1}' \oplus E'(\mathcal{D}_{3h})$ broken symmetry.
      Similarly, in the presence of a perpendicular electric field (labeled $\mathcal{E}_x$ in Table~\ref{tab:densym}), for which the unitary symmetry group is reduced to $\mathcal{C}_{3v}$, the UHF wavefunction now has $A_{1} \oplus E(\mathcal{C}_{3v})$ symmetry, as does the corresponding density.
      The symmetry breaking of the total density in these two cases can be visualized most easily via the density isosurfaces and $yz$-contours: the electron cloud is not equally distributed over the three symmetry-equivalent hydrogen nuclei.

      The external fields can also be applied parallel to the plane of the molecular frame of \ce{H3^{+}}.
      The wavefunction and density symmetries and density isosurfaces resulting from the UHF calculations in these field orientations are summarized in Table~S1 in the Supporting Information.
      In all cases, it can be observed that, whenever the wavefunction is non-degenerate, the density is totally symmetric in the prevailing symmetry group, and whenever the wavefunction exhibits degeneracy, be it because of symmetry breaking or not, the corresponding density becomes non-totally-symmetric.

      We note that it is also possible to obtain density symmetries without having access to any symmetries of the underlying wavefunction, meaning that the same approach can be readily applied to densities from correlated wavefunction methods as well as HF or KS-DFT, with no additional implementation.
      In the context of KS-DFT, this means that symmetry analysis can be carried out directly on the density, rather than the KS orbitals and non-interacting wavefunction as a proxy for the physical wavefunction.

      Examples of symmetry analyses for KS densities obtained with the r\textsuperscript{2}SCAN0 functional are also shown in Table~S1 in the Supporting Information for various external field orientations.
      By considering the symmetries of electron densities, we have a qualitative way to compare HF and KS-DFT calculations: if HF and KS densities have the same symmetry, then there is a likelihood that both calculations describe the same electronic state of the system, but if HF and KS densities differ in their symmetries, then it must be concluded that they are qualitatively different, perhaps because they describe different electronic states, which can occur especially if there are multiple SCF solutions that occur close to one another (see Refs.~\citenum{article:Stanton1968,article:King1969,article:Redondo1989,article:Thom2008,article:Huynh2020} and also Section~\ref{sec:symbreaking} above).
      In fact, for the cases listed in Table~S1, based on the symmetries of densities, both HF and KS-DFT calculations in each field orientation can be said to approximate the same electronic state.

%% file: conclusion/conclusion.tex
\section{Conclusion}
\label{sec:conclusion}

  A new program for quantum symbolic symmetry analysis, \qsymsq{}, has been presented in this work.
  A key feature of the program is its capability to generate character tables symbolically on-the-fly, which endows it with the ability to perform symmetry analysis for general groups automatically.
  This flexibility means that \qsymsq{} can yield reliable symmetry assignments for systems exhibiting degeneracy and symmetry breaking effects, where standard implementations cannot be applied.
  In addition, \qsymsq{} can handle reduced symmetries arising in electric and/or magnetic fields, thus providing a valuable tool for analysis and insight into systems under less chemically intuitive conditions.

  The ability of \qsymsq{} to perform analysis of high-symmetry systems was demonstrated for \ce{C84H64}, \ce{C60}, and \ce{B9-}, where in each case the full molecular symmetry group could be correctly identified and carried through to classify the symmetry of the resulting MOs, including their associated degeneracies.
  The octahedral transition-metal complex \ce{Fe(CN)6^{3-}} was then used to demonstrate how \qsymsq{} is able to deduce representation spaces spanned by symmetry-broken determinants and MOs, giving a way to classify and understand symmetry breaking effects.
  Furthermore, the capability of \qsymsq{} to analyze symmetry in external magnetic fields was demonstrated for the hydrogen fluoride molecule, where the symmetry of the MOs under a magnetic field was shown to provide a rationalization of the behavior of the molecular electric dipole moment as a function of field strength and orientation.

  An important benefit of the generic symmetry-orbit-based representation analysis framework formulated in this article and used in \qsymsq{} is the ability to analyze symmetry of quantities other than wavefunctions and MOs that arise in quantum-chemical calculations.
  As a simple example, the changes in the electron density of the equilateral \ce{H3+} as a function of electric and magnetic field were analyzed, with the density symmetry analysis revealing the symmetry breaking or conservation of the underlying wavefunction.
  This approach can be applied on an equal footing to densities arising from SCF calculations or more elaborate post-HF correlated calculations without the need to explicitly perform symmetry analysis on the correlated wavefunctions.

  The generality in the code design of \qsymsq{} opens up many avenues for future research.
  In particular, the applicability of the symmetry-orbit-based representation analysis to members of general linear spaces makes it possible to directly consider the symmetry of many important quantities in quantum chemistry.
  One group of such quantities includes normal coordinates that describe translational, rotational, and vibrational modes of molecules.
  Another interesting class of such quantities includes functions of electron density and/or density matrix, of which the Fukui function\cite{booksection:Ayers2009}, which encapsulates chemical reactivity information, and the magnetically induced current density\cite{article:Sundholm2016,article:Sundholm2021}, which provides an interpretation for observations in magnetic spectroscopic methods, are prime examples.
  Moreover, \qsymsq{} can already provide a more complete analysis of magnetic symmetry using corepresentation theory\cite{book:Wigner1959,article:Cracknell1966,article:Bradley1968} and is not limited to uniform external fields --- these developments will be reported in future publications.

  Finally, we emphasize that the Rust implementation of \qsymsq{} is also flexible, such that the program can operate either as a tool to be applied subsequent to a quantum-chemical calculation in a stand-alone manner (as used in this work with \textsc{Q-Chem}), or as a library readily integrated into existing programs (as used in this and earlier\cite{article:Wibowo2023,article:Wibowo-Teale2023} work with \textsc{QUEST}\cite{software:Quest2022}).
  Since the program is available open-source, we hope that it will become a useful tool for application in a wide range of chemical simulations.

%% file: suppinfo/symopstruct/symopstruct.tex
\section{Computational design of the \texttt{SymOp} structure}
\label{app:symopstruct}

  \subsection{Fields in the \texttt{SymOp} structure}

    Every finite-order symmetry operation $\hat{g}$ of molecular systems in three dimensions can be written most generally as
    \begin{equation}
      \hat{g} = \hat{t} \hat{s} \hat{C}_n^k(\mathbf{n}).
      \label{eq:gensymop}
    \end{equation}
    In the above expression, $\hat{C}_n^k(\mathbf{n})$ denotes a proper rotation about an axis defined by the normalized vector $\mathbf{n}$ through an angle $2\pi k/n$ as viewed down $\mathbf{n}$ where the positive sign is associated with the \emph{anticlockwise} direction.
    It must be noted that both $k$ and $n$ are constrained to be integers such that
    \begin{equation}
      n \ge 1, \quad
      \left\lfloor -\frac{n}{2} \right\rfloor < k \le \left\lfloor \frac{n}{2} \right\rfloor,\quad \textrm{and} \quad
      \gcd(k, n) = 1\ \textrm{if}\ k \ne 0,
      \label{eq:knconditions}
    \end{equation}
    so that all possible rotation angles can be uniquely represented over the rationals via the irreducible fractions
    \begin{equation}
      \frac{k}{n} \in \interval[open left, scaled]{-\frac{1}{2}}{\frac{1}{2}} \cap \mathbb{Q}.
    \end{equation}
    This choice of $k \in \mathbb{Z}/n\mathbb{Z}$ ensures that the identity operation, which corresponds to $k = 0$, always lies at the middle of the interval.
    Furthermore, the restriction of rotation angle fractions from the reals to the rationals over $\interval[open left]{-1/2}{1/2}$ is possible because of the discrete nature of atomic arrangements in molecular systems.

    The proper rotation $\hat{C}_n^k(\mathbf{n})$ may then be accompanied by a potentially improper involution:
    \begin{equation}
      \hat{s} =
      \begin{cases}
        \hat{e}                   & \textrm{if $\hat{g}$ is proper},                \\
        \hat{i}                   & \textrm{if $\hat{g}$ is an inversion-rotation}, \\
        \hat{\sigma}_{\mathbf{n}} & \textrm{if $\hat{g}$ is a reflection-rotation},
      \end{cases}
      \label{eq:improperterm}
    \end{equation}
    where $\hat{e}$ is the identity, $\hat{i}$ the spatial inversion, and $\hat{\sigma}_{\mathbf{n}}$ the reflection in a plane perpendicular to $\mathbf{n}$.
    If $\hat{s} = \hat{\sigma}_{\mathbf{n}}$, then the composition $\hat{\sigma}_{\mathbf{n}} \hat{C}_n^k$ is commonly written as $\hat{S}_n^k$ where care must be taken when interpreting the exponent $k$ which belongs only to the proper-rotation part of the composition.
    $\hat{C}_n^k(\mathbf{n})$ may also be accompanied by a potentially antiunitary involution:
    \begin{equation}
      \hat{t} =
      \begin{cases}
        \hat{e}      & \textrm{if $\hat{g}$ is unitary}, \\
        \hat{\theta} & \textrm{if $\hat{g}$ is antiunitary},
      \end{cases}
      \label{eq:antiunitaryterm}
    \end{equation}
    where $\hat{\theta}$ is the time-reversal operation introduced in Section~2.1.3 of the main text.

    To faithfully represent any symmetry operation with the general form shown in Equation~\eqref{eq:gensymop}, the \texttt{SymOp} structure is, in its simplest form, a product type over the following constituent fields:
    \begin{enumerate}[label=(\roman*)]
      \item \texttt{t}: an enumerated type representing $\hat{t}$ whose possible variants are \texttt{identity} for $\hat{e}$ or \texttt{time\_reversal} for $\hat{\theta}$,
      \item \texttt{s}: an enumerated type representing $\hat{s}$ whose possible variants are \texttt{identity} for $\hat{e}$, \texttt{inversion} for $\hat{i}$, or \texttt{reflection} for $\hat{\sigma}_{\mathbf{n}}$,
      \item \texttt{frac}: an ordered pair of integers representing the irreducible fraction $k / n$, and
      \item \texttt{axis}: an ordered triplet of floating-point numbers representing the three components of $\mathbf{n}$.
    \end{enumerate}

  \subsection{Conversion between improper-rotation conventions}

\begin{figure*}
      \centering
      \includegraphics[]{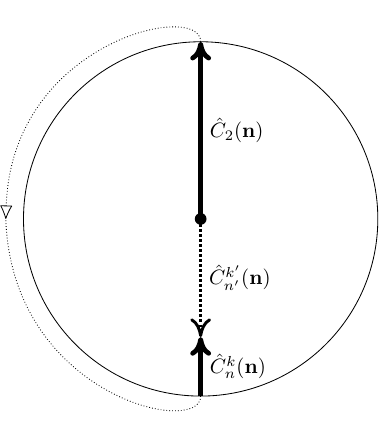}
      \caption{%
        The relation $\hat{C}_2(\mathbf{n}) \hat{C}_n^k(\mathbf{n}) = \hat{C}_{n'}^{k'}(\mathbf{n})$ represented on the parametric ball of rotations in $\mathsf{SO}(3)$.
        The commutativity between $\hat{C}_2(\mathbf{n})$ and $\hat{C}_n^k(\mathbf{n})$ has been exploited to place the path for the $\hat{C}_2(\mathbf{n})$ rotation in a more convenient location.
      }
      \label{fig:improperconversion}
    \end{figure*}

    Both forms of improper rotations are supported in \textsc{QSym\textsuperscript{2}} [Equation~\eqref{eq:improperterm}] since reflection-rotations are more common in chemical applications of symmetry\cite{book:Ceulemans2013}, whereas inversion-rotations are much more natural for comparisons and compositions of symmetry operations\cite{book:Altmann1986}.
    It is therefore essential to be able to convert between the two forms of any improper rotation.
    In other words, given $\hat{\sigma}_{\mathbf{n}}\hat{C}_n^k(\mathbf{n})$, we seek integers $k'$ and $n'$ such that
    \begin{equation}
      \hat{\sigma}_{\mathbf{n}}\hat{C}_n^k(\mathbf{n})
      =
      \hat{i}\hat{C}_{n'}^{k'}(\mathbf{n}),
      \label{eq:improperconversion}
    \end{equation}
    where
    \begin{equation*}
      n' \ge 1, \quad
      \left\lfloor -\frac{n'}{2} \right\rfloor < k' \le \left\lfloor \frac{n'}{2} \right\rfloor,\quad \textrm{and} \quad
      \gcd(k', n') = 1\ \textrm{if}\ k' \ne 0,
    \end{equation*}
    just as $k$ and $n$ in Equation~\eqref{eq:knconditions}.
    By recognising that $\hat{\sigma}_{\mathbf{n}} = \hat{i}\hat{C}_2(\mathbf{n})$, we deduce from Equation~\eqref{eq:improperconversion} the relation
    \begin{equation}
      \hat{C}_2(\mathbf{n}) \hat{C}_n^k(\mathbf{n}) = \hat{C}_{n'}^{k'}(\mathbf{n}),
      \label{eq:improperconversion_rot}
    \end{equation}
    which is represented on the parametric ball of $\mathsf{SO}(3)$ rotations in Figure~\ref{fig:improperconversion} (\textit{cf.} Chapter 10 of Ref.~\citenum{book:Altmann1986} for a detailed description of the rotation parametric ball).
    It is clear from Figure~\ref{fig:improperconversion} that this relation only holds when $k$ and $k'$ are both non-zero and have opposite signs, and that the following equation constraining $k$, $k'$, $n$, and $n'$ must be true:
    \begin{equation}
      \frac{2\pi k}{n} - \frac{2\pi k'}{n'} =
      \begin{cases}
        \hphantom{-}\pi & \quad k > 0, k' < 0 \\
        -\pi            & \quad k < 0, k' > 0
      \end{cases}
      \label{eq:improperconversion_angles}
    \end{equation}

    We consider first the case $k > 0, k' < 0$.
    Equation~\eqref{eq:improperconversion_angles} becomes
    \begin{equation*}
      \frac{2k}{n} - \frac{2k'}{n'} = 1
      \implies
      n' = \frac{2n}{\frac{n-2k}{-k'}}.
    \end{equation*}
    Let $x = \frac{n-2k}{-k'}$.
    For $n'$ to be a positive integer, $x$ must also be a positive integer such that $x \mid 2n$ and $x \mid (n - 2k)$.
    If $n'$ is to be the smallest possible solution, then $x$ must be the largest integer possible that satisfies the above requirements.
    Thus, $x = \gcd(2n, n-2k)$, and consequently,
    \begin{equation*}
      \begin{aligned}
        k' &= -\frac{n - 2k}{\gcd(2n, n - 2k)},\\
        n' &= \hphantom{-}\frac{2n}{\gcd(2n, n - 2k)}.
      \end{aligned}%
      \qquad (k > 0, k' < 0)
    \end{equation*}
    For the second case where $k < 0, k' > 0$, by a similar argument, we obtain
    \begin{equation*}
      \begin{aligned}
        k' &= \frac{n + 2k}{\gcd(2n, n + 2k)},\\
        n' &= \frac{2n}{\gcd(2n, n + 2k)}.
      \end{aligned}%
      \qquad (k < 0, k' > 0)
    \end{equation*}

    The two cases can then be combined to yield a single set of equations giving $k'$ and $n'$ in terms of $k$ and $n$ by
    \begin{equation}
      \begin{aligned}
        k' &= \frac{(2\lvert k \rvert - n)\operatorname{sgn}k}{\gcd(2n, n - 2\lvert k \rvert)},\\
        n' &= \frac{2n}{\gcd(2n, n - 2\lvert k \rvert)}.
      \end{aligned}
      \label{eq:improperconversion_kn}
    \end{equation}
    Furthermore, the conversion in Equation~\eqref{eq:improperconversion} is self-inverse, so if given $\hat{i}\hat{C}_{n'}^{k'}(\mathbf{n})$, one can obtain the values of $k$ and $n$ in $\hat{\sigma}_{\mathbf{n}}\hat{C}_n^k(\mathbf{n})$ using the expressions in Equation~\eqref{eq:improperconversion_kn} but with the primed and unprimed quantities swapped.

  \subsection{Equality of \texttt{SymOp} instances}

\begin{figure*}
      \centering
      \includegraphics[]{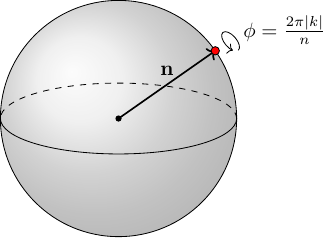}
      \caption{%
        The rotation pole and positive rotation angle of the rotation $\hat{C}_n^k(\mathbf{n})$.
      }
      \label{fig:rotationpole}
    \end{figure*}

    The $2\pi$-periodicity of spatial proper rotations must be accounted for by the equality comparisons of \texttt{SymOp} instances.
    In particular, this periodicity means that $\hat{C}_n^k(\mathbf{n}) = \hat{C}_n^{-k}(-\mathbf{n})$.
    To this end, the concept of rotation poles introduced in Chapter~9 of Ref.~\citenum{book:Altmann1986} is utilized where a unique point on the unit sphere in three dimensions is defined for every rotation.
    The rotation pole for a proper rotation $\hat{C}_n^k(\mathbf{n})$ implemented in \textsc{QSym\textsuperscript{2}} is given by
    \begin{equation*}
      \Pi[\hat{C}_n^k(\mathbf{n})] =
      \begin{cases}
        \hphantom{-}\mathbf{n} & \textrm{if $0 < k < n/2$}, \\
        \hphantom{-}\mathbf{n}_{+} & \textrm{if $n$ is even and $k = n/2$}, \\
        -\mathbf{n} & \textrm{if $k < 0$}, \\
        \hphantom{-}\mathbf{z} & \textrm{if $k = 0$},
      \end{cases}
    \end{equation*}
    where $\mathbf{z}$ is the Cartesian unit vector along the $z$-direction and $\mathbf{n}_{+}$ denotes whichever of $\mathbf{n}$ or $-\mathbf{n}$ that lies on a predefined positive unit hemisphere.
    The \emph{standard} positive unit hemisphere\cite{book:Altmann1986} contains all points that satisfy any one of the following conditions:
    \begin{enumerate}[label=(\roman*)]
      \item $n_z > 0$; or
      \item $n_z = 0, n_x > 0$; or
      \item $n_z = 0, n_x = 0, n_y > 0$.
    \end{enumerate}
    However, other definitions of positive unit hemispheres, including those in which the hemisphere consists of non-adjacent spherical wedges, are possible.
    Geometrically, one interprets the rotation pole of $\hat{C}_n^k(\mathbf{n})$ as the point on the unit sphere that is left invariant by $\hat{C}_n^k(\mathbf{n})$ and such that this rotation appears as anticlockwise through a positive angle $\phi = \frac{2\pi \lvert k \rvert}{n} \in \interval[open left]{0}{\pi}$ when observed from outside the sphere (Figure~\ref{fig:rotationpole})\cite{book:Altmann1986}.
    Thus, comparing $\hat{C}_n^k(\mathbf{n})$ operations where $k \ne 0$ is equivalent to comparing their poles $\Pi[\hat{C}_n^k(\mathbf{n})]$ and positive angles $\phi$.
    For $k = 0$, $\hat{C}_n^0(\mathbf{n})$ is simply the identity, irrespective of the values of $n$ and $\mathbf{n}$, and so all $\hat{C}_n^0(\mathbf{n})$ must be regarded as identical.

    It should also be noted that any improper rotation must be converted to the inversion-rotation form prior to equality comparisons, so that the pole of $\hat{i}\hat{C}_{n}^{k}(\mathbf{n})$ can be identified with that of $\hat{C}_{n}^{k}(\mathbf{n})$, as shown in Chapter~9 of Ref.~\citenum{book:Altmann1986}.
    Once this has been ensured, two symmetry operations $\hat{g} = \hat{t} \hat{s} \hat{C}_n^k(\mathbf{n})$ and $\hat{g}' = \hat{t}' \hat{s}' \hat{C}_{n'}^{k'}(\mathbf{n}')$ can easily be compared term-wise, where the comparison of $\hat{C}_n^k(\mathbf{n})$ and $\hat{C}_{n'}^{k'}(\mathbf{n}')$ is achieved via their poles and positive angles as described above.
    Consequently, the comparison of the corresponding $\texttt{SymOp}(\hat{g})$ and $\texttt{SymOp}(\hat{g}')$ instances is effected through comparisons of the \texttt{t} and \texttt{s} fields and of the poles and positive rotation angles calculated from the \texttt{frac} and \texttt{axis} fields.

  \subsection{Hashability of \texttt{SymOp} instances}

    In light of the above discussions, an equality-compatible hashing algorithm on the \texttt{SymOp} structure must take several factors into account.
    First, spatial improper rotations must be converted to the inversion-rotation form prior to hashing to ensure that the same improper rotation in whichever form is hashed to the same value.
    Then, the fields \texttt{t} and \texttt{s} must constitute two independent arguments in the hash function, since unitary and antiunitary operations must be hashed to different values, as must proper and improper operations.
    Finally, the \texttt{frac} and \texttt{axis} fields are used to compute the pole and positive rotation angle which then constitute two additional arguments in the hash function, while making sure that the identity is hashed uniquely.

    Vectors whose components are real numbers constitute the \texttt{axis} field and thus require a careful treatment in the comparison and hashing of \texttt{SymOp}, since comparing and hashing real numbers is normally discouraged because of the potential violation of transitivity.
    For instance, given a threshold $\epsilon$, two real vectors $\mathbf{u}$ and $\mathbf{v}$ are considered to be equal with respect to this threshold if $\lVert \mathbf{u} - \mathbf{v} \rVert \le \epsilon$.
    However, if there is another real vector $\mathbf{w}$ such that $\lVert \mathbf{v} - \mathbf{w} \rVert \le \epsilon$, but that $\lVert \mathbf{u} - \mathbf{w} \rVert > \epsilon$, then there is ambiguity in ascertaining the equality between $\mathbf{u}$ and $\mathbf{w}$ with respect to the same $\epsilon$.

    Fortunately, owing to the discrete nature of molecular symmetry, suitable choices for $\epsilon$ are possible whereby the above ambiguity is avoided when comparing and hashing poles of \texttt{SymOp}.
    In particular, if one chooses any $\epsilon < d_{\mathrm{min}} / 2$ where $d_{\mathrm{min}}$ is the minimum distance between any two poles of all symmetry operations in the system, then the $\epsilon$-neighborhoods\cite{book:Kolmogorov1970} of symmetry poles are all disjoint.
    Then, all vectors within the same $\epsilon$-neighborhood of a pole are considered equal to one another and hashed to the same value.

  \subsection{Compositability of \texttt{SymOp} instances}

    Let $\hat{g} = \hat{t} \hat{s} \hat{C}_n^k(\mathbf{n})$ and $\hat{g}' = \hat{t}' \hat{s}' \hat{C}_{n'}^{k'}(\mathbf{n}')$ be two symmetry operations.
    It can be seen from Equation~\eqref{eq:antiunitaryterm} that the potentially antiunitary operator $\hat{t}$ always commutes with the spatial unitary operators $\hat{s}$ and $\hat{C}_n^k(\mathbf{n})$ since $\hat{t} = \hat{\theta}$ only acts on spin coordinates.
    However, the potentially improper operator $\hat{s}$ need not commute with $\hat{C}_n^k(\mathbf{n})$ because reflections and rotations in general do not commute.
    Fortunately, spatial inversion does commute with rotations, so commutativity between $\hat{s}$ and $\hat{C}_n^k(\mathbf{n})$ can be guaranteed by ensuring that $\hat{g}$ is expressed as an inversion-rotation if it is an improper operation.
    With these commutativity conditions satisfied, the composition between $\hat{g}$ and $\hat{g}'$ is given by
    \begin{equation*}
      \hat{g}\hat{g}' = (\hat{t}\hat{t}') (\hat{s}\hat{s}') \left[ \hat{C}_n^k(\mathbf{n})  \hat{C}_{n'}^{k'}(\mathbf{n}')\right].
    \end{equation*}
    The compositions $\hat{t}\hat{t}'$ and $\hat{s}\hat{s}'$ are trivial to compute.
    On the other hand, the rotation composition $\hat{C}_n^k(\mathbf{n})  \hat{C}_{n'}^{k'}(\mathbf{n}')$ is handled via quaternion algebra\cite{book:Altmann1986} in which each rotation is represented by a normalized quaternion:
    \begin{equation*}
      \hat{C}_n^k(\mathbf{n})
      \equiv [\![ \lambda, \mathbf{\Lambda} ]\!],
    \end{equation*}
    where
    \begin{equation*}
      \lambda = \cos\frac{k\pi}{n}, \quad
      \mathbf{\Lambda} = \mathbf{n} \sin\frac{k\pi}{n}.
    \end{equation*}
    The composition can then be computed algebraically:
    \begin{align}
      \hat{C}_n^k(\mathbf{n})  \hat{C}_{n'}^{k'}(\mathbf{n}')
      &\equiv [\![ \lambda, \mathbf{\Lambda} ]\!] [\![ \lambda', \mathbf{\Lambda}' ]\!] \nonumber \\
      &= [\![
        \lambda\lambda' - \mathbf{\Lambda} \cdot \mathbf{\Lambda}',
        \lambda\mathbf{\Lambda}' + \lambda'\mathbf{\Lambda} + \mathbf{\Lambda} \times \mathbf{\Lambda}'
      ]\!],
    \end{align}
    and then translated back into the encoding $\hat{C}_{n''}^{k''}(\mathbf{n}'')$ where $\mathbf{n}''$ is normalized and $k'', n''$ are integers satisfying the same conditions as in Equation~\eqref{eq:knconditions}.
    The integrality of $k''$ and $n''$ is guaranteed by the closure of the prevailing finite molecular symmetry group: the composition of two symmetry rotations of a finite molecular system must yield another symmetry rotation of the same system.

%% file: suppinfo/chartabgen/chartabgen.tex
\section{Algorithms for character table generation}
\label{app:chartabgen}

  In this Section, the algorithms for automatic character table computation implemented in \textsc{QSym\textsuperscript{2}} are briefly described to highlight their main ideas.
  The readers are referred to the references therein for more detailed and technical descriptions.

  \subsection{The eigenvalue problem for characters}

    Consider a group $\mathcal{G}$ with $k$ unitary conjugacy classes and hence $k$ irreducible representations.
    For any conjugacy classes $K_r$, $K_s$, and $K_t$, and for \emph{any} $g_t \in K_t$, define
    \begin{equation*}
      n_{rst} = \left\lvert\{
        (g_r, g_s) \in K_r \times K_s : g_r g_s = g_t
      \}\right\rvert,
    \end{equation*}
    which can be shown to be \emph{independent of the choice of $g_t \in K_t$}.
    The numbers $n_{rst}$ can be gathered into $k$ \emph{class matrices} whose entries are all positive integers:
    \begin{equation*}
      \mathbf{N}_r = [n_{rst}] \in \mathbb{N}^{k^2}, \quad r = 1, \ldots, k.
    \end{equation*}
    The $k$ class matrices can be determined efficiently from the group's Cayley table using Schneider's method\cite{article:Schneider1990} and are key to the determination of irreducible representation characters, as shall be seen shortly.

    Next, let $\rho_i$ be the $i$\textsuperscript{th} irreducible representation of $\mathcal{G}$ with the corresponding character function $\chi_i$.
    For the conjugacy class $K_j$, define
    \begin{equation*}
      \omega_{ij} = \frac{\lvert K_j \rvert \chi_i(K_j)}{\chi_i(K_1)},
    \end{equation*}
    where $\lvert K_j \rvert$ denotes the size of $K_j$ and $K_1$ the identity class of $\mathcal{G}$.
    It can then be shown that
    \begin{subequations}
      \begin{equation}
        \sum_{t} n_{rst} \omega_{it} = \omega_{ir} \omega_{is},
      \end{equation}
      or, in matrix notations,
      \begin{equation}
        \mathbf{N}_r \boldsymbol{\omega}_i = \omega_{ir} \boldsymbol{\omega}_i,
      \end{equation}
      \label{eq:classmateigeq}
    \end{subequations}
    and
    \begin{equation}
      \sum_{r=1}^{k} \frac{\omega_{ir}\omega_{jr'}}{\lvert K_r \rvert}
      = \frac{\delta_{ij} \lvert \mathcal{G} \rvert}{\chi_i(K_1)\chi_j(K_1)},
      \label{eq:orthoomega}
    \end{equation}
    where $K_{r'} = K_r^{-1}$ is the inverse conjugacy class of $K_r$ (\textit{i.e.}, the elements in $K_{r'}$ are all inverses of the elements in $K_r$).
    Equation~\eqref{eq:classmateigeq} indicates that the $\boldsymbol{\omega}_i$'s are all eigenvectors of each $\mathbf{N}_r$, and that the $\omega_{ir}$'s are also eigenvalues of $\mathbf{N}_r$.
    Therefore, in principle, as long as any one of the $k$ class matrices has been computed, its linearly independent eigenvectors can be found and, with the help of Equation~\eqref{eq:orthoomega}, characters of all irreducible representations in the group can be determined.
    This is known as the Burnside algorithm\cite{book:Grove1997}.

  \subsection{Diagonalization over a finite field}

    The fact that the $\omega_{ir}$'s are eigenvalues of $\mathbf{N}_r$ means that they must be roots of the characteristic polynomials of the class matrices:
    \begin{equation*}
      p_{\mathbf{N}_r}(\lambda) = \det (\lambda \mathbf{I} - \mathbf{N}_r) = 0.
    \end{equation*}
    But since the entries of the class matrices are all integers, the polynomials $p_{\mathbf{N}_r}(\lambda)$ must all have integer coefficients, which implies that their roots, \textit{i.e.}, the $\omega_{ir}$'s, must be \emph{algebraic integers}.
    Additionally, it can be shown that the characters $\chi_i$ are algebraic integers that can be written as sums of roots of unity.
    Therefore, algorithms that do not make use of this property and instead only na\"{\i}vely diagonalize the class matrices over the complex field $\mathbb{C}$ end up doing more work than needed and thus suffer from an unnecessary reduction of efficiency and accuracy.

    Recognising these properties and problems, Dixon\cite{article:Dixon1967,book:Grove1997} proposed a variation to the Burnside algorithm in which all expensive algebras in the character table construction for the group $\mathcal{G}$ are performed in a finite number field of prime order denoted $\mathsf{GF}(p)$, where $p$ is a prime number determined by the orders of the elements in $\mathcal{G}$.
    $\mathsf{GF}(p)$ can be most simply constructed as the field of integers modulo $p$, denoted $\mathbb{Z}/p\mathbb{Z}$.
    The finiteness of $\mathsf{GF}(p)$ allows all arithmetic operations, especially division, to be performed exactly, since their outcomes must reside within the field due to the field's closure properties.
    In \textsc{QSym\textsuperscript{2}}, modular arithmetic over $\mathsf{GF}(p)$ is carried out efficiently by making use of the Montgomery form\cite{article:Montgomery1985} and eigenvectors of class matrices are determined by Gaussian elimination.
    Once character values have been calculated exactly in $\mathsf{GF}(p)$, they can be stored, represented, and manipulated symbolically as sums of roots of unity, and are only lifted back to $\mathbb{C}$ when necessary.

%% file: suppinfo/densym/densym.tex
\section{Symmetry of electron densities}

  \begin{table}[H]
    \centering
    \caption{%
      Electronic wavefunction and total density symmetries of the lowest $M_S = -1$ state of \ce{H3^{+}}.
      Calculations were performed at the UHF and r\textsuperscript{2}SCAN0 levels of theory in the 6-311++(2+,2+)G** basis set.
      The magnitudes of the applied electric fields and magnetic fields are \SI{0.1}{\atomicunit} and $1.0 B_0$, respectively.
      Isosurfaces of total densities are plotted at $\lvert \rho(\mathbf{r}) \rvert = 0.050$.
    }
    \label{tab:densym-supp}
    \renewcommand*{\arraystretch}{0.4}
    \scalebox{0.95}{
      \begin{tabular}{%
        l M{1.25cm} M{1.95cm} M{2.50cm} M{2.50cm} M{3.25cm}
      }
        \toprule
        & Field
        & Symmetry group
        & Wavefunction symmetry
        & Density symmetry
        & Density isosurface\\
        \midrule
        UHF
        & $\mathbf{0}$
        & $\mathcal{D}_{3h}$
        & $A_{1}^{\prime} \oplus E^{\prime}$
        & $A_{1}^{\prime} \oplus E^{\prime}$
        & \includegraphics[trim=0.0cm 0.15cm 0.0cm 0.15cm, clip, scale=0.9]{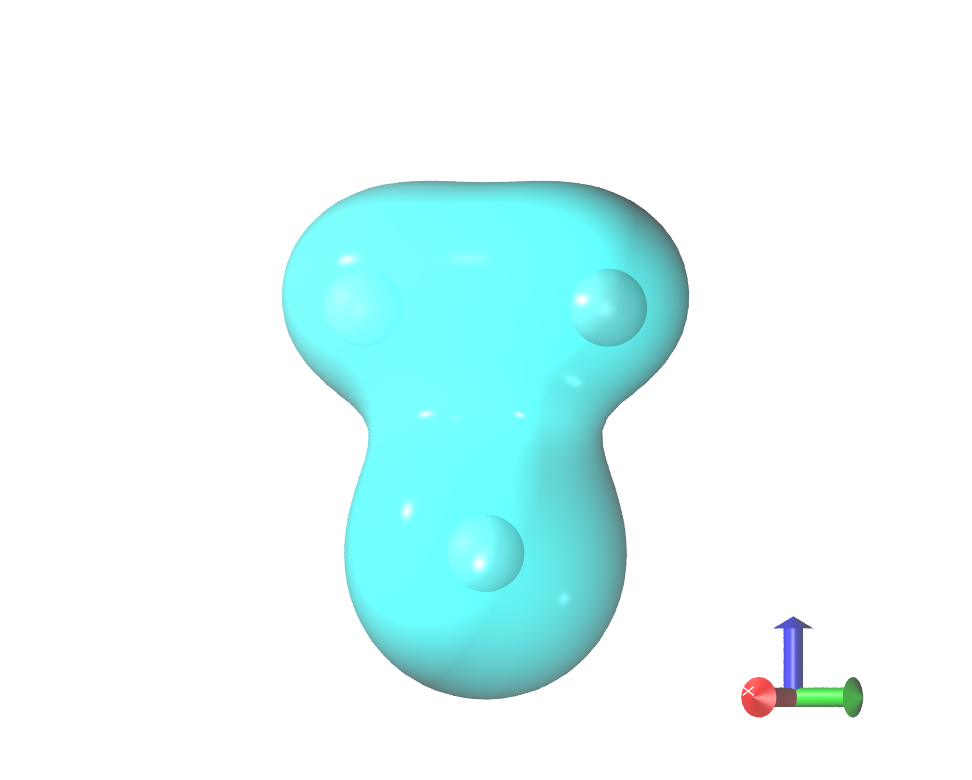}\\
        & $E_{x}$
        & $\mathcal{C}_{3v}$
        & $A_{1} \oplus E$
        & $A_{1} \oplus E$
        & \includegraphics[trim=0.0cm 0.15cm 0.0cm 0.15cm, clip, scale=0.9]{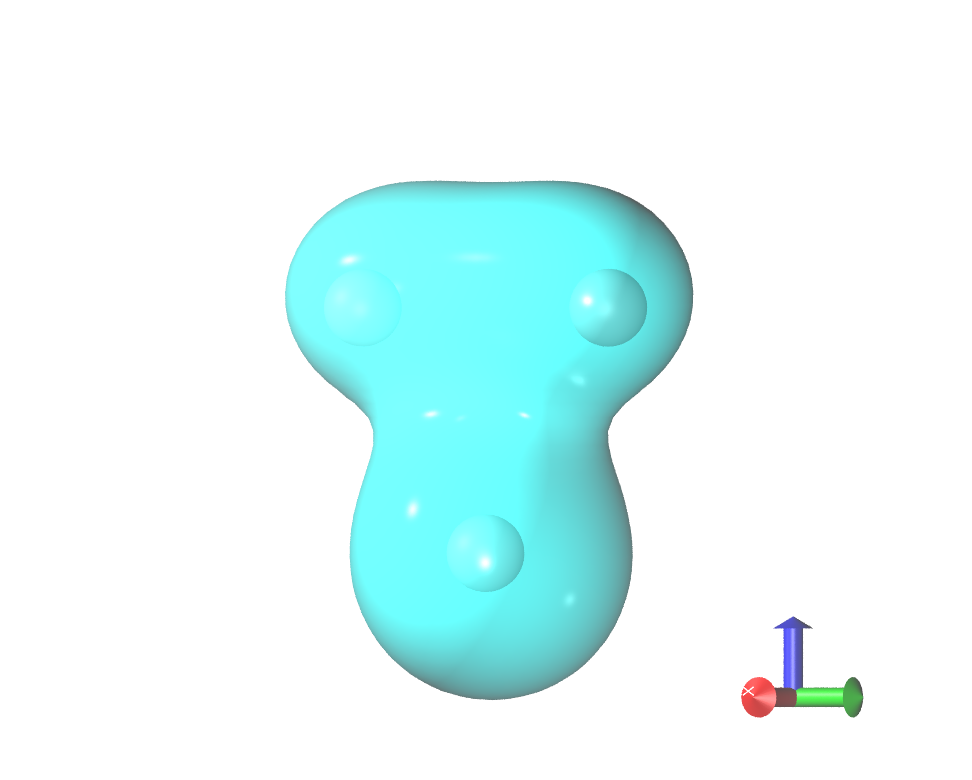}\\
        & $E_{y}$
        & $\mathcal{C}_{s}$
        & $A^{\prime}$
        & $A^{\prime}$
        & \includegraphics[trim=0.0cm 0.15cm 0.0cm 0.15cm, clip, scale=0.9]{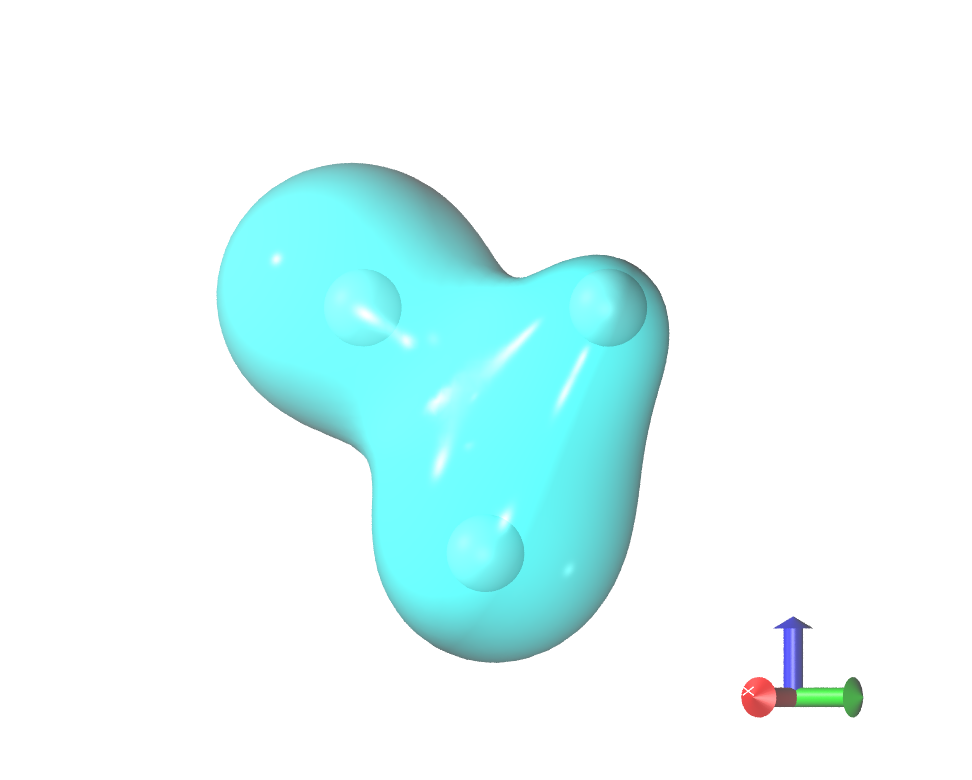}\\
        & $E_{z}$
        & $\mathcal{C}_{2v}$
        & $A_{1}$
        & $A_{1}$
        & \includegraphics[trim=0.0cm 0.15cm 0.0cm 0.15cm, clip, scale=0.9]{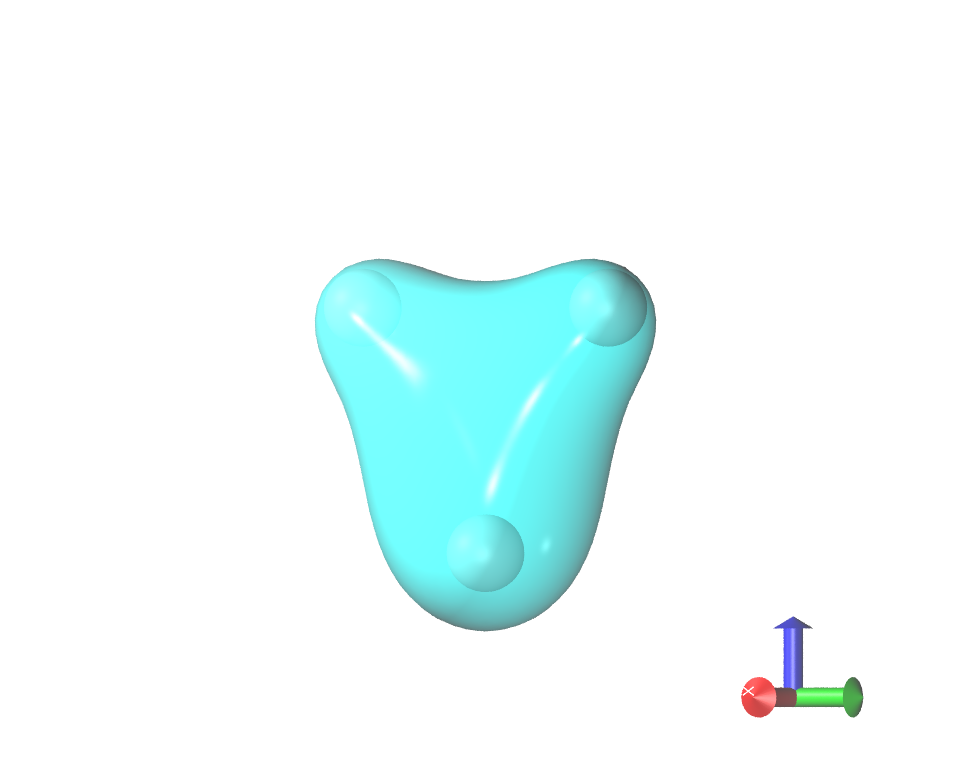}\\
        & $B_{x}$
        & $\mathcal{C}_{3h}$
        & $\Gamma^{\prime}$
        & $A^{\prime}$
        & \includegraphics[trim=0.0cm 0.15cm 0.0cm 0.15cm, clip, scale=0.9]{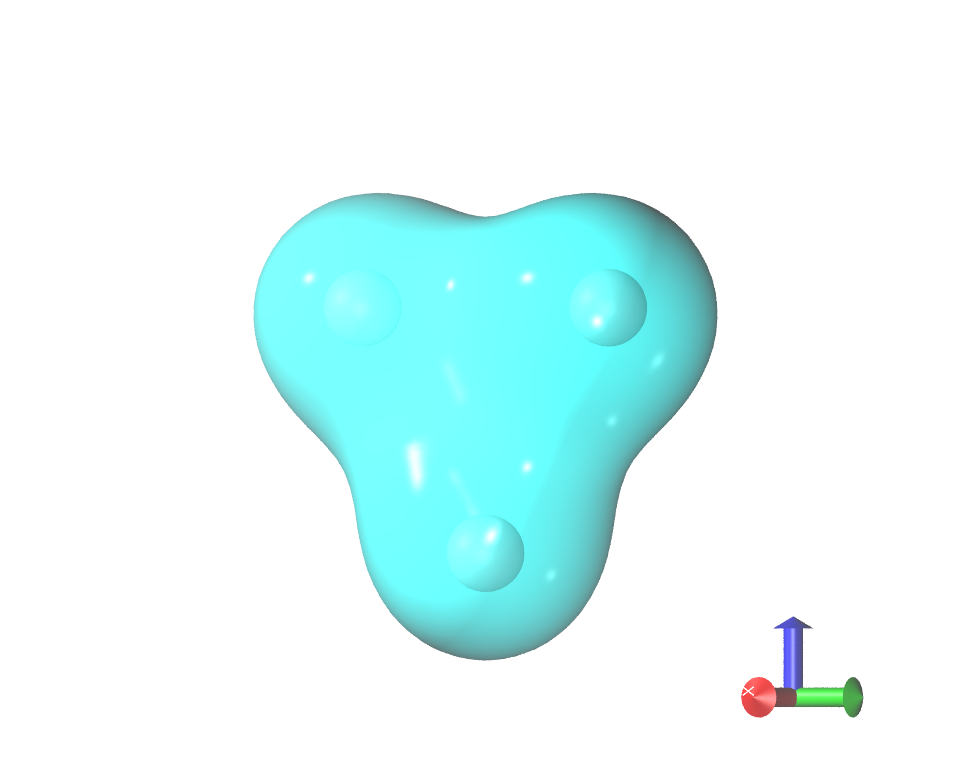}\\
        & $B_{y}$
        & $\mathcal{C}_{s}$
        & $A^{\prime}$
        & $A^{\prime}$
        & \includegraphics[trim=0.0cm 0.15cm 0.0cm 0.15cm, clip, scale=0.9]{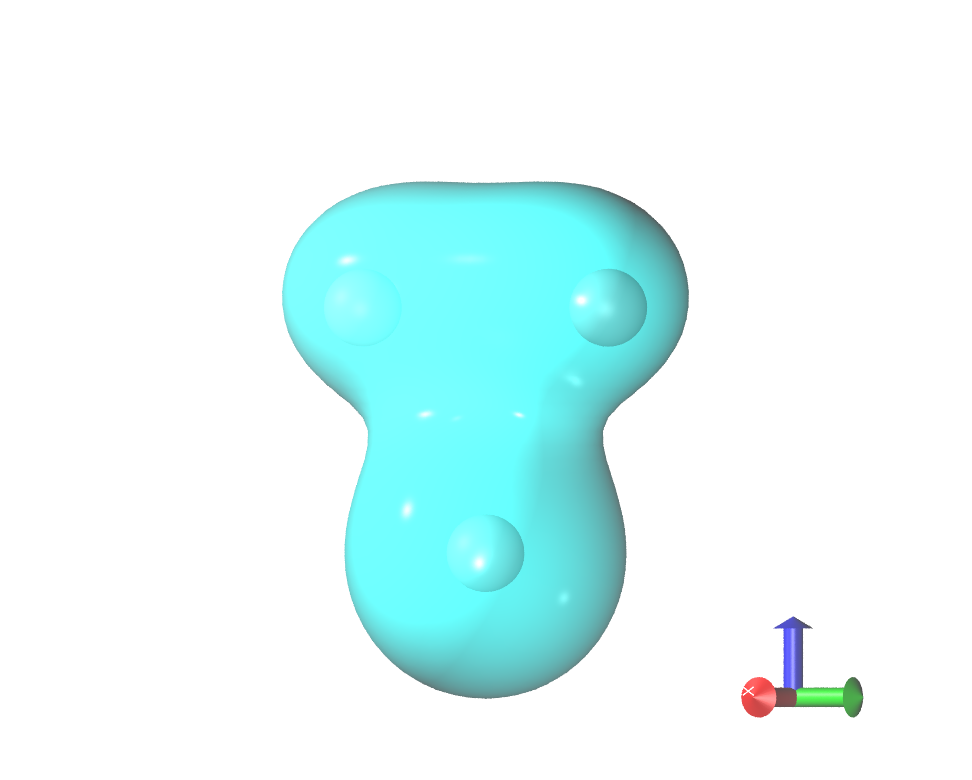}\\
        & $B_{z}$
        & $\mathcal{C}_{2}$
        & $B$
        & $A$
        & \includegraphics[trim=0.0cm 0.15cm 0.0cm 0.15cm, clip, scale=0.9]{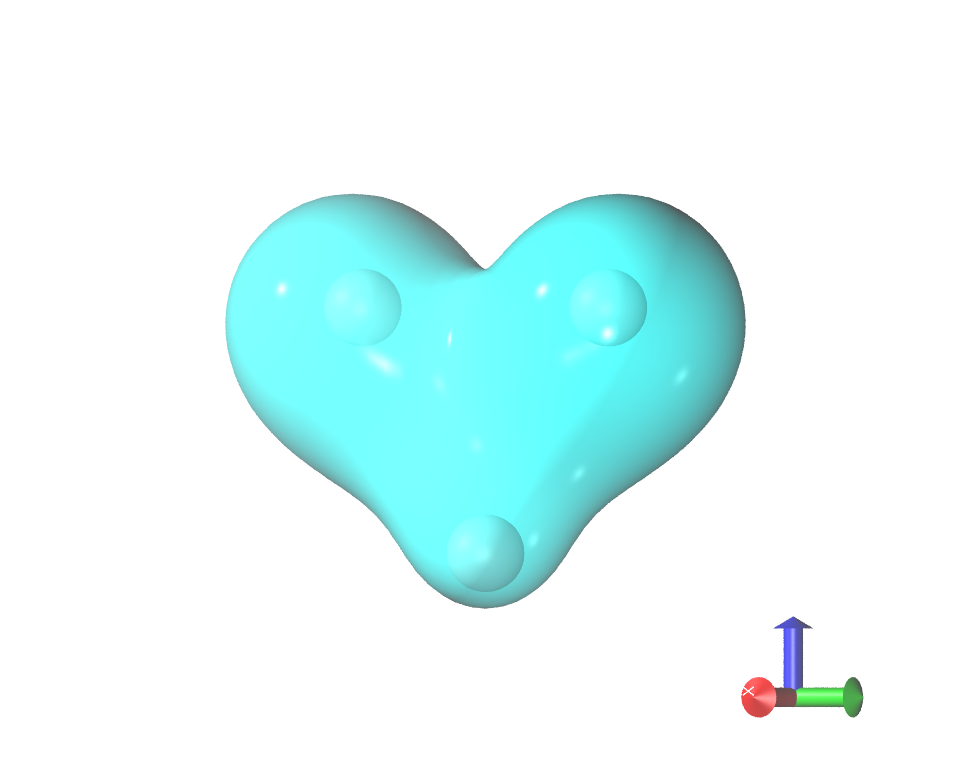}\\
        \midrule
        r$^2$SCAN0
        & $\mathbf{0}$
        & $\mathcal{D}_{3h}$
        & --
        & $A_{1}^{\prime} \oplus E^{\prime}$
        & \includegraphics[trim=0.0cm 0.15cm 0.0cm 0.15cm, clip, scale=0.9]{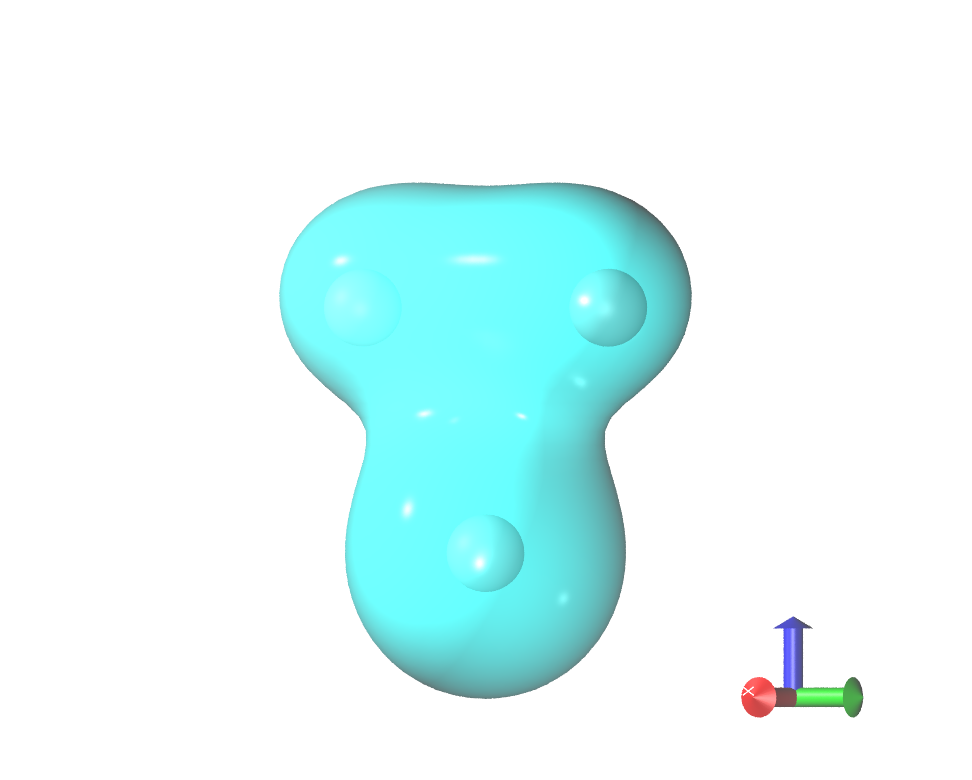}\\
        & $E_{x}$
        & $\mathcal{C}_{3v}$
        & --
        & $A_{1} \oplus E$
        & \includegraphics[trim=0.0cm 0.15cm 0.0cm 0.15cm, clip, scale=0.9]{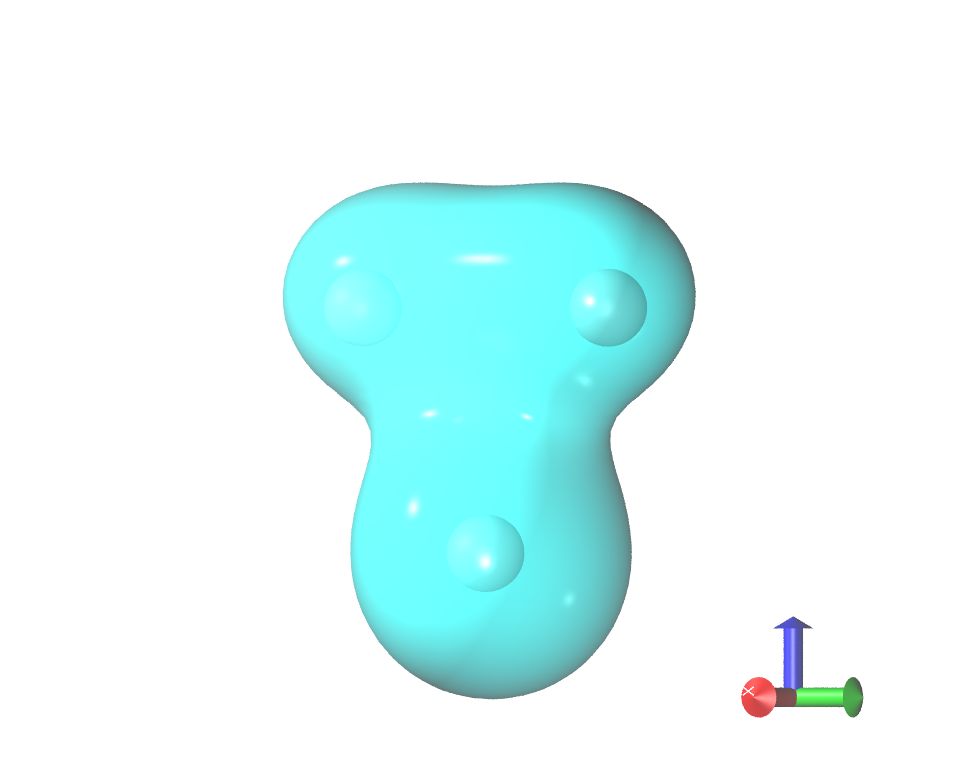}\\
        & $E_{y}$
        & $\mathcal{C}_{s}$
        & --
        & $A^{\prime}$
        & \includegraphics[trim=0.0cm 0.15cm 0.0cm 0.15cm, clip, scale=0.9]{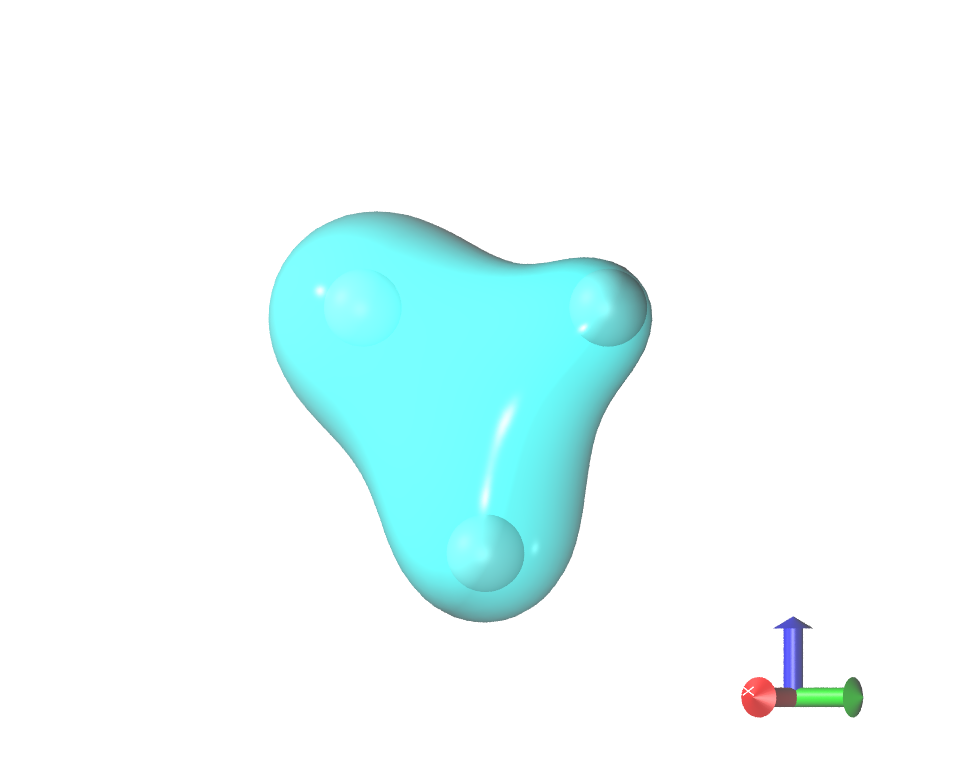}\\
        & $E_{z}$
        & $\mathcal{C}_{2v}$
        & --
        & $A_{1}$
        & \includegraphics[trim=0.0cm 0.15cm 0.0cm 0.15cm, clip, scale=0.9]{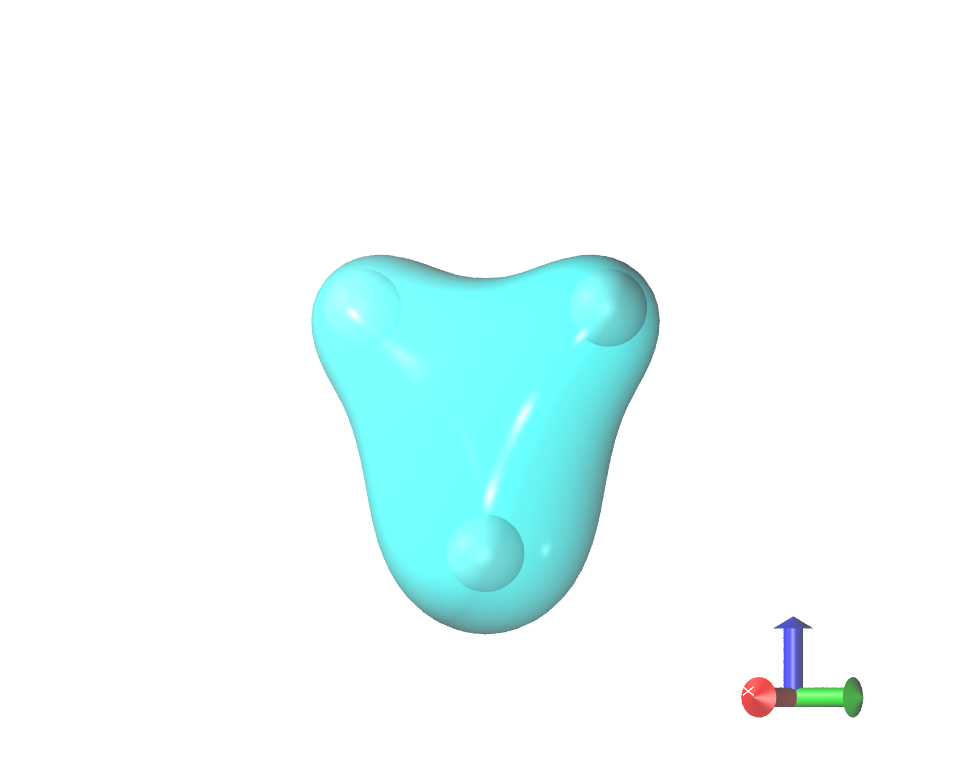}\\
        & $B_{x}$
        & $\mathcal{C}_{3h}$
        & --
        & $A^{\prime}$
        & \includegraphics[trim=0.0cm 0.15cm 0.0cm 0.15cm, clip, scale=0.9]{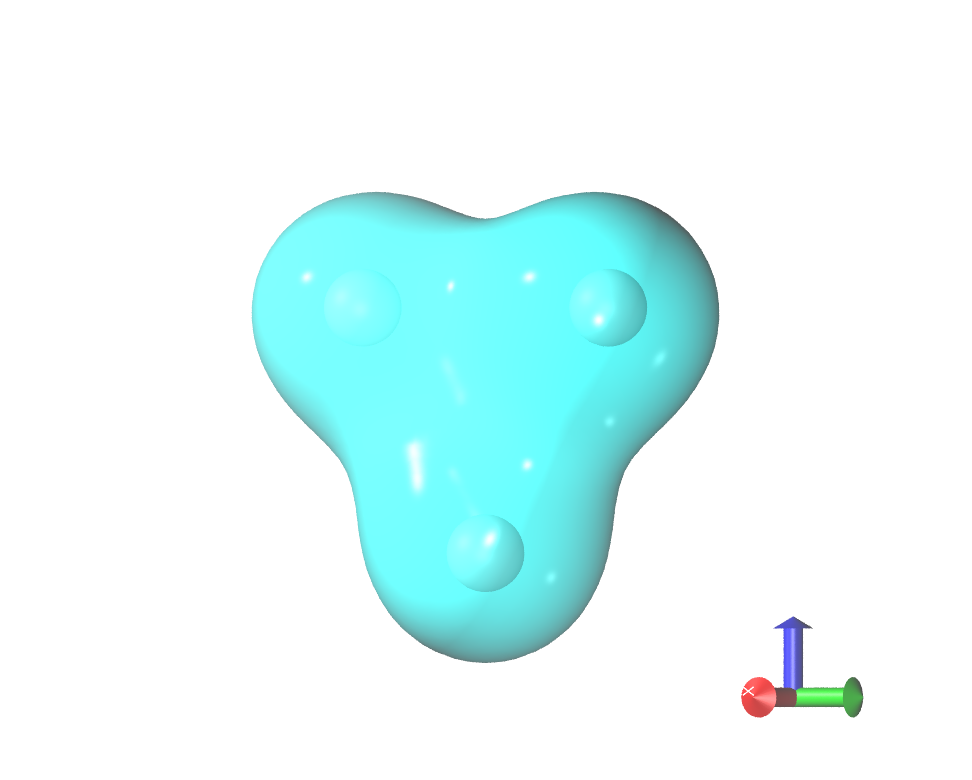}\\
        & $B_{y}$
        & $\mathcal{C}_{s}$
        & --
        & $A^{\prime}$
        & \includegraphics[trim=0.0cm 0.15cm 0.0cm 0.15cm, clip, scale=0.9]{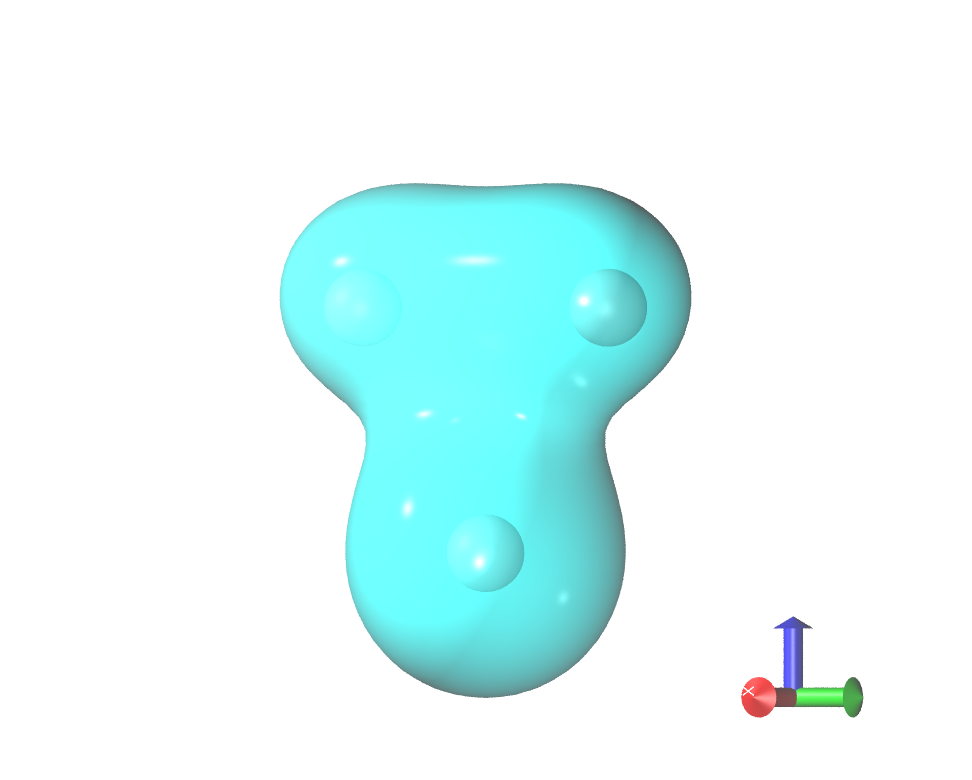}\\
        & $B_{z}$
        & $\mathcal{C}_{2}$
        & --
        & $A$
        & \includegraphics[trim=0.0cm 0.15cm 0.0cm 0.15cm, clip, scale=0.9]{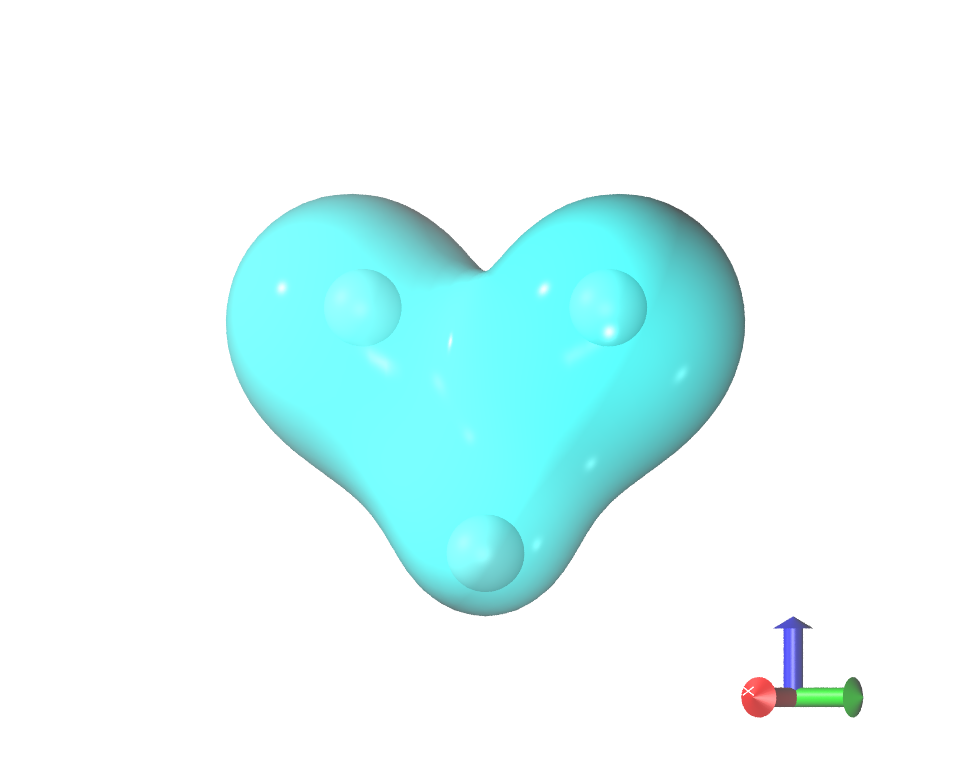}\\
        \bottomrule
      \end{tabular}
    }
  \end{table}


%% file: qsym2.bbl
\begin{mcitethebibliography}{109}
\bibitem[Not()]{Note-1}
Strictly speaking, the integrals obtained from these packages only give the
  spatial part of the spin-spatial integral in Equation~\eqref{eq:twoeov}. The
  spin part is typically handled separately and implicitly, especially for the
  commonly used spin functions $\alpha$ and $\beta$ whose orthogonality is
  known.\relax
\mciteBstWouldAddEndPunctfalse
\mciteSetBstMidEndSepPunct{\mcitedefaultmidpunct}
{}{\mcitedefaultseppunct}\relax
\EndOfBibitem
\bibitem[Not()]{Note-2}
This choice is merely for computational convenience because symmetry properties
  of MOs must be gauge-origin-invariant.\relax
\mciteBstWouldAddEndPunctfalse
\mciteSetBstMidEndSepPunct{\mcitedefaultmidpunct}
{}{\mcitedefaultseppunct}\relax
\EndOfBibitem
\end{mcitethebibliography}
